\documentclass[12pt]{article}%
\usepackage{amsfonts,array}
\usepackage{multicol,subcaption,makecell}
\usepackage{afterpage}
\usepackage{amsmath,bbm,dsfont,mathrsfs,mathtools,appendix}
\usepackage{amssymb}
\usepackage{graphicx}
\usepackage{fullpage}%
\usepackage{framed}
\usepackage[table]{xcolor}
\usepackage[colorlinks=true,linkcolor=blue,citecolor=green,plainpages=false,pdfpagelabels]%
{hyperref}%
\hypersetup{
	colorlinks,
	linkcolor={blue!60!green},
	citecolor={green!50!yellow!75!black},
	urlcolor={blue!80!black},
	linktoc=all
}
\providecommand{\U}[1]{\protect\rule{.1in}{.1in}}

\newtheorem{theorem}{Theorem}

\newtheorem{corollary}[theorem]{Corollary}

\newtheorem{definition}[theorem]{Definition}
\newtheorem{example}[theorem]{Example}

\newtheorem{lemma}[theorem]{Lemma}

\newtheorem{proposition}[theorem]{Proposition}
	
\newtheorem{remark}[theorem]{Remark}

\newenvironment{proof}[1][Proof]{\noindent\textbf{#1.} }{\ \rule{0.5em}{0.5em}}

\newcommand{\cF}{{\mathcal{F}}}

\newcommand{\id}{{\rm{id}}} 
\newcommand{\spec}{{\rm{spec}}}

\newcommand{\cB}{{\mathcal{B}}}

\newcommand{\cH}{\mathcal{H}}
\newcommand{\R}{\mathbbm{R}}
\newcommand{\C}{\mathbb{C}}

\newcommand{\N}{\mathbb{N}}

\newcommand{\nn}{\nonumber}

\newcommand{\cT}{\mathcal{T}}
\newcommand{\cC}{\mathcal{C}}
\newcommand{\cS}{\mathcal{S}}
\newcommand{\cR}{\mathcal{R}}

\newcommand{\cE}{\mathcal{E}}

\newcommand{\cX}{{\cal X}}

\newcommand{\cM}{\mathcal{M}}
\newcommand{\cN}{\mathcal{N}}

\newcommand{\cU}{\mathcal{U}}
\newcommand{\cY}{\mathcal{Y}}
\newcommand{\cZ}{\mathcal{Z}}
\newcommand{\1}{\mathbbm{1}}

\def\>{{\rangle}}
\def\<{{\langle}}
\newcommand{\be}{\begin{equation}}
	\newcommand{\ee}{\end{equation}}
\newcommand{\bea}{\begin{eqnarray}}
	\newcommand{\eea}{\end{eqnarray}}

\newcommand{\eps}{\varepsilon}

\newcommand{\ket}[1]{|#1\rangle} 
\newcommand{\bra}[1]{\langle#1|} 
\newcommand{\kb}[1]{|#1\rangle\!\langle#1|} 

\def\placeholder{\,\cdot\,}
\newcommand{\Tr}{\mathrm{Tr}}

\newcommand{\comment}[1]{}

\newcommand{\ran}{\operatorname{ran}}

\newcommand{\CPTP}{{\rm{CPTP}}}
\newcommand{\Prob}{{\rm{Prob}}}
\newcommand{\Stoch}{{\rm{Stoch}}}

\numberwithin{equation}{section}
\numberwithin{theorem}{section}

\usepackage{color}
\definecolor{colorthree}{rgb}{0.01,0.51,0.93}

\def\smallsection#1{\bigskip\noindent\textbf{#1}}

\usepackage{setspace}

\newcommand{\footremember}[2]{%
    \footnote{#2}
    \newcounter{#1}
    \setcounter{#1}{\value{footnote}}%
}
\newcommand{\footrecall}[1]{%
    \footnotemark[\value{#1}]%
} 

\allowdisplaybreaks
\usepackage{subfiles}

\title{
Robustness of Fixed Points of Quantum Channels and Application to Approximate Quantum Markov Chains}

\author{Robert Salzmann \footremember{Cam}{Department of Applied Mathematics and Theoretical Physics, University of Cambridge, United Kingdom}\footremember{Lyon}{Univ Lyon, Inria, ENS Lyon, UCBL, LIP, F-69342, Lyon Cedex 07, France}  \and Bjarne Bergh\footrecall{Cam}   \and  Nilanjana Datta \footrecall{Cam}}

\usepackage{tablefootnote}
\begin{document}
\maketitle
\begin{abstract}
Given a quantum channel
and a state which satisfy a fixed point equation approximately (say, up to an error $\eps$), can one find a new channel
and a state, which are respectively close to the original ones, such that they satisfy an
{\em{exact}} fixed point equation? It is interesting to ask this question for different choices of constraints on the structures of the original channel and state,  
and requiring that these are also satisfied by the new channel and state. 
We affirmatively answer the above question, under fairly general assumptions on these structures, through a compactness argument. Additionally, for channels and states satisfying certain
specific structures, we find explicit upper bounds on the distances between the pairs of channels (and states) in question. When these distances decay quickly (in a particular, desirable manner) as $\eps\to 0$, we say that the original approximate fixed point equation is {\em{rapidly fixable}}. We establish rapid fixability, not only for general quantum channels, but also when the original and new channels are both required to be unitary, mixed unitary or unital. In contrast, for the case of bipartite quantum systems with channels acting trivially on one subsystem, we prove that approximate fixed point equations are not rapidly fixable. In this case, the distance to the closest channel (and state) which satisfy an exact fixed point equation can depend on the dimension of the quantum system in an undesirable way.
We apply our results on approximate fixed point equations to the question of robustness of quantum Markov chains (QMC) and
establish the following: For any tripartite quantum state, 
there exists a dimension-dependent upper bound on its distance to the set of QMCs, which decays to zero as the conditional
mutual information of the state vanishes.

\end{abstract}

\tableofcontents

\section{Introduction}

In quantum theory, possible preparations of a physical system are mathematically described by so-called \emph{quantum states}. Quantum processes or evolutions of a physical system are described by \emph{quantum channels}.\footnote{In the following we often refer to quantum states as states and quantum channels as channels for brevity.} A quantum channel $\cN$ accepts the initial state $\rho$ of the system as input and yields a new state $\sigma$ as output. Invariance of the system under this quantum process is hence mathematically described by $\rho$ and $\cN$ satisfying the fixed point equation
\begin{align}
\label{eq:FixedPointIntroThesis}
    \cN(\rho) = \rho.
\end{align}
The set of fixed points of a channel exhibits rich mathematical properties which have been extensively studied in the existing literature, see e.g.~\cite{Morozova_StatofStoch_1976,KoashiImoto,hayden_structure_2004,Wolf12,Burgarth_2013,watrous_2018} and references therein.
Remarkably, in certain situations it is possible to deduce interesting mathematical structures that both the state and the channel mutually conform to when they satisfy a fixed point equation. This was the key observation of \cite{KoashiImoto,hayden_structure_2004} providing a characterisation of so-called \emph{(short) Quantum Markov chains} (QMC), i.e.~tripartite quantum states $\rho_{ABC}$ which have zero conditional mutual information $I(A:C|B)_\rho=0$. In particular, there it is shown that QMCs have a certain structure, which generalises in a meaningful way the well-known conditional independence satisfied by classical Markov chains to the quantum case. For a detailed discussion of quantum Markov chains and their approximate versions mentioned below, see Section~\ref{sec:QMCSection}. 

Tripartite states, $\rho_{ABC}$, for which the conditional mutual information $I(A:C|B)_\rho = \eps$ with $\eps$ small but strictly positive, 
 are called \emph{approximate quantum Markov chains} (AQMC). Such states appear naturally as Gibbs states of local Hamiltonians \cite{KatoBrandao_AQMCThermal_19} showcasing their relevance in condensed matter physics. AQMCs have the property that they 
 satisfy a certain fixed point equation approximately, i.e.~up to a small error \cite{FawziRenner_CMIRecovery_2015,Brandao_CMI_2015,Wilde_RecoveryInterpolation_2015,Sutter_UniversalRecoveryCMI_2016,Sutter_Multivariate_2017,JungeSutterwilde_UniversalRecovery_2018,Sutter_approximateThesis_2018}. It is hence natural to ask whether AQMCs in some sense approximately conform to the structure of QMCs mentioned above.

With this motivation, we study in this paper, states and channels for which 
the fixed point equation~\eqref{eq:FixedPointIntroThesis} holds approximately, that is only up to a small error $\eps> 0.$ We say that a state $\rho$ and a channel $\cN$ satisfy an \emph{approximate fixed point equation}, denoted by
\begin{align}
\label{eq:ApproximateFixedIntroThesis}
    \cN(\rho)\approx^{\eps} \rho,
\end{align}
if the trace distance of the output of the channel and $\rho$ is upper bounded as $\frac{1}{2}\left\|\cN(\rho)-\rho\right\|_1\le \eps.$  
\comment{The ultimate goal of Chapter~\ref{chap:FixPoint} is to deduce properties that both a state and a channel must adhere to when assuming that the approximate fixed point equation~\eqref{eq:ApproximateFixedIntroThesis} is valid. }

Firstly, we can readily observe that \eqref{eq:ApproximateFixedIntroThesis} does not necessarily imply that $\rho$ is close to an exact fixed point of $\cN.$ Furthermore, exclusively varying the channel does in general also not lead to an exact fixed point equation: In fact, we can identify instances where $\rho$ and $\cN$ satisfy the approximate fixed point equation \eqref{eq:ApproximateFixedIntroThesis} but any channel $\cM$ which has $\rho$ as an exact fixed point must be far from $\cN$ (cf.~Example~\ref{ex:ChangeBothToFix}).

We are therefore concerned with the following broader question for a state $\rho$ a channel $\cN$ such that \eqref{eq:ApproximateFixedIntroThesis} holds: \emph{Can we find a new state $\sigma$ and a new channel $\cM$ which are respectively close to the original ones,\footnote{Here, we used the notation of approximate equality writing $\rho\approx^{\eps}\sigma$ for two states $\rho,\sigma$ if $\frac{1}{2}\|\rho-\sigma\|_1\le\eps$ and $\cN\approx^{\eps}\cM$ for two channels $\cN,\cM$ if $\frac{1}{2}\|\cN-\cM\|_\diamond\le\eps$.}
\begin{align}
\label{eq:SigmaMThesisIntro}
    \sigma\approx^{f(d,\eps)} \rho,\quad\quad\cM\approx^{g(d,\eps)}\cN
\end{align}
such that the exact fixed point equation 
\begin{align}
    \cM(\sigma) =\sigma
\end{align}
holds?} Here, $f$ and $g$ denote 
\emph{approximation functions} 
which, in general, can depend on the dimension $d$ of the considered quantum system and should satisfy $\lim_{\eps\to 0}f(d,\eps)=\lim_{\eps\to 0}g(d,\eps)=0.$\footnote{Furthermore, the functional form of $f$ and $g$ should not depend on the particular state $\rho$ and channel $\cN$ considered as the question otherwise becomes meaningless. For a more precise formulation of the problem, please refer to Definition~\ref{def:FixingApproxFixPoints}. } 

This question can be asked 
under the condition that $\rho$ and $\cN$ satisfy certain natural constraints, with the requirement that the new state and channel need to satisfy these constraints as well. For instance, we might want to explore the scenario where the channels under consideration are unitary channels, unital channels or local channels on some bipartite quantum system. These constraints can be represented by subsets $\cS$ and $\cC$ of the full set of states and channels and we demand that both the original and new states (resp.~channels) are elements of $\cS$ (resp.~$\cC$). We say \emph{approximate fixed points are fixable} for a specific choice of $\cS$ and $\cC$ if we can always find such a $\sigma$ and $\cM$ (cf.~Definition~\ref{def:FixingApproxFixPoints}). It is important to note that the constraints given by the sets $\cS$ and $\cC$ influence the difficulty of the problem in two ways: Firstly, as a promise that the original state and channel satisfy $\rho\in\cS$ and $\cN\in\cC$, and secondly, as a requirement that the new state and channel should also satisfy $\sigma\in\cS$ and $\cM\in\cC$. Therefore, proving that approximate fixed point equations are fixable for a specific choice $\cS_1$ and $\cC_1$ does not necessarily imply the same for another choice $\cS_2\subseteq \cS_1$ or $\cC_2\subseteq \cC_1,$ or vice versa. Furthermore,  the approximation functions in \eqref{eq:SigmaMThesisIntro} for $\cS_1$ and $\cC_1$ and  $\cS_2\subseteq \cS_1$ and $\cC_2\subseteq \cC_1$ are in general not related, i.e. we cannot bound the approximation functions for $\cS_2$ and $\cC_2$ by the ones for $\cS_1$ and $\cC_1.$ 

\comment{The motivation to study this question is to demonstrate that states and channels satisfying an approximate fixed point equation inherit certain structures, up to small errors, which are present in the case of an exact fixed point equation. }

\subsection{Summary of main results}

For finite dimensional quantum systems and under the fairly general assumption that $\cS$ and $\cC$ are closed sets which contain a fixed point pair,\footnote{I.e.~a state $\sigma_\star\in\cS$ and a channel $\cM_\star\in\cC$ such that $\cM_\star(\sigma_\star)=\sigma_\star$} we show that approximate fixed point equations are always fixable in the above sense (Proposition~\ref{prop:AbstractFixingApproxFix}). Our argument is based on a compactness argument and our result implies the existence of approximation functions $f$ and $g$ satisfying \eqref{eq:SigmaMThesisIntro}.

This result can be applied to offer a partial answer to a long-standing open problem in the robustness theory of quantum Markov chains: For every tripartite state $\rho_{ABC}$ 
for which $I(A:C|B)_\rho\le \eps$, for some $\eps>0,$ we can find an {\em{exact}} quantum Markov chain $\sigma_{ABC}$ close to it (Proposition~\ref{prop:RobustnessQMC}). In particular, we show the existence of a function $f(d_{ABC},\eps)$ which satisfies $\lim_{\eps\to 0}f(d_{ABC},\eps)=0,$ and upper bounds the minimal distance of $\rho_{ABC}$ to the set of quantum Markov chains as
 \begin{align}
 \label{eq:RobQMCIntro}
  \min_{\sigma_{ABC}\ \text{QMC}}\frac{1}{2}\left\|\rho_{ABC}-\sigma_{ABC}\right\|_1 \le f(d_{ABC},\eps), 
 \end{align}
 for every tripartite state $\rho_{ABC}$ for which $I(A:C|B)_\rho\le\eps$. In the above, $d_{ABC}$ denotes the dimension of the tripartite system $ABC$.
 Providing a more explicit form for this upper bound remains, however, an open problem. From specific examples examined in the literature  \cite{IbinsonWinter_RobustnessQMC_2008,christandl_entanglement_2012,Sutter_approximateThesis_2018}, it is nonetheless evident that the upper bound necessarily  depends on the dimensions of the systems.

\bigskip

We then focus on examples of $\cS$ and $\cC$ for which we can provide explicit control of the approximation functions $f$ and $g$ in \eqref{eq:SigmaMThesisIntro}. If $f$ and $g$ can be shown to be of a specific form, for which the dimension $d$ and the error parameter $\eps$ interact rather mildly and the decay as $\eps\to 0$ is rather fast (as detailed in Definition~\ref{def:RapidFixingApproxFixedPoint}), we say that the approximate fixed point equations are \emph{rapidly fixable}.

We show rapid fixability of approximate fixed points for a variety of interesting constraints on the states and channels as summarised in the following table:

\begin{center}
\begin{tabular}{c|c|c|c|c}
     \cellcolor{gray!35}$\cS$ &\cellcolor{gray!35}$\cC$ & \cellcolor{gray!35}$f(d,\eps)$ & \cellcolor{gray!35}$g(d,\eps)$ & \cellcolor{gray!35}\makecell{\textbf{Shown in}}\\ \hline \makecell{$\mathfrak{S}(\cH)$} &\makecell{$\CPTP(\cH)$} & \makecell{$\sqrt{\eps}$} & \makecell{$\sqrt{\eps}$} & 
     \makecell{Thm.~\ref{thm:FixedPointQuantum} }\\\hline \makecell{$\Prob(\cX)$} &\makecell{$\Stoch(\cX)$} &\makecell{$\sqrt{\eps}$} & \makecell{$\sqrt{\eps}$} & \makecell{Thm.~\ref{thm:FixClassical}}\\\hline \makecell{$\mathfrak{S}(\cH)$} & \makecell{Unitary channels} & \makecell{$4d^{5/4}\sqrt{\eps}$} &  \makecell{$4d^{5/4}\sqrt{\eps}$  } & \makecell{Thm.~\ref{thm:FixUnitary}}\\\hline\makecell{$\mathfrak{S}(\cH)$} & \makecell{Mixed unitary channels} & \makecell{$4d^{2}\eps^{1/5}$} &  \makecell{$7d^2\eps^{1/5}$  } & \makecell{Thm.~\ref{thm:FixingMixtureUnitaries}} \\\hline\makecell{$\mathfrak{S}(\cH)$} & \makecell{Unital channels} & \makecell{$7d^{5/3}\eps^{1/6}$} &  \makecell{$7d^{5/3}\eps^{1/6}$  } & \makecell{Thm.~\ref{thm:FixingUnitalChannels}} \\\hline\makecell{$\operatorname{Pure}(\cH_{AB})$} & \makecell{$\id_A\otimes{\rm{CPTP}}(\cH_B)$} & \makecell{$7\min\{d_A,d_B\}\eps^{1/3}$} &  \makecell{$7\min\{d_A,d_B\}\eps^{1/3}$  } & \makecell{Thm.~\ref{thm:ApproxLocalFixPure}} 
\end{tabular}
\captionof{table}{Summary of results on rapid fixability for different choices of subsets of states $\cS$ and channels $\cC$ with corresponding approximation functions $f(d,\eps)$ and $g(d,\eps).$ Here, the set of quantum states on $\cH$ is denoted by $\mathfrak{S}(\cH),$ the set of pure states by $\operatorname{Pure}(\cH),$ the set of quantum channels by $\CPTP(\cH)$ and for some bipartite Hilbert space $\cH_{AB}\equiv \cH_A\otimes\cH_B$ we defined the set $\id_A\otimes{\rm{CPTP}}(\cH_B)\coloneqq\left\{\id_A\otimes\cN_B\,\big|\,\cN_B\in{\rm{CPTP}}(\cH_B)\right\}. $ Furthermore, $\Prob(\cX)$ denotes the set of probability distributions (or classical states)  and $\Stoch(\cX)$ the set of stochastic mappings (or classical channels) on on some countable alphabet $\cX.$}
\end{center}

The first result in that direction 
pertains to the case in which $\cS$ and $\cC$ are respectively given by the full set of quantum states and channels on some separable, possibly infinite dimensional Hilbert space (Theorem~\ref{thm:FixedPointQuantum}). 
The key insight for the proof is to compose the original channel $\cN$ with a particular choice of generalised depolarising channel. We then show the resulting channel $\cM$ has a unique fixed point state $\sigma$ which is close to the original state $\rho.$  Notably, the new state depends on both the original state $\rho$ and channel $\cN$ in a highly non-linear way. 
The same proof technique can also be applied in the classical case where the state and channel can respectively be represented by a probability distribution and a stochastic mapping on some countable alphabet (see Theorem~\ref{thm:FixClassical}). The scaling of the approximation functions found in Theorems~\ref{thm:FixedPointQuantum} and~\ref{thm:FixClassical} is essentially optimal as we show in Remark~\ref{rem:LowerBoundApprox}.

Additionally, we establish in the finite dimensional case rapid fixability for a range of other interesting choices for $\cS$ and $\cC$. Specifically, for $\cS$ being the full set of quantum states and $\cC$ being the set of \emph{1)} unitary channels, \emph{2)} mixed-unitary channels or \emph{3)} unital channels (see Theorems~\ref{thm:FixUnitary},~\ref{thm:FixingMixtureUnitaries} and~\ref{thm:FixingUnitalChannels}). Furthermore, we consider the setting of bipartite systems with $\cS$ 
being the set of pure bipartite states, and $\cC$ the set of local channels which act trivially on the first system (Theorem~\ref{thm:ApproxLocalFixPure}). All of these results follow by similar proof techniques, which include the application of what we call \emph{spectral clustering techniques}, \emph{rotation lemmas} and the \emph{generalised depolarising trick}. Here, spectral clustering techniques are used to define the new state $\sigma$ and the latter two methods are used to define the new channel $\cM,$ which fix the corresponding approximate fixed point equation. Interestingly, in contrast to the above-mentioned case of general states and channels, here the new fixed-point state is independent of the original channel.

\bigskip

Lastly, we provide an example of $\cS$ and $\cC$ where rapid fixability of approximate fixed points is impossible (Corollary~\ref{cor:ImpossiLocalFix}): We consider bipartite quantum systems with $\cS=\mathfrak{S}(\cH_{AB})$ being the set of general mixed states and $\cC=\id_A\otimes\CPTP(\cH_{B})$
 being once again the set of local channels which act trivially on the first system. In that case we provide an explicit counterexample for which the corresponding approximate fixed point equation can only be fixed with poor control on the respective approximation functions $f$ and $g.$ In particular they cannot be written in the form of 
$f(d,\eps) = c\,d^{b}\eps^{a}$, and similarly for $g(d,\eps)$, for some constants $a>0,$ $b,c\ge0$ independent of $d$ and $\eps.$ This shows, by contrasting with the statement of Theorem~\ref{thm:ApproxLocalFixPure} for pure states and local channels in the bipartite setting mentioned above, that the situation dramatically changes when widening the set $\cS$ from pure states on $\cH_{AB}$ to the set of general mixed states. Interestingly, the mentioned counterexample is essentially classical and therefore also disproves the possibility of rapid fixability in the corresponding classical case. 
 
  To prove this result we naturally extend the question of (rapid) fixability of approximate fixed points to the situation of multiple approximate fixed point equations. Here, we consider multiple states and one channel, which all satisfy an approximate fixed point equation, and we ask whether there exist close-by new states and one universal channel which satisfies fixed point equations for all of the states. We then provide a concrete counter example showing that already in the case of two approximate fixed point equations rapid fixability is not possible in general (Theorems~\ref{thm:MultipleFixCounterEx} and \ref{thm:MultipleFixCounterExQuant}). This example is then encoded into the bipartite setting proving the above-mentioned Corollary~\ref{cor:ImpossiLocalFix}.

\smallsection{Outline of the rest of the paper}\\ 
\noindent In Section~\ref{chap:Preliminaries} we provide some basic notations and facts which are need for the derivations of the results of this paper.
In Section~\ref{sec:Setup} we give formal definitions for the fixability and rapid fixability of approximate fixed points (Definitions~\ref{def:FixingApproxFixPoints} and~\ref{def:RapidFixingApproxFixedPoint}) and discuss some elementary examples and observations. In Section~\ref{sec:ClosedFixAbstract} we show that for general closed sets $\cS\subseteq\mathfrak{S}(\cH)$ and $\cC\subseteq{\rm{CPTP}}(\cH)$ such that a fixed point pair exists, approximate fixed point equations are fixable in the sense of Definition~\ref{def:FixingApproxFixPoints} (Proposition~\ref{prop:AbstractFixingApproxFix}). We then review, as a motivating example for studying the fixability of approximate fixed points, the robustness theory of quantum Markov chains. We apply Proposition~\ref{prop:AbstractFixingApproxFix} in this context and obtain a robustness result for approximate quantum Markov chains of the form \eqref{eq:RobQMCIntro} in Proposition~\ref{prop:RobustnessQMC}.
In Section~\ref{sec:GeneralQuantandClass} we prove rapid fixability of approximate fixed points in the sense of Definition~\ref{def:RapidFixingApproxFixedPoint} for  general quantum states and quantum channels, i.e.~for the choice $\cS=\mathfrak{S}(\cH)$ and $\cC = {\rm{CPTP}}(\cH)$ (Theorem~\ref{thm:FixedPointQuantum}) and
 for classical states and classical channels, i.e.~for the choice $\cS={\rm{Prob}}(\cX)$ and $\cC = {\rm{Stoch}}(\cX)$ (Theorem~\ref{thm:FixClassical}).

In Section~\ref{sec:FixStrictSubUnitaryEtc} we show rapid fixability of approximate fixed points for a variety of different natural choices of $\cS$ and $\cC,$ which are all strict subsets of the full sets of states and channels. In particular, we consider $\cS=\mathfrak{S}(\cH)$ together with $\cC$ being the set of unitary channels (Theorem~\ref{thm:FixingUnitalChannels}), mixed-unitary channels (Theorem~\ref{thm:FixingMixtureUnitaries}) and unital channels (Theorem~\ref{thm:FixingUnitalChannels}). All of these results rely on what we call  \emph{spectral clustering techniques} and \emph{rotation lemmas} studied in Sections~\ref{sec:SpectralClustering} and~\ref{sec:TurnSubspaces} respectively. We then end this section by proving rapid fixability in the bipartite setting with Hilbert space $\cH_{AB} = \cH_A\otimes\cH_B,$ $\cS$ being the set of pure states on $\cH_{AB}$ and $\cC= \id_A\otimes{\rm{CPTP}}(\cH_{AB})$ the set of local channels (Theorem~\ref{thm:ApproxLocalFixPure}). This result relies on similar proof techniques as the others in this section. In particular we again make use of the rotation lemmas mentioned above.

In Section~\ref{sec:MultipleImpossLocal} we show the impossibility of rapid fixability of approximate fixed points in the bipartite setting for the pairs of sets $\cS=\mathfrak{S}(\cH_{AB})$ and $\cC=\id_A\otimes{\rm{CPTP}}(\cH)$ and also $\cS={\rm{Prob}}(\cX\times\cY)$ and $\cC=\id_\cX\otimes{\rm{Stoch}}(\cY)$ (Corollary~\ref{cor:ImpossiLocalFix}). 

We then end the paper with a summary of the results and a discussion of open problems for interesting future research in Section~\ref{sec:SumandOpen}.

	\section{Preliminaries}
    \label{chap:Preliminaries}
    
    In this section we recall some well-known definitions and facts which are used for the derivations of this paper.   
    
    \smallsection{Basic notation:} For $n\in\N$ we denote the set $[n]\coloneqq\{1,\cdots,n\}.$ The set of non-negative real numbers is written as $\R_+.$ In this thesis, if not explicitly stated otherwise, $\log$ denotes the natural logarithm with base $e.$ 
    
	\label{sec:Mathprelim}

\subsection{States and channels}
In this section we review some basic facts about quantum states and channels on some separable Hilbert space $\cH,$ for a reference consider \cite{holevo_statistical_2001,wilde_2013,Wolf12,watrous_2018,khatri2020principles,ReedSimon_FunctionalAnalysis_1976}. Here and in the rest of the thesis we focus on quantum channels in the so-called \emph{Schrödinger picture.}

\medskip

We denote by $\cB(\cH)$ the set of bounded linear operators and by $\cT(\cH)$ the set of trace class operators on $\cH.$ We say an operator $\rho\in\cT(\cH)$ is a \emph{quantum state} (or often simply state in the following) if $\rho\ge 0$ and it has unit trace $\Tr(\rho) =1.$ We denote the set of states on $\cH$ by $\mathfrak{S}(\cH).$

We say a state $\rho$ is \emph{pure} if it is a rank-1 projection and can hence be written as $\rho=\kb{\psi}$ for some normalised $\ket{\psi}\in\cH.$ The set of pure states on $\cH$ is denoted by $\operatorname{Pure}(\cH).$ We refer to the vectors corresponding to pure states (i.e. $\ket{\psi} \in \cH$ such that $\langle\psi|\psi\rangle = 1$) as \emph{state vectors}. On the other hand, we call states $\rho,$ which are not pure, \emph{mixed.} 

We can write any state vector $\ket{\psi}_{AB}$ on some bipartite Hilbert space $\cH_{AB}=\cH_A\otimes\cH_B$ in \emph{Schmidt decomposition} as
\begin{align*}
    \ket{\psi}_{AB} = \sum_{i=1}^{d_*} \sqrt{\lambda_i}\ket{e_i}_A\ket{f_i}_B
\end{align*}
with $d_* = \min\{\dim(\cH_A),\dim(\cH_B)\}$ (and $\dim(\cH_A),\dim(\cH_B)=\infty$ if the underlying Hilbert spaces are infinite dimensional), $\left(\lambda_i\right)_{i=1}^{d_*}$ being a probability distribution which we refer to as \emph{Schmidt coefficients} and $\left(\ket{e_i}_A\right)_{i=1}^{d_*}$ and $\left(\ket{f_i}_B\right)_{i=1}^{d_*}$ being orthonormal systems on $\cH_A$ and $\cH_B$ respectively which we refer to as \emph{Schmidt vectors.}

For $\rho$ and $\sigma$ being states, we denote the \emph{trace distance} by $\frac{1}{2}\left\|\rho-\sigma\right\|_1\in[0,1],$ where the trace norm is given by $\left\|x\right\|_1 \coloneqq \Tr|x| = \Tr\sqrt{x^*x}$ for $x\in\cB(\cH)$. For a  hermitian operator $A\in\cT(\cH)$ with $\Tr(A)=0$ we have the variational expression 
\begin{align}
\label{eq:VarExpTraceLess}
    \frac{1}{2}\left\|A\right\|_1 = \sup_{0\le\Lambda\le\1} \Tr\left(\Lambda A\right) = \Tr\left(\pi_+A\right),
\end{align}
where we denoted the orthogonal projection onto the support of the positive part of $A$ by $\pi_+.$ This in particular implies the following variational expression of the trace distance of two states $\rho$ and $\sigma$ 
\begin{align}
\label{eq:VarExpTraceDistance}
    \frac{1}{2}\left\|\rho-\sigma\right\|_1 = \sup_{0\le\Lambda\le\1} \Tr\left(\Lambda\left(\rho-\sigma\right)\right).
\end{align}
Moreover, for two pure states $\kb{\psi}$ and $\kb{\varphi}$ the trace distance has the form
\begin{align}
\label{eq:PureTraceDistance}
    \frac{1}{2}\left\|\kb{\phi}-\kb{\psi}\right\|_1 = \sqrt{1-|\langle\psi,\varphi\rangle|^2}.
\end{align}

\medskip

We say a linear map $\cN:\cT(\cH)\to\cT(\cH)$ is a \emph{quantum channel}, or often also just channel for simplicity, if it is completely positive and trace preserving. 
The set of quantum channels on $\cH$\footnote{Here and henceforth we often refer to a channel $\cN$ on some Hilbert space $\cH$, although strictly speaking it is linear map on $\cT(\cH)$ the set of trace class operators on $\cH$.} is denoted by ${\rm{CPTP}}(\cH).$

 Every $\cN\in {\rm{CPTP}}(\cH)$ can by \emph{Stinespring's dilation theorem} be written in the form
    \begin{align*}
         \cN(x) = \Tr_E\left(VxV^*\right),
     \end{align*}
     for some isometry $V:\cH\to\cH\otimes\cH_E,$ which we henceforth call the \emph{Stinespring isometry} of $\cN,$ $\cH_E$ being some \emph{environment Hilbert space} and $\Tr_E$ denoting the \emph{partial trace} on $\cH_E$.

   For a bounded linear map $\cN:\cT(\cH)\to\cT(\cH)$ we can define the \emph{diamond norm}  by
   \begin{align*}
       \left\|\cN\right\|_\diamond \coloneqq \sup_{d\in\N, x\in\cT(\cH\otimes\C^d)\, \|x\|_1=1}\left\|(\cN\otimes\id_d)(x)\right\|_1.
   \end{align*}

\noindent\textbf{Classical states and channels}
\label{sec:PrelimChannels}
 
\noindent In the following we discuss so-called \emph{classical states and channels}, which are a special case of quantum states and channels discussed previously. We focus on the case of discrete classical alphabets here as this is all we need for the remainder of this paper.

Consider for some countable set $\cX,$ refered to as \emph{classical alphabet} in the following, the Banach space $l^1(\cX)$ consisting of absolutely summable sequences $s\equiv\left(s_x\right)_{x\in\cX}$ with norm
\begin{align}
    \left\|s\right\|_1 \coloneqq \sum_{x\in\cX}|s_x|.  
\end{align}

We say $s\in l^1(\cX)$ is \emph{positive semi-definite} if $s$ is non-negative component wise. A \emph{classical state} is hence a positive semi-definite sequence $P\in l^1(\cX)$ which sums to 1 or, in other words, a probability distribution over $\cX.$ The set of probability distributions on $\cX$ is denoted by ${\rm{Prob}}(\cX).$ Furthermore, we refer to probability distributions on a finite classical alphabet $\cX$ as \emph{probability vectors}. 

A \emph{classical channel} is a linear map $\cN:l^1(\cX)\to l^1(\cX)$ which maps probability distributions to probability distributions. Denoting the matrix elements $T\equiv \left(T_{xy}\right)_{x,y\in\cX}\coloneqq\left(\cN(\delta_y)_x\right)_{x,y\in\cX}$ with $\delta_y$ being the delta distribution with $y^{th}$ element equal to 1 and all other elements equal to zero, $\cN$ is a classical channel if and only if $T$ is a \emph{stochastic mapping}, i.e.~$T_{xy}\ge 0$ for all $x,y\in\cX$ and 
\begin{align}
    \sum_{x\in\cX} T_{xy} =1
\end{align}
for all $y\in\cX.$ Furthermore, we call a stochastic mapping on a finite classical alphabet $|\cX|<\infty$ a \emph{stochastic matrix}.
We denote the set of classical channels or stochastic mappings on $\cX$ by ${\rm{Stoch}}(\cX).$ 

Note that $l^1(\cX)$ can be embedded into $\cT(\cH)$ for $\cH$ being a separable Hilbert space, which can be taken to have finite dimension if $|\cX|<\infty$. For that we fix an orthonormal basis $\left(\ket{x}\right)_{x\in\cX}$ of $\cH$ and consider the map
\begin{align*}
    l^1(\cX) &\to \cT(\cH)\\
    s&\mapsto \sum_{x\in\cX} s_x \kb{x}.
\end{align*}
Classical states can hence be seen as states on $\cH$ which are diagonal in the fixed basis, in particulur we introduce for $P=\left(p_x\right)_{x\in\cX}\in{\rm{Prob}}(\cX)$ the notation $\rho_P\in\mathfrak{S}(\cH)$ with
\begin{align}
\label{eq:ProbClassicalState}
    \rho_{P} \coloneqq \sum_{x\in\cX} p_x \kb{x}.
\end{align}
Furthermore, for $T\in\Stoch(\cX)$ we can define the channel on $\cH$ by
\begin{align}
\label{eq:StochMatrixClassicalChannel}
\cN_T \coloneqq \sum_{x,y\in\cX} T_{xy}\kb{x}\bra{y}(\,\cdot\,)\ket{y}.
\end{align}
Note for two stochastic mappings $T$ and $S$ we have the relation for the diamond norm
\begin{align}
\label{eq:StochNormEquv}
    \left\|\cN_T-\cN_S\right\|_\diamond = \|T-S\| \coloneqq \sup_{y\in\cY}\sum_{x\in\cX}\left|T_{xy}-S_{xy}\right|.
\end{align}

If the fixed basis $\left(\ket{x}\right)_{x\in\cX}$ is clear from the context, we interchangeably refer to states on $\cH$ of the form \eqref{eq:ProbClassicalState} and elements of $\Prob(\cX)$ as classical states and to channels on $\cH$ of the form \eqref{eq:StochMatrixClassicalChannel} and elements in $\Stoch(\cX)$ as classical channels.

\subsection{Fixed points of channels}
\label{sec:FixedPointsChannel}
In this section we review some basic facts about fixed points of quantum channels. In fact, the following section applies to all positive and trace preserving maps $\cN:\cT(\cH)\to\cT(\cH)$ without requiring complete positivity. For a more complete overview including also several facts which exclusively hold for quantum channels, consider \cite{Wolf12,Burgarth_2013,watrous_2018}.

The space of fixed points of $\cN$ is denoted by \begin{align*}
    \cF(\cN)\coloneqq\left\{x\in \cT(\cH)\Big|\, \cN(x) =x\right\} = \ker\left(\id-\cN\right).
\end{align*}
Furthermore, we can define the \emph{Ces\`aro mean} of $\cN$ by
	\begin{align}
 \label{eq:CesaroMean}
	A_n =\frac{1}{n}\sum_{k=1}^{n}\cN^k.
\end{align}
If the limit $P\coloneqq \lim_{n\to\infty}A_n$ exists, $P$ given by the projection onto the fixed point space $\cF(\cN)$ and for $\cN\in\CPTP(\cH)$ we also have $P\in\CPTP(\cH).$

First, note that on a finite dimensional Hilbert space, the fixed point space of a positive and trace preserving map $\cN$ is non-trivial. This is a direct consequence of the fact that $\cN$ maps the convex and compact set of states on $\cH$ onto itself and therefore there exists, by Brouwer's fixed point theorem, a fixed point state \begin{align}
    \rho\in\mathfrak{S}(\cH)\cap\cF(\cN).
\end{align} 
Again, in finite dimensions, the limit of the Ces\`aro mean $A_n$ as $n\to\infty$ always exists by a standard compactness argument.

For infinite dimensional separable Hilbert spaces $\cH$ these arguments based on compactness break down. In fact, one can consider as an example the shift channel $\cN_{\text{shift}}$ which, for $\left(\ket{i}\right)_{i\in\N}$ being an orthonormal basis of $\cH$, is defined by 
 \begin{align*}
     \cN_{\text{shift}}(x) = \sum_{i=1}^{\infty} \bra{i} x \ket{i}\kb{i+1}
 \end{align*}
 for all $x\in\cT(\cH).$ Assuming $\cN_{\text{shift}}(x) =x$ immediately gives $\bra{i}x\ket{j}=0$ for all $i\neq j$ and furthermore $\bra{i}x\ket{i} =\bra{i+1}x\ket{i+1}$ for all $i\in\N.$ This already shows that $x =c \1$ for some $c\in\C$ and as $x$ is supposed to be a trace class operator, we have $c=0.$ Therefore, $\cN_{\text{shift}}$ cannot have any non-trivial fixed points, i.e.~\begin{align*}
     \cF(\cN_{\text{shift}})=\{0\}.
 \end{align*}
 However, we still have that $1$ is in the spectrum of $\cN_{\text{shift}}$ which follows as the pure state $\kb{1}$ does not lie in the image of $\id -\cN_{\text{shift}}.$

Another useful fact for $\cN:\cT(\cH)\to\cT(\cH)$ being a positive and trace preserving is that for any fixed point $A\in\cF(\cN)$ we have that also the corresponding real- and imaginary- and furthermore their respective positive- and negative parts are fixed points as well\footnote{Here, we introduced the notation $\Re(A)\coloneqq \frac{A+A^*}{2}$ and $\Im(A) \coloneqq\frac{A-A^*}{2i}$ for real and imaginary part of an operator $A\in\cB(\cH).$ And furthermore the positive and negative parts $[A]_+ \coloneqq\frac{|A|+A}{2}$ and $[A]_- \coloneqq\frac{|A|-A}{2}$ which are positive semi-definite for $A$ self-adjoint.}
\begin{align}
\label{eq:ReIMPosNegFix}
    [\Re(A)]_+,[\Re(A)]_-,[\Im(A)]_+,[\Im(A)]_-\in\cF(\cN).
\end{align}
An approximate version of this result, which includes \eqref{eq:ReIMPosNegFix} as a special case, is shown in Lemma~\ref{lem:ApproxBau} below.

Lastly, for any non-zero and positive semi-definite operator $\rho\in\cF(\cN)$, its support is an \emph{invariant subspace} of $\cN.$ More precisely, denoting by $\pi$ the orthogonal projection onto the support of $\rho,$ we have
\begin{align}\
\label{eq:InvarianteSubspaceFix}
    \Tr((\1-\pi)\cN(\pi)) =0.
\end{align}
Furthermore, for any state $\sigma$ with support in $\ran(\pi),$ also the image $\cN(\sigma)$ has support in $\ran(\pi),$ which gives $\cN(\sigma)\le\pi.$
Interestingly, unlike the fact~\eqref{eq:ReIMPosNegFix}, \eqref{eq:InvarianteSubspaceFix} is not robust\footnote{With that we mean that $\frac{1}{2}\left\|\cN(\rho)-\rho\right\|_1\le \eps$ does not imply $\Tr((\1-\pi)\cN(\pi))=f(\eps)$ for some function $f(\eps)\xrightarrow[\eps\to 0]{}0.$} against introducing small errors in the fixed point equations, i.e.~when only demanding $\frac{1}{2}\left\|\cN(\rho)-\rho\right\|_1\le \eps$ for some small $\eps>0.$ A construction where exactly this non-robustness is employed can be found in Section~\ref{sec:MultipleImpossLocal}.

\subsection{Notation for approximate equality}
\label{sec:NotationApproxEquality}
In the following we introduce some notation for \emph{approximate equality} between elements of normed vector spaces. Here, the meaning of the approximate equality, denoted by $\approx$, depends on the context as follows: Let $\eps\ge0.$
For two quantum or classical states $\rho$ and $\sigma$ we write
\begin{align}
\label{eq:StateApprox}
    \rho \approx^\eps\sigma\quad \text{if}\quad \frac{1}{2}\left\|\rho -\sigma\right\|_1 \le\eps.
\end{align}
Similarly, we write for two quantum channels $\cN$ and $\cM$
\begin{align}
\label{eq:ChannApprox}
    \cN \approx^\eps\cM\quad\text{if}\quad \frac{1}{2}\left\|\cN -\cM\right\|_\diamond \le\eps.
\end{align}
For two stochastic mappings $T$ and $S$ we write 
\begin{align}
\label{eq:ChannApprox}
    T\approx^\eps S\quad\text{if}\quad \frac{1}{2}\left\|T -S\right\|\le\eps,
\end{align}
with corresponding norm being defined in \eqref{eq:StochNormEquv}.
For two unitaries $U$ and $V$ on $\cH$ we write
\begin{align}
\label{eq:UnitaryApprox}
    U \approx^\eps V\quad\text{if}\quad \left\|U -V\right\| \le\eps,
\end{align}
and similarly for two orthogonal projections $E$ and $F$ 
\begin{align}
\label{eq:UnitaryApprox}
    E \approx^\eps F\quad\text{if}\quad \left\|E -F\right\| \le\eps,
\end{align}
with the norm being the operator norm in both cases.
For two Hilbert space vectors $\psi,\varphi\in\cH$ we write 
\begin{align}
 \psi \approx^\eps \varphi\quad\text{if}\quad \left\|\psi -\varphi\right\| \le\eps.
\end{align}
with norm being the usual Hilbert space norm.
Lastly, we write for two complex numbers $z_1,z_2\in\C$ 
\begin{align}
 z_1 \approx^\eps z_2\quad\text{if}\quad |z_1 -z_2| \le\eps.
\end{align}

 \nopagebreak
\section{Fixability of approximate fixed point equations}
\label{sec:Setup}
As explained in the Introduction, in this paper we adress the following question:

\smallskip

{\em{Given a quantum channel and a quantum state which almost satisfy a fixed point equation, can we find a new channel and a new state, which are respectively close to the original ones, such that they satisfy an exact fixed point equation?}}

\smallskip

This question can be asked under many interesting constraints in which the original channel and state are assumed to have certain structures which the new channel and state are supposed to satisfy as well. 

In precise mathematical terms we consider, on some separable Hilbert space $\cH$, subsets of states $\cS\subseteq \mathfrak{S}(\cH)$ and subsets of channels $\cC\subseteq {\rm{CPTP}}(\cH)$ which correspond to certain structures of interest. For example $\cS$ could be the set of pure states, general mixed states, or, on a bipartite Hilbert space $\cH_{AB} =\cH_A\otimes\cH_B,$ the set of entangled states. Similarly, $\cC$ could for example be the set of general quantum channels, classical channels, unitary channels, unital channels or, on a bipartite Hilbert space $\cH \equiv\cH_A\otimes\cH_B$, local channels of the form $\cN_{AB} \equiv\id_A\otimes\cN_B.$

For error parameter $\eps\ge 0,$ $\rho\in\cS$ and $\cN\in\cC$
 we say $\rho$ is an \emph{approximate fixed point} of $\cN$ if
\begin{align}
\label{eq:ApproxFixedPointIntroFirst}
    \cN(\rho)\approx^{\eps}\rho.
\end{align}

For a given pair $(\cS,\cC)$ we are hence concerned with whether approximate fixed point equations like \eqref{eq:ApproxFixedPointIntroFirst} can be fixed in the sense of the following definition:
\begin{definition}[Fixability of approximate fixed points]
\label{def:FixingApproxFixPoints}
We say approximate fixed point equations\footnote{In the following we often interchangeably refer to an approximate fixed point being fixable meaning fixability of the corresponding approximate fixed point equation.} are fixable for the pair $\cS\subseteq\mathfrak{S}(\cH)$ and $\cC\subseteq{\rm{CPTP}}(\cH),$ with $\cH$ being some separable Hilbert space, if the following holds true: There exist approximation functions $f,\,g:\R_+\to\R_+$ such that we have
\begin{align}
    \lim_{\eps\to 0}f(\eps)=\lim_{\eps\to 0}g(\eps)=0
\end{align}
and for all $\rho\in\cS,\,\cN\in\cC$ and $\eps\ge 0$ such that $\cN(\rho)\approx^{\eps}\rho$, we can find a new state and a channel of the same structure
\begin{align}
 \sigma\in\cS\quad\quad\text{and}\quad\quad\cM\in\cC,\quad
\end{align}
which are close to original ones, i.e.
\begin{align}
\label{eq:IntroFixfgAbstract}
\sigma\approx^{f(\eps)} \rho \quad\quad\text{and}\quad\quad\cM\approx^{g(\eps)}\cN,
\end{align}
 and which are a fixed point pair, i.e.~they satisfy the fixed point equation\footnote{Putting this differently, approximate fixed point equations are fixable for the pair $(\cS,\cC)$ if and only if we have for the worst case error $\sup_{\substack{(\rho,\cN)\in\cS\times \cC \\\cN(\rho)\approx^{\eps}\rho}}\inf_{\substack{(\sigma,\cM)\in \cS\times\cC\\\cM(\sigma)=\sigma}}\max\{\|\rho-\sigma\|_1,\|\cN-\cM\|_\diamond\}\xrightarrow[\eps\to 0]{}0.$}
\begin{align}
\label{eq:FixedEquationIntro}
\cM(\sigma)=\sigma.
\end{align}
\end{definition}

It is important to note that the functional form of $f$ and $g$ in the above definition solely depends on the sets $\cS$ and $\cC$ (and hence also on the underlying Hilbert space $\cH$), but not on the specific $\rho\in\cS,$ $\cN\in\cC$ and $\eps\ge0.$
Note also that for finite dimensional Hilbert spaces with dimension $d,$ the approximation functions $f$ and $g$ can hence implicitly depend on $d,$ i.e. $f(d,\eps)\equiv f(\eps)$ and $g(d,\eps)\equiv g(\eps)$ as denoted in the Introduction.

To illustrate Definition~\ref{def:FixingApproxFixPoints} we first note that for general sets $\cS$ and $\cC,$ approximate fixed point equations might not be fixable, and hence the above is not trivially true in general. For that we could consider sets with just one element $\cS =\{\rho\}$ and $\cC=\{\cN\}$ such that $\cN(\rho)\neq\rho.$ In that case approximate fixed point equations are not fixable, simply as there exists no pair $(\sigma,\cM)\in\cS\times\cC$ satisfying an exact fixed point equation. More interesting counterexamples occur when one or both of the sets are non-closed. For example, take as Hilbert space $\cH=\C^2$, and then the convex but open set $\cS = \{(1-\eps)\kb{0}+\eps\kb{1}\ \big|\ 0<\eps<1\}$ and furthermore $\cC=\{\Tr(\,\cdot\,)\kb{0},\Tr(\,\cdot\,)\1/2\}.$ First of all, as $\sigma = \1/2\in\cS$ and $\cM=\Tr(\,\cdot\,)\1/2\in\cC,$ we have, unlike in the previously mentioned counterexample, a pair of state and channel satisfying $\cM(\sigma)=\sigma.$ However, for every $\eps>0$ there exists $\rho\in\cS$ such that for $\cN=\Tr(\,\cdot\,)\kb{0}\in\cC$ we have
\begin{align}
\label{eq:ApproxFixNotFixable}
    \cN(\rho )\approx^{\eps}\rho.
\end{align}
As $\frac{1}{2}\|\cN-\Tr(\,\cdot\,)\1/2\|_\diamond=1/2$ and the unique fixed point state of $\cN$ given by $\kb{0}$ is not contained in $\cS,$ the approximate fixed point equation \eqref{eq:ApproxFixNotFixable} can not be fixed in this setup.

It turns out that in the finite dimensional case the above counterexamples illustrate the only two situations in which approximate fixed point equations might not be fixable, i.e.~non-existence of a fixed point pair or non-closedness of the sets $\cS$ and $\cC.$ In fact, in Proposition~\ref{prop:AbstractFixingApproxFix} below we show by a compactness argument that for $\cS$ and $\cC$ being closed and such that a fixed point pair exists, approximate fixed point equations are always fixable. 

Under the named assumptions of Proposition~\ref{prop:AbstractFixingApproxFix}, the existence of approximation functions $f$ and $g$ satisfying Definition~\ref{def:FixingApproxFixPoints} is merely ensured by compactness and their precise form is rather implicit. Our focus therefore lies in providing good control of the functions $f$ and $g$ for concrete examples of $\cS$ and $\cC.$ Especially, we are interested in cases where the interplay of the dimension $d$ and $\eps$ in $f(d,\eps)\equiv f(\eps)$ and $g(d,\eps)\equiv g(\eps)$ is rather mild, and the convergence in $\lim_{\eps\to 0}f(d,\eps)=\lim_{\eps\to 0}g(d,\eps)$ happens rather fast and we therefore have good approximation by the fixed point pair in \eqref{eq:IntroFixfgAbstract}. To make this more precise, we consider $\cS =\left(\cS_d\right)_{d\in\N}$ and $\cC =\left(\cC_d\right)_{d\in\N}$ to be sequences of sets with $\cS_d\subseteq\mathfrak{S}(\cH_d)$ and $\cC_d\subseteq\CPTP(\cH_d)$ and $\cH_d\cong \C^d.$ These could, for example, consist of the full set of quantum states and channels or alternatively unitary channels on Hilbert spaces for different dimensions. We are then interested in whether approximate fixed point equations are \emph{rapidly fixable} for a pair $(\cS,\cC)$ in the sense of the following definition:
\begin{definition}[Rapid fixability of approximate fixed points]
\label{def:RapidFixingApproxFixedPoint}
Let $\cS =\left(\cS_d\right)_{d\in\N}$ and $\cC =\left(\cC_d\right)_{d\in\N}$  with $\cS_d\subseteq\mathfrak{S}(\cH_d)$ and $\cC_d\subseteq\CPTP(\cH_d)$ and $\cH_d$ being a $d$-dimensional Hilbert space. We say approximate fixed point equations are rapidly fixable for the pair $(\cS,\cC)$ if for all $d\in \N$ they are fixable for $(\cS_d,\cC_d)$ in the sense of Definition~\ref{def:FixingApproxFixPoints} with corresponding approximation functions $f$ and $g$ satisfying
\begin{align}
f(d,\eps) = c\,d^{b}\eps^{a}\quad\quad\text{and}\quad\quad g(d,\eps) = c'd^{b'}\eps^{a'},
\end{align}
for some $a,a'>0$ (and typically $a,a'\le 1 $) and $b',b',c,c'\ge 0$ all independent of $d$ and $\eps.$
\end{definition}

In the following, we often drop for simplicity the difference between the sequences $\cS =\left(\cS_d\right)_{d\in\N}$ and $\cC =\left(\cC_d\right)_{d\in\N}$ and the corresponding sets $\cS_d$ and $\cC_d$ when the full sequence is clear from the context.

For the majority of the remainder of this paper we are concerned with proving that for certain examples of $\cS$ and $\cC$ approximate fixed point equations are rapidly fixable and to provide concrete numerical values for $a,a',b',b',c$ and $c'.$ In particular we show that approximate fixed point equations are rapidly fixable for  $\cS=\mathfrak{S}(\cH)$ being the full set of quantum states on $\cH$ and
\begin{enumerate}
    \item $\cC={\rm{CPTP}}(\cH)$ being the full set of quantum channels (Theorem~\ref{thm:FixedPointQuantum}),
    \item$\cC$ \ being the set of unitary channels (Theorem~\ref{thm:FixUnitary}),
    \item $\cC$ \ being the set of mixed-unitary channels (Theorem~\ref{thm:FixingMixtureUnitaries}),
    \item  $\cC$ \ being the set of unital channels (Theorem~\ref{thm:FixingUnitalChannels}).
 \end{enumerate}
Furthermore, we also show rapid fixabilitity of approximate fixed point equations in the following cases:
\begin{enumerate}
    \setcounter{enumi}{4}
    \item $\cS={\rm{Prob}}(\cX)$ being the set of classical states, i.e.~probability vectors, and $\cC={\rm{Stoch}}(\cX)$ being the set of classical channels, i.e.~stochastic matrices, on some finite classical alphabet $\cX$ (Theorem~\ref{thm:FixClassical}),
    \item $\cS$ being the set of pure states on a bipartite Hilbert space $\cH_{AB}=\cH_A\otimes\cH_B$ and $\cC=\id_A\otimes{\rm{CPTP}}(\cH_B)\coloneqq\left\{\id_A\otimes\cN_B\,\big|\,\cN_B\in{\rm{CPTP}}(\cH_B)\right\},$ which we call \emph{local channels} in the following  (Theorem~\ref{thm:ApproxLocalFixPure}).
\end{enumerate}
On the other hand, by providing a concrete counterexample, we show that approximate fixed points for $\cS =\mathfrak{S}(\cH_{AB})$ being the full set of quantum states on a bipartite Hilbert space and the set of local channels $\cC=\id_A\otimes{\rm{CPTP}}(\cH_B)$ are not rapidly fixable (Corollary~\ref{cor:ImpossiLocalFix}). In fact, this counterexample already rules out rapid fixability of approximate fixed points in the corresponding classical case with $\cS = \Prob(\cX\times\cY)$ and  $\cC=\id_X\otimes{\rm{Stoch}}(\mathcal{Y})\coloneqq\left\{\id_X\otimes T_Y\,\big|\,T_Y\in{\rm{Stoch}}(\mathcal{Y})\right\}\,$
\bigskip

To illustrate the above question of fixability of approximate fixed points further, let us discuss the following variant: Instead of allowing to change both state $\rho$ to $\sigma$ and channel $\cN$ to $\cM$ in \eqref{eq:IntroFixfgAbstract}, one could ask whether the approximate fixed point equation $\cN(\rho)\approx^{\eps}\rho$ can be fixed by only changing one of them. The following example shows that this is not possible in general:
\begin{example}[Necessity to change both state and channel]
\label{ex:ChangeBothToFix}
We take $\cS=\mathfrak{S}(\cH)$ and $\cC = {\rm{CPTP}}(\cH)$ the full set of quantum states and channels on some finite-dimensional Hilbert space $\cH.$ Consider first $\cH_1=\C^2$ with orthonormal basis $\left\{\ket{0},\ket{1}\right\}$, $\rho_1 = \kb{0}$ and for $\eps > 0$ the quantum channel
\begin{align*}
    \cN_{1,\eps} = (1-\eps)\id + \eps\Tr(\,\cdot\,)\,\kb{1}.
\end{align*}
This pair of state and channel surely satisfies the approximate fixed point equation 
\begin{align}
\label{eq:ApproxFixExample1}
    \cN_{1,\eps}(\rho_1)\approx^{\eps}\rho_1.
\end{align}
However, $\rho_1$ is far from the fixed point space of $\cN_{1,\eps}$ given by $\cF(\cN_{1,\eps}) = {\rm{span}}\left\{\kb{1}\right\},$ i.e.~\begin{align*}
    \min_{x\in\cF(\cN_{1,\eps})}\left\|\rho_1-x\right\|_1 = \left\|\kb{0}\right\|_1=1
\end{align*}
Therefore, the approximate fixed point equation \eqref{eq:ApproxFixExample1} cannot be fixed in the sense of \eqref{eq:IntroFixfgAbstract} and \eqref{eq:FixedEquationIntro} by solely changing the state $\rho_1$ but not the channel $\cN_{1,\eps}.$\footnote{Another interesting example where this phenomenon occurs is taking $\cN_{1,\eps}$ to be a unitary channel $U(\,\cdot\,)U^*$ for some unitary $U$ satisfying $0<\|U-\1\|\le \eps$ but with eigenbasis given by $\ket{+}$ and $\ket{-}$ and therefore $\rho_1=\kb{0}$ being far from the fixed point space $\cF(\cN_{1,\eps})={\rm={span}}\{\kb{+},\kb{-}\}.$} However, it is possible to fix the approximate fixed point equation \eqref{eq:ApproxFixExample1} by only changing the channel, as we can simply take $\sigma=\rho_1$ and $\cM = \id\approx^{\eps}\cN_{1,\eps}.$

An example where the opposite occurs is given by the following: Take as Hilbert space $\cH_2 =\C^3$ and the unitary channel $\cN_2=U(\,\cdot\,)U^*$ with unitary 
\begin{align*}
    U = \sigma_x\oplus 1 = \begin{pmatrix}
        0 & 1& 0\\
        1&0&0\\
        0&0&1
    \end{pmatrix},
\end{align*}
where we denoted by $\sigma_x$ the Pauli-$x$ matrix. Then take for every $\eps >0$ the state
\begin{align*}
    \rho_{2,\eps} = \eps\kb{0}+(1-\eps)\kb{2} = \begin{pmatrix} \eps&\ \ 0& 0\\
        0&\ \ 0&0\\
        0&\ \ 0&1-\eps
    \end{pmatrix}.
\end{align*}
Clearly, we have the approximate fixed point equation
\begin{align}
\label{eq:ApproxFixExample2}
    \cN_2(\rho_{2,\eps}) = \eps\kb{1} + (1-\eps)\kb{2} \approx^{\eps}\rho_2.
\end{align}
Assume now there exist a function $g:\R_+\to\R_+$ such that $\lim_{\eps\to 0}g(\eps)=0$, and, for all $\eps\ge 0$, a channel\footnote{Note that we only used linearity and positivity of $\cM_\eps$ showing that in fact the approximate fixed point equation cannot be fixed by only changing $\cN_2$ to $\cM_\eps$ even if we allow for the wider class of all positive maps.} $\cM_\eps$ approximating $\cN_2,$ i.e.
\begin{align}
\label{eq:ApproxChangeStateExample}
\cM_\eps\approx^{g(\eps)}\cN_2
\end{align}
and satisfying the exact fixed point equation
\begin{align}
\label{eq:FixedExampleChangeChannel}
    \cM_\eps(\rho_{2,\eps}) =\rho_{2,\eps}.
\end{align}
From \eqref{eq:ApproxChangeStateExample} we however already see 
\begin{align*}
\bra{1}\cM_\eps(\kb{0})\ket{1} \ge 1-g(\eps)>0
\end{align*}
for $\eps$ small enough
and hence
\begin{align*}
    \bra{1}(\cM_\eps(\rho_{2,\eps})-\rho_{2,\eps})\ket{1} = \bra{1}\cM_\eps(\rho_{2,\eps})\ket{1} \ge \eps (1-g(\eps))>0
\end{align*}
which is a contradiction to the fixed point equation \eqref{eq:FixedExampleChangeChannel}. Hence, we have shown that the approximate fixed point equation \eqref{eq:ApproxFixExample2} cannot be fixed in the sense of \eqref{eq:IntroFixfgAbstract} and \eqref{eq:FixedEquationIntro} by only changing the the channel $\cN_2$ but not the state $\rho_{2,\eps}.$ On the other hand, for this particular example \eqref{eq:ApproxFixExample2} can be fixed by doing the reverse, i.e.~only changing the state to $\sigma = \kb{2}\approx^{\eps}\rho_{2,\eps}$ but leaving the channel invariant $\cM= \cN_2.$

Building both examples above on top of each other gives an example of an approximate fixed point equation which can only be fixed by both varying state and channel. More, precisely, we take as Hilbert space $\cH = \cH_1\oplus\cH_2 =\C^5$ and for $\eps>0$ the state
\begin{align*}
    \rho_\eps = \frac{1}{2}\left(\rho_1\oplus\rho_{2,\eps}\right)
\end{align*}
and channel
\begin{align*}
    \cN_\eps = \cN_{1,\eps}\oplus\cN_{2}
\end{align*}
This pair satisfies the approximate fixed point equation $\cN_\eps(\rho_\eps)\approx^{\eps}\rho_\eps$ but, by the above arguments, cannot be fixed by soley changing either state or channel.\footnote{Here, for a more detailed argument, denote by $\pi_1$ and $\pi_2$ the orthogonal projections on $\cH$ onto the subspaces $\cH_1$ and $\cH_2$ and note that for every $\sigma_\eps\in\cF(\cN_\eps)$ we have necessarily $ \sigma_\eps =\pi_1\sigma_\eps\pi_1 + \pi_2\sigma_\eps\pi_2.$  Therefore, $\pi_1\sigma_\eps\pi_1\in\cF(\cN_{1,\eps})={\rm{span}}\{\kb{1}\}$ and hence $\|\sigma_\eps-\rho_\eps\|_1\ge\|\pi_1\sigma_\eps\pi_1-\frac{1}{2}\rho_1\|_1 \ge\frac{1}{2}.$ On the other hand, any channel $\cM_\eps$ having $\rho_\eps$ as fixed point must be far from $\cN_\eps$ by exactly the same argument as for $\cN_{2}$ and $\rho_{2,\eps}.$}
\end{example}

Lastly, we want to point out that the difficulty of the problem outlined above lies in the fact that we demand certain structures (meaning membership of the sets $\cS$ and $\cC$) to be preserved, e.g.~that the new $\sigma$ and $\cM$ satisfying $\cM(\sigma)=\sigma$ are a state and a channel respectively.  To illustrate this, let $V$ be some $d$-dimensional, normed and complex vector space, $\eps\ge0,$ $v\in V$ with $\|v\|=1$ and $A:V\to V$ some  linear map such that $\|Av-v\|\le\eps.$ We would like to fix this approximate fixed point equation by changing $A$ to $B$ while only demanding that $B$ must be linear as well. For that consider the following construction: Let $\left(v_i\right)_{i=1}^d$ be a basis of  $V$ with $v_1=v.$  We can now define $B$ simply on each basis vector and then extend linearly. Taking hence 
\begin{align*}
    Bv_1 = v\quad\quad\text{and}\quad\quad Bv_i=Av_i\quad\text{for all $i=2,\cdots,d$}
\end{align*}
and then extending linearly, gives a well-defined linear map $B:V\to V$ satisfying the fixed point equation $Bv=v$ and furthermore
\begin{align}
    \label{eq:FixingOnlyLinearStructure}\left\|B -A\right\| &= \max_{\|w\|=1}\left\|(B-A)w\right\| = \max_{\substack{\|w\|=1}}\left\|\sum_{i=1}^d\omega_i(B-A)v_i\right\| \nn\\&= \max_{\substack{\|w\|=1}} |\omega_1|\left\|v-Av\right\|\le c_d \,\eps.
\end{align}
Here, $\left(\omega_i\right)_{i=1}^d\subseteq\C$ denote the coefficients of the vector $w$ in the given basis and $c_d\ge0$ is some constant whose precise form may depend on the dimension $d,$ the norm on $V$ and the choice of basis $\left(v_i\right)_{i=1}^d.$ For example, if the norm $\|\,\cdot\,\|$ is induced by a scalar product on $V,$ we can pick the basis to be orthonormal with respect to this scalar product and $c_d=1$ follows. For general norms the existence of such finite $c_d$ follows by equivalence of all norms in finite dimensions. This shows that fixing approximate fixed point equations becomes a lot easier when only demanding to preserve linearity of the map when changing $A$ to $B$.  In particular here, unlike in Example~\ref{ex:ChangeBothToFix} where also positivity of the map was supposed to be preserved,  it suffices to only change the map $A$ to $B$ but leave the the vector $v$ invariant. In the sections below we see that for fixing approximate fixed points equations in terms of quantum states and channels more sophisticated techniques are necessary compared to the purely linear case.

\section{Fixability for closed sets}
\label{sec:ClosedFixAbstract}
In this section we first state Proposition~\ref{prop:AbstractFixingApproxFix} which provides for the finite dimensional case that approximate fixed point equations are always fixable in the sense of Definition~\ref{def:FixingApproxFixPoints} as long as the sets $\cS\subseteq \mathfrak{S}(\cH)$ and $\cC\subseteq{\rm{CPTP}}(\cH)$ are closed and contain a fixed point pair. The proof relies on a compactness argument and hence we merely prove existence of approximation functions $f$ and $g$ satisfying the conditions in Defintion~\ref{def:FixingApproxFixPoints} but do not provide any specific form they need to satisfy.

We then review some of the theory of quantum Markov chains and their approximate analogs. We see that they can be defined by a certain (approximate) fixed point equation which they need to satisfy. Applying Proposition~\ref{prop:AbstractFixingApproxFix} to fix this particular approximate fixed point equation enables us to solve an open problem in the robustness theory of quantum Markov chains. In particular, we show that for every approximate Markov chain there is an exact quantum Markov chain close to it (Proposition~\ref{prop:RobustnessQMC}). Here, the approximation error in mentioned closeness is measured in terms of a function $f(d_{ABC},\eps)$ depending necessarily on the intrinsic dimension of the problem captured in $d_{ABC}$ and the error parameter $\eps\ge 0$ of the approximate fixed point equation and which decays as $\eps\to 0.$

\begin{proposition}[Fixability for closed sets $\cS$ and $\cC$]
\label{prop:AbstractFixingApproxFix}
 Let $\cH$ be a finite dimensional Hilbert space. Further, let $\cS\subseteq \mathfrak{S}(\cH)$ and $\cC\subseteq{\rm{CPTP}}(\cH)$ be closed and such that there exist $\sigma_\star\in\cS$ and $\cM_\star\in\cC$ satisfying the fixed point equation:
    \begin{align}
    \cM_\star(\sigma_\star) =\sigma_\star.
    \end{align}
    Then, there exist functions $f,g:\R_+\to\R_+$ such that  \begin{align}\lim_{\eps\to 0}f(\eps)=\lim_{\eps\to0}g(\eps)=0
    \end{align}
    and the following holds: For all $\rho\in\cS$, $\cN\in\cC$ and $\eps\ge 0$ with\begin{align}
        \cN(\rho)\approx^{\eps}\rho,
    \end{align}
    there exist $\sigma\in\cS$ and $\cM\in\cC$ with
   \begin{align}
\sigma\approx^{f(\eps)} \rho \quad\quad\text{and}\quad\quad\cM\approx^{g(\eps)}\cN,
\end{align}
 such that the exact fixed point equation
\begin{align}
\cM(\sigma)=\sigma
\end{align}
holds. In other words, all approximate fixed point equations are fixable for the pair $(\cS,\cC)$ in the sense of Definition~\ref{def:FixingApproxFixPoints}.
\end{proposition}
Note that the approximation functions $f$ and $g$ in Proposition~\ref{prop:AbstractFixingApproxFix} can implicitly depend on the dimension of the underlying Hilbert space, as remarked below Definition~\ref{def:FixingApproxFixPoints}. 
We defer the proof of Proposition~\ref{prop:AbstractFixingApproxFix} in Subsection~\ref{sec:ProfofPropClosed} and first discuss its application to the robustness theory of quantum Markov chains.
\subsection{Application to the robustness of quantum Markov chains}
\label{sec:QMCSection}
 We review some basic definitions and well-known facts regarding quantum Markov chains and their approximate analogs (see \cite{Sutter_approximateThesis_2018} for a nice overview of all the mentioned results): We consider tripartite Hilbert spaces $\cH_{ABC}=\cH_A\otimes\cH_B\otimes \cH_C$ with finite dimensions $d_A\coloneqq\dim(\cH_A)$, $d_B\coloneqq\dim(\cH_B)=d_B$ and $d_C\coloneqq\dim(\cH_C)$. For a state $\rho_{ABC}$ on $\cH_{ABC}$ we can define the \emph{conditional mutual information}
\begin{align}
\label{eq:CMI}
    I(A:C|B)_\rho \coloneqq S(\rho_{AB}) + S(\rho_{BC}) - S(\rho_{ABC}) -S(\rho_B)
\end{align}
where we denoted the \emph{von Neumann entropy} of a state $\sigma$ by $S(
\sigma) \coloneqq -\Tr(\sigma\log(\sigma))$ and denoted the reduced states of $\rho_{ABC}$ by $\rho_{AB} \coloneqq\Tr_C(\rho_{ABC})$ and analogously for the other combinations. Note that the conditional mutual information is always non-negative, $I(A:C|B)_\rho\ge 0,$ which is due to the strong subadditivity of the von Neumann entropy \cite{LiebRuskai_StrongSubadittivity_1973}.

A state $\rho_{ABC}$ is a \emph{(short) quantum Markov chain (QMC)}  if it satisfies strong subadditivity
with equality, i.e.
\begin{align}
\label{eq:QMCConditional}
    I(A:C|B)_\rho = 0.
\end{align}
 This entropic characterisation is equivalent to the existence of a \emph{recovery channel} \cite{Petz_Recovery_2003}, i.e.~a completely positive and trace preserving linear map $\cR_{B\to BC} : \cB(\cH_{B})\to \cB(\cH_{BC})$ such that
\begin{align}
\label{eq:QMCRecovery}
    (\id_{A}\otimes\cR_{B\to BC})(\rho_{AB}) = \rho_{ABC}.
\end{align}
Intuitively this means that system $C$ can be recovered exclusively by focusing on system $B$ while ignoring $A.$ Note that a quantum Markov chain is hence characterised by satisfying a certain fixed point equation
\begin{align}
\label{eq:FixedPointQMC}
    \cN(\rho_{ABC}) = \rho_{ABC}
\end{align}
for $\cN$ being a channel of the specific form $\cN=(\id_A\otimes\cR_{B\to BC})\circ\Tr_C.$ In \cite{hayden_structure_2004} the authors realised that this implies an interesting structure of $\rho_{ABC}$: There exists an orthogonal sum decomposition 
\begin{align*}
    \cH_B = \bigoplus_{j=1}^n\cH_{b^L_j}\otimes\cH_{b^L_j} 
\end{align*}
with $n\in[d_B]$ and Hilbert spaces $(\cH_{b^L_j})_{j=1}^n$ and $(\cH_{b^R_j})_{j=1}^n$ and such that 
\begin{align}
\label{eq:QMCOrthSum}
    \rho_{ABC} = \bigoplus_{j=1}^n q_j\, \rho_{Ab^L_j}\otimes\rho_{b^R_jC}
\end{align}
for some probability distribution $\left(q_j\right)_{j=1}^n$ and states $\rho_{Ab^L_j}$ on $\cH_{Ab^L_j} = \cH_A\otimes \cH_{b_j^L}$ and $\rho_{b^R_jC}$ on $\cH_{b^R_jC} =  \cH_{b_j^R}\otimes \cH_C$ for all $j\in[n].$ An essential component in the proof of \cite{hayden_structure_2004} is the Koashi-Imoto theorem \cite{KoashiImoto} which implies a comparable orthogonal sum structure for one channel and multiple states, which are all fixed points of the channel. On the other hand, it is also easy to see that \eqref{eq:QMCOrthSum} implies \eqref{eq:QMCConditional} or \eqref{eq:QMCRecovery}, which gives that  \eqref{eq:QMCConditional}, \eqref{eq:QMCRecovery} and \eqref{eq:QMCOrthSum} are all equivalent characterisations of quantum Markov chains.

It is important to note that the above is a sensible quantum generalisation of (classical) Markov chains: A probability distribution $P_{XYZ}$ on some classical alphabet $\cX\times\cY\times\cZ$ is called \emph{(short) Markov chain (MC)} if 
\begin{align}
\label{eq:ClassicalMarkov}
P_{XYZ} = P_{XY}P_{Z|Y},
\end{align}
with $P_{XY}$ denoting the marginal distribution of $X$ and $Y$ and  $P_{Z|Y}$ the conditional distribution of $Z$ given $Y.$ Note that \eqref{eq:ClassicalMarkov} can be seen as classical analog of \eqref{eq:QMCRecovery} as the full distribution $P_{XYZ}$ can be recovered by application of a classical channel which only acts on $Y.$\footnote{More precisely, we can define the stochastic matrix $T\in\R_+^{|\cZ||\cY|\times|\cY|}$ by denoting $w=(z,y)$ and $T_{w\, y'} := \delta_{yy'}P_{Z|Y}(z|y').$  Equation \eqref{eq:ClassicalMarkov} can then by written as the recovery relation $(\id_X\otimes T)(P_{XY})=P_{XYZ}.$} Similarly to the quantum case discussed above, \eqref{eq:ClassicalMarkov} is equivalent to the  conditional mutual information satisfying
\begin{align}
\label{eq:MCCMI}
    I(X:Z|Y)_P = 0,
\end{align}
where  $I(X:Z|Y)_P$ is defined analogously to \eqref{eq:CMI} with von Neumann entropies replaced by the corresponding Shannon entropies. 

An interesting question is whether the equivalences above are robust against small errors, i.e.~when the statements \eqref{eq:QMCConditional}-\eqref{eq:QMCOrthSum} in the quantum case, and \eqref{eq:ClassicalMarkov} and \eqref{eq:MCCMI} in the classical case, does not hold exactly but only holds up to a small error $\eps\ge 0.$ Classically, this is the case as we have a very strong robustness theory of Markov chains due to the equality \cite{horodecki_information_2005}
\begin{align}
\label{eq:MCRob}
I(X:Z|Y)_P = \min_{Q_{XYZ}\  \text{MC}} D(P_{XYZ}\|Q_{XYZ}) = D(P_{XYZ}\|P_{XY}P_{Z|Y}),
\end{align}
where we have denoted the Kullback-Leibler divergence of two distributions $P=\left(P_i\right)_{i=1}^m$ and $Q = \left(Q_i\right)_{i=1}^m$ by $D(P\|Q)\coloneqq \sum_{i=1}^m P_i\log(\frac{P_i}{Q_i})$ if $P_i=0$ for all $i\in[m]$ for which $Q_i=0$, and $D(P\|Q)=\infty$ otherwise. This can be combined with Pinsker's inequality \cite[Chapter I.3.]{csiszar_korner_2011} giving
\begin{align}
\label{eq:PinskerCMI}
    \min_{Q_{XYZ}\  \text{MC}}\,\frac{1}{2}\left\|P_{XYZ} - Q_{XZY}\right\|_1 \le \sqrt{\frac{I(X:Z|Y)_P }{2}}.
\end{align}
Therefore, \eqref{eq:MCRob} and \eqref{eq:PinskerCMI} show that if the conditional mutual information of a distribution $P_{XYZ}$ is small, it is close to a Markov chain, and hence can be written approximately in the form \eqref{eq:ClassicalMarkov}.

In the quantum case, the robustness theory of quantum Markov chains is far more involved: For an error parameter $\eps\ge0$ we call a state $\rho_{ABC}$ an \emph{approximate quantum Markov chain} if 
\begin{align}
\label{eq:CMISmall}
    I(A:C|B)_\rho \le \eps.
\end{align}
 In \cite{FawziRenner_CMIRecovery_2015}\footnote{Note that in \cite{FawziRenner_CMIRecovery_2015} the authors consider logarithms base 2 and not base $e$ as in this thesis, which is why they state \eqref{eq:ApprQMCRecovery} with a different prefactor in the error parameter.} it was shown, and then extended in \cite{Brandao_CMI_2015,Wilde_RecoveryInterpolation_2015,Sutter_UniversalRecoveryCMI_2016,Sutter_Multivariate_2017,JungeSutterwilde_UniversalRecovery_2018,Sutter_approximateThesis_2018},   that this implies the existence of a recovery channel $\cR_{B\to BC}$ such that \eqref{eq:QMCRecovery} holds up to $\sqrt{\eps}$ error, i.e.~precisely
\begin{align}
\label{eq:ApprQMCRecovery}
    (\id_{A}\otimes\cR_{B\to BC})(\rho_{AB}) \approx^{\sqrt{\eps}} \rho_{ABC}.
\end{align}
On the other hand, as argued in \cite{FawziRenner_CMIRecovery_2015} using the Alicki-Fannes inequality \cite{alicki_continuity_2004}, \emph{approximate recovery} of the form \eqref{eq:ApprQMCRecovery} implies, up to changes in the error parameters, smallness of the conditional mutual information \eqref{eq:CMISmall}. Using the same argument as around \eqref{eq:FixedPointQMC}, an approximate quantum Markov chain is hence characterised by satisfying a certain approximate fixed point equation.

In the quantum case, upper bounding the distance of a tripartite state to the set of quantum Markov chains in terms of some decaying function in the conditional mutual information similarly to \eqref{eq:PinskerCMI} is, however, still an open problem. This is an interesting question as a positive answer would imply that an approximate quantum Markov chain can also, up to error terms, be written in an orthogonal sum decomposition of the form \eqref{eq:QMCOrthSum}. It is, however, known that a possible answer in the quantum case must be far more complicated than in the classical case as concrete counterexamples of the quantum analogs of \eqref{eq:MCRob} and \eqref{eq:PinskerCMI} have been found in \cite{IbinsonWinter_RobustnessQMC_2008,christandl_entanglement_2012,Sutter_approximateThesis_2018}. In particular, it has been shown that any possible upper bound of the trace distance to the closest quantum Markov chain in terms of some decaying function in the conditional mutual information must necessarily be dimension dependent.

Applying our result on fixability of approximate fixed point equations for closed sets $\cS$ and $\cC$, Proposition~\ref{prop:AbstractFixingApproxFix}, we are able to close this gap to some extent: In the proposition below we see that any approximate quantum Markov chain is close to an exact one, in the sense that if the conditional mutual information approaches zero, the corresponding state approaches the set of quantum Markov chains. The upper bound on the minimal distance to the set of quantum Markov chains is, as discussed above, dimension dependent. Moreover, as the proof relies on the compactness argument from Proposition~\ref{prop:AbstractFixingApproxFix}, we can only show existence of such an approximation function but are not able to show that it must be of a specific form.  

\comment{In particular

\begin{align*}
    \frac{1}{2}\left\|(\id_{A}\otimes\cR_{B\to BC})(\rho_{AB}) -\rho_{ABC}\right\|_1 \le \sqrt{I(A:C|B)}
\end{align*}}

\begin{proposition}[Robustness of quantum Markov chains]
\label{prop:RobustnessQMC}
    There exists a function $f:\N\times\R_+\to\R_+$ such that for all $d\in\N$ we have
    \begin{align}
        \lim_{\eps\to 0}f(d,\eps)=0
    \end{align}
    and for all $\cH_{ABC}=\cH_A\otimes\cH_B\otimes\cH_C$ being a tripartite Hilbert space with finite dimensions $d_A\coloneqq\dim(\cH_A),$ $d_B\coloneqq \dim(\cH_B),$ $d_C\coloneqq\dim(\cH_C),$ and $d_{ABC}\coloneqq\dim(\cH_{ABC})=d_Ad_Bd_C$ and tripartite states $\rho_{ABC}\in\mathfrak{S}(\cH_{ABC})$ such that
    \begin{align}
        I(A:C|B)_\rho \le\eps
    \end{align}
    for some $\eps\ge 0$ there exists a quantum Markov chain $\sigma_{ABC}$ such that
    \begin{align}
        \sigma_{ABC}\approx^{f(d_{ABC},\eps)}\rho_{ABC},
    \end{align}
    In particular, $\rho_{ABC}$ can be written approximately in an orthogonal sum decomposition of the form \eqref{eq:QMCOrthSum}.
\end{proposition}
\begin{proof}
By assumption and the results mentioned around \eqref{eq:ApprQMCRecovery}, there exists a recovery channel $\cR_{B\to BC}$ such that \begin{align}
    \rho_{ABC} \approx^{\sqrt{\eps}} (\id_A\otimes\cR_{B\to BC})(\rho_{AB}) = (\id_A\otimes\cR_{B\to BC})\circ\Tr_C(\rho_{ABC}). 
\end{align}
We define the sets $\cS \coloneqq \mathfrak{S}(\cH_{ABC})$ and as set of channels 
\begin{align}
    \cC \coloneqq \left\{\Phi\in\CPTP(\cH_{ABC})\,\Big|\,\Phi= (\id_A\otimes\cR'_{B\to BC})\circ\Tr_C,\ \, \cR'_{B\to BC}\,\text{ channel }\right\}.
\end{align}
 Note that $\cS$ and $\cC$ are closed and furthermore the corresponding set of fixed point pairs is non-empty as it precisely consists of the quantum Markov chains on $\cH_{ABC}$ and their respective recovery channels. Hence, by Proposition~\ref{prop:AbstractFixingApproxFix} there exist an approximation function\footnote{The approximation function in Proposition~\ref{prop:AbstractFixingApproxFix} does only implicitly depend on the dimension of the underlying Hilbert space. As we want to proof a result for all finite dimensional tripartite Hilbert spaces, we make this dimension dependence explicit in the following.} $f_1:\N\times\R_+\to\R_+$ such that
\begin{align*}
    \lim_{\eps\to 0}f_1(d_{ABC},\eps) = 0,
\end{align*}
 a state $\sigma_{ABC}$ on $\cH_{ABC}$ satisfying
\begin{align*}
    \sigma_{ABC}\approx^{f_1(d_{ABC},\eps)}\rho_{ABC},
\end{align*} 
 and a recovery channel $\cR'_{B\to BC}$ such that
\begin{align}
    (\id_A\otimes\cR'_{B\to BC})\circ\Tr_C(\sigma_{ABC}) = (\id_A\otimes\cR'_{B\to BC})(\sigma_{AB}) = \sigma_{ABC}.
\end{align}
Hence, $\sigma_{ABC}$ is by the results stated around \eqref{eq:QMCRecovery} a quantum Markov chain and we can take $f(d_{ABC},\eps) \coloneqq f_1(d_{ABC},\sqrt{\eps}),$ which finishes the proof.\comment{there exits a fixed point pair, e.g.~take $\sigma_{\star} = \omega_{AB}\otimes\omega_C$ and $\Phi_\star=\cS_{B\to BC}\circ\Tr_C$ with $\cS_{B\to BC} = (\placeholder)\otimes\omega_C,$ with $\omega_{AB}\in\mathfrak{S}(\cH_{AB})$ and $\omega_{C}\in\mathfrak{S}(\cH_{C})$, which clearly satisfies
\begin{align}
\id_A\otimes\Phi_\star(\sigma_{\star}) =\sigma_{\star}.
\end{align}}
\end{proof}

\medskip 

It would be desirable to obtain better control on the approximation function $f$ in Propositon~\ref{prop:RobustnessQMC}. In particular, similarly to the definition of rapid fixability of approximate fixed points (cf.~Defintion~\ref{def:RapidFixingApproxFixedPoint}), we could hope for a bound of the form
\begin{align}
\label{eq:QMCROBPLEASE}
  \min_{\sigma_{ABC}\  \text{QMC}}\,\frac{1}{2}\left\|\rho_{ABC} - \sigma_{ABC}\right\|_1 \stackrel{?}{\le} c\,d^{\,b_A}_Ad^{\,b_B}_Bd^{\,b_C}_C \,\eps^a
\end{align}
for tripartite states $\rho_{ABC}$ satisfying 
\begin{align}
\label{eq:SmallCMIAfterMath}
    I(A:C|B)_\rho\le \eps
\end{align} and some constants $a>0$ and $c,b_A,b_B,b_C\ge 0$ all independent of the dimensions $d_A,d_B,d_C$ and the state $\rho_{ABC}.$ As used in the proof of Proposition~\ref{prop:RobustnessQMC}, we know by \cite{FawziRenner_CMIRecovery_2015} that \eqref{eq:SmallCMIAfterMath} implies the existence of a channel
\begin{align}
\label{eq:RecoverySetQMC}
    \cN\in\cC = \left\{\Phi\in\CPTP(\cH_{ABC})\,\Big|\,\Phi= (\id_A\otimes\cR'_{B\to BC})\circ\Tr_C,\ \, \cR'_{B\to BC}\,\text{ channel }\right\}.
\end{align}
such that
\begin{align*}
    \cN(\rho_{ABC})\approx^{\sqrt{\eps}}\rho_{ABC}.
\end{align*}
Hence, \eqref{eq:QMCROBPLEASE} could be established by proving rapid fixability of approximate fixed points for the sets $\cS = \mathfrak{S}(\cH_{ABC})$ and $\cC$ as above.
However, unfortunately, this is impossible by Corollary~\ref{cor:ImpossiLocalFix} below. In particular,  here, already for the case of system $C$ being trivial, i.e.~$\cH_C=\C$ for which the above reduces to the pair $\cS =\mathfrak{S}(\cH_{AB})$ and $\cC = \id_A\otimes\CPTP(\cH_B),$ rapid fixability of approximate fixed points is ruled out.

Note, however, that this does not exclude the possibility of the bound \eqref{eq:QMCROBPLEASE}. The reason is that for proving \eqref{eq:QMCROBPLEASE} we only need to find 
\begin{align}
    \sigma_{ABC}\approx^{c\,d^{b_A}_Ad^{b_B}_Bd^{b_C}_C \,\eps^a} \rho_{ABC} 
\end{align}
and a channel $\cM$ in the set denoted in \eqref{eq:RecoverySetQMC}, which crucially is allowed to be far from $\cN,$ such that
\begin{align}
    \cM(\sigma_{ABC}) =\sigma_{ABC}
\end{align}
(cf.~\eqref{eq:FixedPointQMC}).\footnote{Note that following the proof in \cite{hayden_structure_2004}, one can show that the knowledge of explicit approximation functions $f$ and $g$ satisfying Definition~\ref{def:FixingApproxFixPoints} for the pair $\cS =\mathfrak{S}(\cH_{AB})$ and $\cC = \id_A\otimes\CPTP(\cH_B)$ gives an explicit bound on  $\min_{\sigma_{ABC}\  \text{QMC}}\,\frac{1}{2}\left\|\rho_{ABC} - \sigma_{ABC}\right\|_1$ in terms of $f$ and $g.$ For that specific proof technique it is important to also have control on the function $g$ which measures the distance of the original and the new channel in Definition~\ref{def:FixingApproxFixPoints}. Therefore, by the impossibility statement Corollary~\ref{cor:ImpossiLocalFix}, proving \eqref{eq:QMCROBPLEASE} with that line of arguing is in fact excluded.}

\subsection{Proof of Proposition~\ref{prop:AbstractFixingApproxFix}}
\label{sec:ProfofPropClosed}
We finish the section by giving a proof of Proposition~\ref{prop:AbstractFixingApproxFix}:

\begin{proof}[Proof of Proposition~\ref{prop:AbstractFixingApproxFix}]
Define the function
\begin{align*}
    F: \cS\times\cC &\to \R_+\\
        (\rho,\cN)&\mapsto \frac{1}{2}\left\|\cN(\rho)-\rho\right\|_1.  
\end{align*}
Firstly, note that $F$ is continuous. Moreover, the preimage $F^{-1}(\{0
\})$ is exactly the set of pairs of states and channels in $\cS\times\cC$ which satisfy an exact fixed point equation and is non-empty by assumption.

Now, since the set $\cS\times\cC$ is closed and hence compact, we see in the following
\begin{align}
\label{eq:CompactArgument}
    \forall \delta>0 \,\exists \eps>0 \,\text{ s.t}\ \forall (\rho,\cN)\in\cS\times\cC \text{ with}\ F(\rho,\cN)\le\eps \ \text{ we have }{\rm{dist}}((\rho,\cN),F^{-1}(\{0\}))\le \delta,
\end{align}
where $${\rm{dist}}((\rho,\cN),F^{-1}(\{0\})) = \min_{(\sigma,\cM)\in F^{-1}(\{0\})}\left(\frac{1}{2}\left\|\rho-\sigma\right\|_1+\frac{1}{2}\left\|\cN-\cM\right\|_\diamond\right).$$
To prove that \eqref{eq:CompactArgument} holds, we assume the opposite to arrive at a contradiction: In that case there exists $\delta>0$ such that for all $n\in\N$ there exists $(\rho_n,\cN_n)\in\cS\times\cC$ with $F(\rho_n,\cN_n)\le\frac{1}{n}$ and ${\rm{dist}}((\rho,\cN),F^{-1}(\{0\}))> \delta.$ But, by compactness of $\cS\times\cC,$ the sequence $\left((\rho_n,\cN_n)\right)_{n\in\N}$ is, up to taking a subsequence, convergent. This already gives a contradiction as by continuity of $F$ the limit of named subsequence must lie in $F^{-1}(\{0\})$.

Define now for $\eps\ge 0$
\begin{align*}
\delta_\star(\eps) = \sup_{(\rho,\cN)\in F^{-1}([0,\eps])}{\rm{dist}}((\rho,\cN),F^{-1}(\{0\})).
\end{align*}
Since $\emptyset\neq F^{-1}(\{0\}) \subseteq F^{-1}([0,\eps]),$ we have $\delta_\star(\eps)\ge 0.$
Furthermore, we can show that
\begin{align}
\label{eq:DeltaStarTo0}
    \lim_{\eps\to0} \delta_\star(\eps)=0.
\end{align}
To see this, let $\delta'>0$ and use that by \eqref{eq:CompactArgument} there exists an $\eps'>0$ such that for all $(\rho,\cN)\in F^{-1}([0,\eps'])$ we have ${\rm{dist}}((\rho,\cN),F^{-1}(\{0\}))\le \delta'$ and therefore 
\begin{align*}
\delta_\star(\eps)\le \delta_\star(\eps')\le \delta'.
\end{align*}
for all $0\le\eps\le \eps',$ where the first inequality follows since by definition $\eps\mapsto \delta_\star(\eps)$ is monotonically increasing.
As $\delta'>0$ was arbitrary, this shows \eqref{eq:DeltaStarTo0}. 

Furthermore, simply by definition of $\delta_\star(\eps)$, we have for all $(\rho,\cN)\in \cS\times\cC$ and $\eps\ge 0$ such that $\cN(\rho)\approx^{\eps}\rho$ that there exists a fixed point pair, i.e. $(\sigma,\cM)\in F^{-1}(\{0\})$ such that
\begin{align*}
\frac{1}{2}\left\|\rho-\sigma\right\|_1+\frac{1}{2}\left\|\cN-\cM\right\|_\diamond\le \delta_\star(\eps).
\end{align*}
We can hence pick $f(\eps)=g(\eps)= \delta_\star(\eps)$ which finishes the proof.
\end{proof}

\comment{The following result from \cite{hayden_structure_2004}. It is essentially a consequence of the Koashi-Imoto theorem~\cite{KoashiImoto} (also see \cite[Theorem 9]{hayden_structure_2004} for an elegant algebraic version of the proof).
\begin{lemma}[\cite{KoashiImoto,hayden_structure_2004}]
Let $\sigma_{AB}$ be a state on a finite-dimensional bipartite Hilbert space $\cH_{AB} = \cH_A\otimes\cH_B$ and $\cM_B\in{\rm{CPTP}}(\cH_B)$ such that the exact fixed point equation holds
\begin{align}
(\id_A\otimes\cM_B)(\sigma_{AB})=\sigma_{AB}.
\end{align}
Then there exists an orthogonal sum decomposition
\begin{align}
\cH_{B} = \bigoplus_{j}\cH_{b^L_j}\otimes\cH_{b^L_j} 
\end{align}
such that
\begin{align}
    \sigma_{AB} = \bigoplus_{j}\,q_j\sigma_{Ab^L_j}\otimes\sigma_{b^R_j}
\end{align}
and for any Stinespring isometry $V:\cH_B\to\cH_B\otimes\cH_E$ of $\cM_B$ we have the form
\begin{align}
    V = \bigoplus_{j} \1_{{b^L_j}}\otimes V_j
\end{align}
with isometries $V_j:\cH_{b^R_j}\to\cH_{b^R_j}\otimes\cH_E.$ 
\end{lemma}
\begin{proof}
The proof follows directly from the proof of Theorem 6 in \cite{hayden_structure_2004}, in particular from the arguments given from Equation (12)-(15) therein. 
\end{proof}

\begin{proof}
By assumption we know
\begin{align}
    (\id_A\otimes\cN_B)(\rho_{AB})\approx^{\eps}\rho_{AB}.
\end{align}
Hence, there exists  
\end{proof}}

\section{Rapid fixability for general quantum and classical channels }
\label{sec:GeneralQuantandClass}
In this section we show rapid fixability of approximate fixed points in the sense of Definition~\ref{def:RapidFixingApproxFixedPoint} for
the following:
\begin{enumerate}
    \item general quantum states and quantum channels, i.e.~for the choice $\cS=\mathfrak{S}(\cH)$ and $\cC = {\rm{CPTP}}(\cH)$ (Theorem~\ref{thm:FixedPointQuantum}),
    \item classical states and classical channels, i.e.~for the choice $\cS={\rm{Prob}}(\cX)$ and $\cC = {\rm{Stoch}}(\cX)$ (Theorem~\ref{thm:FixClassical}).
\end{enumerate}
Here, we explicitly construct for $\rho\in\cS$ and $\cN\in\cC$ and $\eps\ge 0$ satisfying the approximate fixed point equation
\begin{align}
\label{eq:ApproxFixedGeneralQuantSection}
    \cN(\rho) \approx^{\eps} \rho,
\end{align}
a state $\sigma\in\cS$ and a channel $\cM\in\cC$ close to the original ones such that
\begin{align}
\cM(\sigma) =\sigma.
\end{align}
The aim is to have good control on the approximation errors in $\sigma\approx\rho$ and $\cM\approx \cN$ and hence provide explicit forms of the approximation functions $f$ and $g$ in Definitions~\ref{def:FixingApproxFixPoints} and~\ref{def:RapidFixingApproxFixedPoint}.

Both these results follow the same lines of proof. The key insight is to compose the original channel $\cN$ with a particular choice of generalised depolarising channel given by
\begin{align*}
    \Phi = (1-\sqrt{\eps})\, \id + \sqrt{\eps}\,\Tr(\placeholder)\rho.
\end{align*}
The main technical tool for proving Theorem~\ref{thm:FixedPointQuantum} and~\ref{thm:FixClassical} is then given by Lemma~\ref{lem:PertUniqueFix} below. Here, it is shown that the resulting channel $\cM=\Phi\circ\cN$ 
has a unique fixed point state $\sigma,$ which is close to $\rho.$ In particular we show that $\sigma\approx^{\sqrt{\eps}}\rho.$

The results in this section hold true for general separable, possibly infinite dimensional Hilbert spaces or, in the classical case, for general countable classical alphabets $\cX$. In the following we first state the above mentioned results for the quantum case and then for the classical case:
\begin{theorem}[Fixability for general quantum channels]
\label{thm:FixedPointQuantum}
Let $\rho$ be state and $\cN$ be a channel on some separable Hilbert space $\cH$ such that for $\eps\ge 0$ we have
\begin{align}
\label{eq:RhoApproxN}
\cN(\rho) \approx^{\eps} \rho.
\end{align}
Then, there exists a state $\sigma$ and a channel $\cM$ satisfying \begin{align}\sigma\approx^{ \sqrt{\eps}}\rho,\quad\quad \cM\approx^{\sqrt{\eps}}\cN\end{align} and
\begin{align}
\cM(\sigma) =\sigma.
\end{align}
\end{theorem}
\begin{theorem}[Fixability for classical channels]
\label{thm:FixClassical}
Let $P$ be a probability distribution and $T$ be a stochastic mapping on some countable classical alphabet $\cX$ satisfying for some $\eps\ge0$
\begin{align}
 T P \approx^{\eps}  P.
\end{align}
Then, there exists a $S$ stochastic mapping and a  probability distribution $Q$ such that
\begin{align}
  Q \approx^{\sqrt{\eps}} P,\quad\quad S \approx^{\sqrt{\eps}} T
\end{align}
and
\begin{align}
 S Q = Q.
\end{align}
\end{theorem}  
\begin{remark}
Note that Theorems~\ref{thm:FixedPointQuantum} and~\ref{thm:FixClassical} can be seen as improved versions of the results which appeared in the thesis of the first author \cite[Theorem 3.2.1 and 3.2.2]{Salzmann_PhDthesis_2023}. In particular the approximations functions provided by Theorems~\ref{thm:FixedPointQuantum} and~\ref{thm:FixClassical}, i.e. $f(d,\eps)=g(d,\eps) =\sqrt{\eps}$ are unlike the ones in the older versions in \cite{Salzmann_PhDthesis_2023} not dimension dependent. Furthermore, the proofs outlined here are conceptually a lot simpler and shorter. In particular they do not rely on the iterative procedure called $p$-iteration defined and analysed in \cite{Salzmann_PhDthesis_2023}.
\end{remark}
\begin{remark}[Optimality of approximation functions in Theorems~\ref{thm:FixedPointQuantum} and~\ref{thm:FixClassical}]
\label{rem:LowerBoundApprox}
The scaling of the approximation functions in Theorems~\ref{thm:FixedPointQuantum} and~\ref{thm:FixClassical} is essentially optimal: For all $0<\eps\le1$ we can find a state $\rho_\eps$ and a channel $\cN_\eps$ with $\cN_\eps(\rho_\eps)\approx^{\eps} \rho_\eps$ but for every new state $\sigma_\eps$ and new channel\footnote{The statement in fact also holds true when only requiring $\cM_\eps$ to be a positive linear map.} $\cM_\eps$ which satisy
\begin{align}
\label{eq:ApproxLowerBound}
    \sigma_\eps \approx^{f(\eps)}\rho_\eps, \quad\quad\cM_\eps \approx^{g(\eps)}\cN_\eps
\end{align}
and 
\begin{align}
    \label{eq:FixedPointLowerBound}
    \cM_\eps(\sigma_\eps) =\sigma_\eps
\end{align}
we have the lower bound on the approximation functions
\begin{align}
\label{eq:LowerBoundGeneralChannel}
    \max\{f(\eps),g(\eps)\} = \Omega\left(\sqrt{\eps}\right),
\end{align}
i.e. $\max\{f(\eps),g(\eps)\} \ge c \sqrt{\eps}$ for some $c>0$ independent of $0<\eps\le 1$.\footnote{Putting differently this means that we have the worst case bound\\ $\sup_{\substack{\rho,\cN\text{ s.t. }\\\cN(\rho)\approx^{\eps}\rho}}\inf_{\substack{\sigma,\cM\text{ s.t. }\\\cM(\sigma)=\sigma}}\max\{\|\rho-\sigma\|_1,\|\cN-\cM\|_\diamond\}=\Omega(\sqrt{\eps}).$} 
The state $\rho_\eps$ and $\cN_\eps$ can in fact taken to be classical, which hence also shows optimality of the scaling in Theorem~\ref{thm:FixClassical} in the classical case.
Note that this necessary $\sqrt{\eps}$-scaling of the approximation functions in the case of general states and channels is in sharp contrast to the $\eps$-scaling which was observed in \eqref{eq:FixingOnlyLinearStructure} when only demanding that linearity of the map $\cN$ is preserved when changing to $\cM.$

In the following we explicitly construct named state $\rho_\eps$ and channel $\cN_\eps$ on the Hilbert space $\cH=\C^3$ with orthonormal basis $\{\ket{0},\ket{1},\ket{2}\}$.\footnote{The example can then be embedded into any separable Hilbert space $\cH$ with $\dim(\cH)\ge 3$.} Consider
\begin{align*}
    \rho_\eps = \sqrt{\eps}\kb{0} +(1-\sqrt{\eps}) \kb{2}.
\end{align*} and furthermore the stochastic matrix
\begin{align*}
    T_\eps = \begin{pmatrix}
        1-\sqrt{\eps} & \sqrt{\eps} & 0\\ \sqrt{\eps} & 1-\sqrt{\eps} & 0 \\ 0 & 0& 1
    \end{pmatrix}
\end{align*}
with corresponding (classical) channel $\cN_\eps \equiv \cN_{T_\eps}$ where we used the notation \eqref{eq:StochMatrixClassicalChannel}. Note that we have the approximate fixed point equation
\begin{align*}
    \cN_\eps(\rho_\eps) = \left(\sqrt{\eps}-\eps\right)\kb{0} + \eps \kb{1} + (1-\sqrt{\eps}) \kb{2} \approx^{\eps} \rho_\eps.
\end{align*}
Assume for contradiction that there exist a new quantum state $\sigma_\eps$ and a new quantum channel $\cM_\eps$ which satisfy \eqref{eq:ApproxLowerBound} and \eqref{eq:FixedPointLowerBound} but with approximation functions satisfying\footnote{To really consider the negation of \eqref{eq:LowerBoundGeneralChannel} we should consider general sequences $(\eps_n)_{n\in\N}\subseteq(0,1]$ such that $\lim_{n\to \infty}\frac{\max\{f(\eps_n),g(\eps_n)\}}{\sqrt{\eps_n}} = 0.$ However, also in this case all of the following calculations just through and we hence restrict to limits $\eps\to 0$ for notational simplicity.}
\begin{align}
\label{eq:CounterExampleSpeed}
\lim_{\eps\to 0}\frac{\max\{f(\eps),g(\eps)\}}{\sqrt{\eps}} = 0.
\end{align}
Denoting the diagonal and off-diagonal part of $\sigma_\eps$ by $\sigma^{\text{diag}}_\eps$ and $\sigma^{\text{off}}_\eps$ respectively, we see by \eqref{eq:ApproxLowerBound} and the fact that $\rho_\eps$ is diagonal that $\frac{1}{2}\left\|\sigma^{\text{off}}_\eps\right\|_1 \le f(\eps).$ Using this, \eqref{eq:VarExpTraceLess}, \eqref{eq:ApproxLowerBound} again and furthermore the fact that $\cN_\eps(\sigma^{\text{off}}_\eps)=0$ gives
\begin{align*}
    \left|\bra{1}\cM_\eps(\sigma^{\text{off}}_\eps)\ket{1}\right| &=\left|\bra{1}\left(\cM_\eps(\sigma^{\text{off}}_\eps)-\cN_\eps(\sigma^{\text{off}}_\eps)\right)\ket{1}\right| \le \frac{f(\eps)}{2}\left\|\cM_\eps-\cN_\eps\right\|_\diamond \le f(\eps)g(\eps)\\&
\end{align*}
From this, positivity of $\cM_\eps$ and the definition of $\rho_\eps$ and $\cN_\eps$ combined with \eqref{eq:ApproxLowerBound} and \eqref{eq:FixedPointLowerBound}  we see
\begin{align*}
    0&=\bra{1}(\cM_\eps(\sigma_\eps) -\sigma_\eps)\ket{1} =\bra{1}\cM_\eps(\sigma^{\text{diag}}_\eps)\ket{1} +\bra{1}\cM_\eps(\sigma^{\text{off}}_\eps)\ket{1}  -\bra{1}\sigma_\eps)\ket{1} \\ &\ge \bra{1}\cM_\eps(\kb{0})\ket{1}\,\bra{0}\sigma_\eps\ket{0} +\left(\bra{1}\cM_\eps(\kb{1})\ket{1} -1\right)\,\bra{1}\sigma_\eps\ket{1}- f(\eps)g(\eps)\\&\ge \left(\sqrt{\eps} -g(\eps)\right)\left(\sqrt{\eps}-f(\eps)\right) -\left(\sqrt{\eps} +g(\eps)\right)f(\eps)- f(\eps)g(\eps)\\&=\eps -\sqrt{\eps}\left(2f(\eps) +g(\eps)\right) - f(\eps)g(\eps)
\end{align*}
Dividing by $\eps$ this gives
\begin{align}
\label{eq:ContradictionCounter}
    \frac{\left(2f(\eps) +g(\eps)\right)}{\sqrt{\eps}} +\frac{f(\eps)g(\eps)}{\eps}\ge 1
\end{align}
which gives a contradiction with \eqref{eq:CounterExampleSpeed} for $\eps>0$ small enough as the left hand side of \eqref{eq:ContradictionCounter} converges to $0$ as $\eps\to 0.$ Therefore, we have proven \eqref{eq:LowerBoundGeneralChannel}.
\end{remark}
We continue to give the proof of Theorems~\ref{thm:FixedPointQuantum} and~\ref{thm:FixClassical} and start with the above mentioned key lemma:
\begin{lemma}
\label{lem:PertUniqueFix}
Let $\cN$ be a channel on some separable Hilbert space $\cH.$ For all states $\tau$ and $\lambda \in(0,1]$ we can define the channel
\begin{align}
    \cM = (1-\lambda)\cN +\lambda\Tr(\placeholder)\tau,
\end{align}
which satisfies that \begin{align}
    P = \lim_{k\to\infty} \cM^k
\end{align}
exists and $\dim(\ran(P)) = 1,$ i.e. $\cM$ has a unique fixed point state. Furthermore, we have for all states $\rho$
\begin{align}
\label{eq:PerturbChannelBound}
    \left\|P(\rho) - \rho\right\|_1 \le \frac{\left\|\cM(\rho) - \rho\right\|_1}{\lambda}.
\end{align}
\end{lemma}
\begin{proof}
Let $\omega_1,\omega_2$ be states and note that for $j\in\N$ we have 
\begin{align}
\label{eq:IterativeChannelBound}
\nn\left\|\cM^{j+1}(\omega_1) -\cM^j(\omega_2) \right\|_1 &= (1-\lambda) \left\|\cN\left(\cM^{j}(\omega_1)-\cM^{j-1}(\omega_2)\right)\right\|_1 \\&\le\nn (1-\lambda) \left\|\cM^{j}(\omega_1)-\cM^{j-1}(\omega_2)\right\|_1 \\&\le \cdots \le (1-\lambda)^j \left\|\cM(\omega_1) -\omega_2\right\|_1 \le 2 (1-\lambda)^j .
\end{align}
From this we see that for $k,k'\in\N$ with $k' > k$ 
\begin{align}
\label{eq:FixpointCauchy}
    \left\|\cM^{k'}(\omega_1) -\cM^{k}(\omega_2)\right\|_1 \le \sum_{j=k}^{k'-1}\left\|\cM^{j+1}(\omega_1) -\cM^{j}(\omega_2)\right\|_1 \le 2\sum_{j=k}^{k'-1}\left(1-\lambda\right)^j \xrightarrow[k',k\to\infty]{}0,
\end{align}
where the limit follows by $\lambda>0$ and hence finiteness of the geometric series.
From that we see that $\left(\cM^k\right)_{k\in\N}$ is a Cauchy sequence and hence converges to a quantum channel $P = \lim_{k\to\infty}\cM^k.$  Note that $P$ is hence equal to the limit of the Ces\`aro means of $\cM$ (cf. \eqref{eq:CesaroMean}) and therefore equal to the projection onto the fixed point space of $\cM.$ Furthermore, \eqref{eq:FixpointCauchy} also gives that
\begin{align*}
    P(\omega_1) = \lim_{k\to\infty}\cM^{k}(\omega_1) = \lim_{k\to\infty}\cM^{k}(\omega_2) = P(\omega_2)  
\end{align*}
which shows $\dim(\ran(P)) = 1$ and hence that $\cM$ has a unique fixed point state. Furthermore, to prove \eqref{eq:PerturbChannelBound} we note by using \eqref{eq:IterativeChannelBound} again for $\rho$ being a state that
\begin{align*}
    \left\|P(\rho)-\rho\right\|_1 &= \lim_{k\to\infty}\left\|\cM^k(\rho)-\rho\right\|_1\le\lim_{k\to\infty}\sum_{j=1}^{k-1}\left\|\cM^{j+1}(\rho)-\cM^{j}(\rho)\right\|_1 \\&\le \sum_{j=1}^{\infty}\left(1-\lambda\right)^j\left\|\cM(\rho)-\rho\right\|_1 \le \frac{\left\|\cM(\rho)-\rho\right\|_1}{\lambda}.
\end{align*}
\end{proof}

\begin{proof}[Proof of Theorem~\ref{thm:FixedPointQuantum}]
 We apply Lemma~\ref{lem:PertUniqueFix} for the specific choice $\tau=\rho$ and $\lambda =\sqrt{\eps},$ i.e. for the channel
\begin{align*} 
\cM = (1-\sqrt{\eps})\,\cN + \sqrt{\eps}\,\Tr(\placeholder)\rho.
\end{align*}
Note that from \eqref{eq:RhoApproxN} it is immediate that also $\cM$ and $\rho$ satisfy an approximate fixed point equation with 
\begin{align}
\label{eq:MAlsoGoodAprox}
\cM(\rho) \approx^{\eps} \rho.
\end{align}
Furthermore, we have 
\begin{align*}
\frac{1}{2}\left\|\cM-\cN\right\|_\diamond \le \frac{\sqrt{\eps}}{2}\left(\left\|\cN\right\|_\diamond+\left\|\Tr(\placeholder)\rho\right\|_\diamond\right) \le \sqrt{\eps}.
\end{align*}
Define the state $\sigma = P(\rho) = \lim_{k\to\infty}\cM^{k}(\rho)$ which by definition is a fixed point of $\cM.$ Moreover, combining now \eqref{eq:MAlsoGoodAprox}  with \eqref{eq:PerturbChannelBound} finishes the proof of the theorem as we have
\begin{align*}
    \frac{1}{2}\left\|\sigma -\rho\right\|_1 \le \frac{1}{2}\frac{\left\|\cM(\rho)-\rho\right\|_1}{\sqrt{\eps}}\le \sqrt{\eps}.
\end{align*}

\end{proof}

\begin{proof}[Proof of Theorem~\ref{thm:FixClassical}]
The proof follows exactly the same lines as the one of Theorem~\ref{thm:FixedPointQuantum} by noting that the new channel $\cM$ in there is also classical and hence corresponds to a specific choice of a stochastic mapping $S.$
\end{proof}

\section{Rapid fixability for strict subsets of states and channels}
\label{sec:FixStrictSubUnitaryEtc}
In this section we prove for finite dimensional quantum systems rapid fixability of approximate fixed points in the sense of Definition~\ref{def:RapidFixingApproxFixedPoint} for a variety of choices $\cS$ and $\cC.$
In particular, here we focus on $\cS =\mathfrak{S}(\cH)$ the full set of quantum states
and \begin{enumerate}
    \item\label{it:Unitary} $\cC$ being the set of unitary channels (Theorem~\ref{thm:FixUnitary}),
    \item \label{it:MixUnitary} $\cC$ being the set of mixed-unitary channels (Theorem~\ref{thm:FixingMixtureUnitaries}),
    \item \label{it:Unital} $\cC$ being the set of unital channels (Theorem~\ref{thm:FixingUnitalChannels}).
\end{enumerate}
Furthermore, in Theorem~\ref{thm:ApproxLocalFixPure} we show rapid fixiability of approximate fixed points for 
\begin{enumerate}
\setcounter{enumi}{3}
    \item \label{it:LocalFix} $\cS$ being the set of pure states on some bipartite Hilbert space $\cH_{AB}=\cH_A\otimes\cH_B$ and $\cC$ being the set of local channels of the form $\id_A\otimes\cN_B$ with $\cN_B\in{\rm{CPTP}}(\cH_B).$
\end{enumerate}

All these results rely on similar proof techniques, which is why we present them together in this section. For $\cS$ and $\cC$ being one of the choices above and an approximate fixed point equation
\begin{align}
\label{eq:ApproxUnitaryLocalSection}
    \cN(\rho) \approx^{\eps} \rho
\end{align}
with $\eps\ge0$ and $\rho\in\cS$ and $\cN\in\cC,$ we explicitly construct a fixed point pair $\sigma\in\cS$ and $\cM\in\cC$ approximating the original state and channel. The focus lies on bounding the approximation errors in $\sigma\approx\rho$ and $\cM\approx\cN$ and by that providing explicit forms of the approximation functions $f$ and $g$ from Definitions~\ref{def:FixingApproxFixPoints} and~\ref{def:RapidFixingApproxFixedPoint}.

Notably, in the results of this section, the corresponding new state $\sigma$ only depends on the original state $\rho$ and the approximation error $\eps$ but not on the channel $\cN.$ For the results \ref{it:Unitary}-\ref{it:Unital} the choice of $\sigma$ relies on spectral clustering techniques which we outline in Section~\ref{sec:SpectralClustering} below. 
Here, spectral points of $\rho,$ which are close to each other, get clustered together in such a way that the resulting state has significant spectral gaps between different eigenvalues. From the approximate fixed point equation \eqref{eq:ApproxUnitaryLocalSection} and a standard matrix analytic result for continuity of spectral subspaces under spectral gaps (see Lemma~\ref{lem:BhathiaSpectralProj} below), we then see that the resulting spectral projections are in some sense approximate fixed points of the channel $\cN$ as well. It then remains to fix named approximate fixed point equations for the spectral projections by changing the channel $\cN$ to $\cM.$ For that we establish in Section~\ref{sec:TurnSubspaces} a 
plethora of results, which we call \emph{rotation lemmas} and which are concerned with the question of how to rotate two or more close-by subspaces onto each other. 

For the result~\ref{it:LocalFix}, instead of spectral clustering techniques, we define the new state by setting small Schmidt coefficients of the original pure state to zero. By that, similarly as due to the spectral gaps mentioned above, we find approximate fixed point equations for each the remaining Schmidt vectors. Using again a rotation lemma then enables us to find a close-by channel which has the new state as a fixed point.

The size of the spectral gaps or, in the case of result~\ref{it:LocalFix}, the size of the neglected Schmidt-coefficients is parametrised by some
\begin{align}
     \delta \gg \eps
\end{align}
and relates to the approximation error of the state in $\sigma\approx\rho.$ Choosing a smaller $\delta$ leads to a better approximation of the state but worse approximation of the channel in $\cM\approx\cN$ as the error in the above-mentioned approximate fixed point equations for the remaining spectral projections and Schmidt vectors becomes larger. The last step of all of the proofs in this section is then given by an optimisation over $\delta$ in this trade-off to guarantee good approximation of both state and channel.

In the remainder of this section we first discuss named spectral clustering technique, then the rotation lemmas and conclude with the precise statements and proofs of the results~\ref{it:Unitary} \,-
\ref{it:LocalFix}.

\subsection{Spectral clustering}
\label{sec:SpectralClustering}
In the following we discuss a technique to separate the spectrum of a quantum state into \emph{(spectral) clusters,}  such that spectral points within a cluster are close to each other whereas different clusters are far from another. This technique is then used for proving Theorems~\ref{thm:FixUnitary},~\ref{thm:FixingMixtureUnitaries} and~\ref{thm:FixingUnitalChannels}. 

Before discussing the spectral clustering technique, let us recall some well-known facts about the continuity of spectral points and spectral projections of self-adjoint operators on some $d$-dimensional Hilbert space:  For $A_1,A_2$ being self-adjoint we have by \cite[Equation (IV.62)]{Bhatia_MatrixAnalysis_1997}\footnote{In fact in \cite[Equation (IV.62)]{Bhatia_MatrixAnalysis_1997} the inequality \eqref{eq:ContOfSpectrum} is stated for all unitarily invariant norms.} the inequality 
\begin{align}
\label{eq:ContOfSpectrum}
    \left\|\overrightarrow{\spec}(A_1)-\overrightarrow{\spec}(A_2)\right\|_1\le \left\|A_1-A_2\right\|_1, 
\end{align}
where we denoted by $\overrightarrow{\spec}(A_i)$ the vector containing the ordered spectral points of $A_i$. This can be seen as a continuity statement of the spectral points of self-adjoint operators. 

Such a continuity statement does, however, not hold for the respective spectral projections. To see this consider as an example the family of operators defined for $\eps> 0$ by
\begin{align*}
    A_1(\eps)\coloneqq \frac{(1+\eps)}{2}\kb{0} +\frac{(1-\eps)}{2}\kb{1},\quad\quad A_2(\eps)\coloneqq \frac{(1+\eps)}{2}\kb{+} +\frac{(1-\eps)}{2}\kb{-}
\end{align*}
where $\{\ket{0},\ket{1}\}$ denotes some ortonormal set, $\ket{+} \coloneqq \left(\ket{0}+\ket{1}\right)/\sqrt{2}$ and $\ket{-} \coloneqq \left(\ket{0}-\ket{1}\right)/\sqrt{2}.$ It is easy to see that $A_1(\eps)$ and $A_2(\eps)$ are close as 
\begin{align*}
    \left\|A_1(\eps) -A_2(\eps)\right\| \le \eps,
\end{align*}
whereas the corresponding spectral projections are far apart, i.e.
\begin{align*}
    \left\|\kb{0}\,-\,\kb{+}\right\| =  \frac{1}{\sqrt{2}},\quad\quad \left\|\kb{1}\,-\,\kb{-}\right\| =  \frac{1}{\sqrt{2}}.
\end{align*}

Assuming, however, gaps in the spectrum of $A_1$ and $A_2,$ it is well-known that some form continuity of spectral projections can be recovered. This is stated as Theorem VII.3.1 in \cite{Bhatia_MatrixAnalysis_1997}, which we recall in the following lemma.

\begin{lemma}[Theorem VII.3.1 in \cite{Bhatia_MatrixAnalysis_1997}]
\label{lem:BhathiaSpectralProj}
Let $A_1,A_2$ be two self-adjoint operators on a $d$-dimensional Hilbert space $\cH$ with spectral decompositions
\begin{align}
    A_1 = \sum_{i=1}^d \lambda_i\kb{\varphi_i}\quad\text{and}\quad A_2 = \sum_{i=1}^d \mu_i\kb{\psi_i}.
\end{align}
Let $J\subseteq\R$ be an interval separating parts of the spectrum of $A_1$ and $A_2$ respectively, i.e.
\begin{align*}
    {\rm{dist}}(J\cap{\spec(A_i)},J^c\cap{\spec(A_i)})\ge\delta
\end{align*}
for $i=1,2$ and some $\delta>0.$ Then for the spectral projections of $A_1$ and $A_2$ given by
\begin{align}
    P_{A_1}(J) \coloneqq \sum_{i\text{ s.t. }\lambda_i\in J}\kb{\varphi_i}  \quad\text{and}\quad  P_{A_2}(J) \coloneqq \sum_{i\text{ s.t. }\mu_i\in J}\kb{\psi_i}
\end{align}
we have\footnote{Note again that in  \cite[Theorem VII.3.1]{Bhatia_MatrixAnalysis_1997} the inequality \eqref{eq:ProjSpectralGapClosBhat} is stated for all unitarily invariant norms but we only need it for the operator norm in the following.}
\begin{align}
\label{eq:ProjSpectralGapClosBhat}
\left\|P_{A_1}(J)-P_{A_2}(J)\right\| \le\frac{2\|A_1-A_2\|}{\delta}.
\end{align}
\end{lemma}
\begin{proof}
The lemma follows by noting
\begin{align*}
  \left\|P_{A_1}(J)-P_{A_2}(J)\right\| \le   \left\|(\1-P_{A_2}(J))P_{A_1}(J)\right\| + \left\|P_{A_2}(J)(\1-P_{A_1}(J))\right\| \le \frac{2\|A_1-A_2\|}{\delta}.
\end{align*}
where the last inequality is true due to Theorem VII.3.1 in \cite{Bhatia_MatrixAnalysis_1997}.
\end{proof}

The spectral clustering technique defined below is a way to ensure the existence of spectral gaps larger than $\delta$ between two respective spectral clusters. We can hence combine it with Lemma~\ref{lem:BhathiaSpectralProj} to provide continuity bounds for the spectral projections of the respective clusters. 
\\\indent Let us now introduce the spectral clustering technique: We write a state $\rho$ on a $d$-dimensional Hilbert space $\cH$ in spectral decomposition
\begin{align*}
    \rho = \sum_{i=1}^m\lambda_i\pi_i,
\end{align*}
with $m\in[d]$, spectral points $\lambda_1>\lambda_2>\cdots>\lambda_n\ge 0$ and $\left(\pi_i\right)_{i=1}^m$ non-zero orthogonal projections summing to $\1$. For $\delta\ge 0$ fixed, the idea is to cluster the spectral points in a way that each point of a cluster is $\delta$-close to the next one in the cluster and different clusters are separated from each other with distance strictly larger than $\delta.$
For that, to define the first cluster, let $k_1\in[d]$ be the smallest number such that 
\begin{align*}
 \lambda_{k_1} - \lambda_{k_1+1} > \delta.
\end{align*} 
The first cluster is then defined as $I_1 \coloneqq [k_1]$ and consequently we have 
\begin{align*}
    \lambda_{i} -\lambda_{i+1} \le \delta\quad \text{ for all $i\in I_1$.}
\end{align*}
By iterating the above, we obtain $k_1 <k_2 < k_3< \cdots < k_n \in[m-1]$ such that $k_l$ is the $l^{th}$ smallest number with
\begin{align}
\label{eq:SpectralCluster}
 \lambda_{k_l} - \lambda_{k_l+1} > \delta
\end{align} 
and moreover $k_n$ is additionally also the largest number such that \eqref{eq:SpectralCluster} occurs. The $l^{th}$ cluster is then defined to be $I_l \coloneqq \{k_{l-1}+1,\cdots, k_l\}$ and by definition we have
\begin{align}
\label{eq:Clustering}
    \nn\lambda_{i} -\lambda_{i+1} &\le \delta \quad\text{ for all $i\in I_l$ and}\\|\lambda_i - \lambda_j| &> \delta \quad\text{ for all $i\in I_{l}, j\in I_{l'}$ with $l\neq l'.$} 
\end{align}
We then define for each cluster the corresponding spectral projection
\begin{align}
\label{eq:ClusterProjection}
    E_l \coloneqq \sum_{i\in I_l} \pi_i  = \sum_{i=k_{l-1}+1}^{k_l}\pi_i
 \end{align}
and denote the average spectral point by
\begin{align}
\label{eq:AverageSpecCluster}
    \mu_l \coloneqq \frac{1}{\Tr(E_l)}\sum_{i\in I_l}\lambda_i\Tr(\pi_i)
\end{align}
With that we can define a new quantum state 
\begin{align}
\label{eq:SpecClusterState}
    \sigma \coloneqq \sum_{l=1}^n \mu_l E_l.
\end{align}
Note that $\sigma$ indeed defines a quantum state as positive semi-definiteness is immediate from the definition and normalisation follows by
\begin{align*}
    \Tr(\sigma) = \sum_{l=1}^m \mu_l \Tr(E_l) = \sum_{l=1}^m\sum_{i\in I_l}  \lambda_i\Tr(\pi_i) = \Tr(\rho) =1.
\end{align*}
The following lemma shows that for $\delta$ small, $\sigma$ is a good approximation of $\rho.$
\begin{lemma}
\label{lem:ApproxClusterState}
Let $\rho$ be a state on $d$-dimensional Hilbert space $\cH$ and $\delta\ge 0.$ Denote by $\sigma$ the state defined through the spectral clustering for $\rho$ and $\delta$ as outlined above, i.e.~\eqref{eq:SpecClusterState}. 
Then we have 
\begin{align}
    \frac{1}{2}\left\|\rho-\sigma\right\|_1\le \frac{d^2\delta}{2}.
\end{align}
\end{lemma}
\begin{proof}
Note that by construction of the clustering \eqref{eq:Clustering}, the averages $\mu_l$ satisfy for all $l\in[n]$ and $i\in I_l$
\begin{align*}
|\lambda_i-\mu_l| \le \sum_{j\in I_l} \frac{\Tr(\pi_j)}{\Tr(E_l)}|\lambda_i-\lambda_j| \le \sum_{j\in I_l}\frac{\Tr(\pi_j)}{\Tr(E_l)}|I_l|\delta =|I_l|\delta.
\end{align*}
By that we see
\begin{align*}
    \frac{1}{2}\left\|\rho -\sigma\right\|_1 &= \frac{1}{2}\left\|\sum_{l=1}^m\sum_{i\in I_l}(\lambda_i -\mu_l)\pi_i\right\|_1 \le \frac{1}{2}\sum_{l=1}^m\sum_{i\in I_l}\Tr(\pi_i)|\lambda_i - \mu_l| \\&\le \frac{\delta}{2}\sum_{l=1}^m\sum_{i\in I_l}\Tr(\pi_i)|I_l| \le\frac{d^2\delta}{2},
\end{align*}
where we have used that $\sum_{l=1}^n\sum_{i\in I_l}\Tr(\pi_i)=d.$
\end{proof}

\subsection{Rotation lemmas}
\label{sec:TurnSubspaces}

In this section we state and prove the rotation lemmas mentioned above, which are a main ingredient in the proofs of Theorems~\ref{thm:FixUnitary}, \ref{thm:FixingMixtureUnitaries}, \ref{thm:FixingUnitalChannels} and \ref{thm:ApproxLocalFixPure}. All of the rotation lemmas are concerned with providing an unitary close to the identity which rotates nearby subspaces onto each other. To be more precise we find unitaries $U\approx \1$ which in the case of \begin{enumerate}
    \item \label{it:RotLem1} $\left(\ket{v_i}\right)_{i=1}^n$ and $\left(\ket{w_i}\right)_{i=1}^n$ being orthonormal systems with $\ket{v_i}\approx\ket{w_i}$ satisfy $U\ket{v_i} =\ket{w_i}$ (Lemma~\ref{lem:UnitaryCloseVectors}),
    \item  \label{it:RotLem2}$\left(\ket{\psi_i}\right)_{i=1}^n$ being an orthonormal system  and  $F$ being an orthogonal projection with $F\ket{\psi_i}\approx\ket{\psi_i}$ satisfy $U\ket{\psi_i} \in\ran(F)$ (Lemma~\ref{lem:TurnVectorsSubspace}),
    \item \label{it:RotLem3}$\left(E_i\right)_{i=1}^n$ and $\left(F_i\right)_{i=1}^n$ being orthogonal projections with $E_i\approx F_i$ satisfy $UE_iU^* =F_i$ (Lemma~\ref{lem:TurningSubspaces}).
\end{enumerate}
We first state and prove Lemmas~\ref{lem:CloseProjSameDim} and~\ref{lem:Turning1Subspace}, which are at the core of the proofs of the results mentioned above and then continue to show the rotation lemmas.

\begin{lemma}
	\label{lem:CloseProjSameDim}
    Let $A$ be a linear operator and $E$ be a projection on some finite dimensional Hilbert space $\cH$  satisfying
    \begin{align}
		\|A-E\|<1.
	\end{align}
 Then $\dim(\ran(A))\ge \dim(\ran(E)).$
 In particular, this gives by symmetry for $E$ and $F$ being projections, satisfying
	\begin{align}
		\|E-F\|<1,
	\end{align}
that $\dim(\ran(E)) = \dim(\ran(F)).$
\end{lemma}
\begin{proof}
    We write $d_E = \dim(\ran(E)).$
	Assume for contradiction $\dim(\text{ran}(E)) > \dim(\text{ran}(A))$ and take $\left(\varphi_i\right)_{i=1}^{d_E}$ to be a basis of $\text{ran}(E)$. As $\left(A\varphi_i\right)_{i=1}^{d_E}$ cannot be linearly independent, there exists $(\lambda_i)_{i=1}^{d_E} \neq 0\in\C^{d_E}$ such that for $\varphi = \sum_{i=1}^{d_E} \lambda_i \varphi_i\neq 0\in\cH$ we have $A\phi=0$.
	But this gives
	\begin{align*}
		\|E-A\| \ge \frac{\|(E-A)\varphi\|}{\|\varphi\|} = \frac{\|E\varphi\|}{\|\varphi\|} = \frac{\|\varphi\|}{\|\varphi\|}=1
	\end{align*}
	contradicting the assumption of the lemma and therefore giving $\dim(\text{ran}(E)) \le \dim(\text{ran}(A)).$ 
\end{proof}
\begin{lemma}
\label{lem:Turning1Subspace}
Let $E$ and $F$ be orthogonal projections on some finite dimensional Hilbert space $\cH$ satisfying
\begin{align*}
\|E-F\| \le \eps
\end{align*}
for some $0\le\eps< 1$. Then there exists an unitary $U$ with $\|U-\1\| \le 2\eps $ and 
\begin{align}
\label{eq:TurnedProj}
UEU^* = F.
\end{align}
\end{lemma}
\begin{proof}
By assumption and Lemma~\ref{lem:CloseProjSameDim} we have $\dim(\ran(E))=\dim(\ran(F))$ and we write in the following $d_E=\dim(\ran(E)).$
Define now
\begin{align*}
A \coloneqq FE + (\1-F)(\1-E)
\end{align*}
and note that 
\begin{align}
	\label{eq:Acloseness}
	\|A- \1\|= \|(F-E)E - (F-E)(\1-E)\| =\|F-E\|\le \eps,
	\end{align}
    where for the second equality we have used that $\1-2E$ is unitary.
	By singular value decomposition (SVD) we can write
	\begin{align}
	\label{eq:SVDProjectors}
	A = \sum_{i=1}^{d_E} a_i\ket{v_i}\!\bra{w_i} + \sum_{i=d_E+1}^{d} a_i\ket{v_i}\!\bra{w_i}
	\end{align}
	 with $a_i\ge 0$ being the singular values of $A$ and $\left(\ket{v_i}\right)_{i=1}^{d_E}$ and $\left(\ket{w_i}\right)_{i=1}^{d_E}$ being orthonormal bases of $\text{ran}(F)$ and $\text{ran}(E)$  respectively and $\left(\ket{v_i}\right)_{i=d_E+1}^{d}$ and $\left(\ket{w_i}\right)_{i=d_E+1}^{d}$ being orthonormal bases of $\text{ran}(\1-F)$ and $\text{ran}(\1-E)$  respectively. That the SVD of $A$ has this form follows by block diagonality of $A$ or equivalently by considering SVDs of $FE$ and $(\1-F)(\1-E)$ individually leading to the first and second summand in \eqref{eq:SVDProjectors}. 
	 
	 Denoting the smallest singular value of $A$ by  $a_{i'}$ we find
	 \begin{align*}
	 \eps\ge \|A-\1\| &\ge \|a_{i'}\ket{v_{i'}} - \ket{w_{i'}}\| = \sqrt{a^2_{i'}+1 -2a_{i'}\Re\langle v_{i'},w_{i'}\rangle} \ge \sqrt{a^2_{i'} +1 -2a_{i'}} \\&=1-a_{i'} = \max_{i=1,\dots,d}|1-a_i|,
	 \end{align*}
	 where we have used in the second line that $a_i\le1$ for all $i=1,\dots,d$ which holds since $A^*A = EFE +(\1-E)(\1-F)(\1-E)\le \1.$
	  Define now the unitary 
 \begin{align*}
U = \sum_{i=1}^{d} \ket{v_i}\!\bra{w_i}.
\end{align*} By the above we have
 \begin{align}
 \label{eq:Ucloseness}
\|U-\1\|\le \|U-A\| + \|A-\1\| = \max_{i=1,\dots,d}|a_i-1| + \|A-\1\| \le 2\eps.
\end{align}
Moreover, by construction we have $UEU^* =F$. 
\end{proof}

We now state and prove the the rotations lemmas mentioned in points~\ref{it:RotLem1}-\ref{it:RotLem3} above.
\begin{lemma}
	\label{lem:UnitaryCloseVectors}
	Let $\left(\ket{v_i}\right)_{i=1}^n$ and $\left(\ket{w_i}\right)_{i=1}^n$ be orthonormal systems on some finite dimensional Hilbert space $\cH$ such that
	\begin{align*}
	\|\ket{v_i} -\ket{w_i}\| \le \eps
	\end{align*}
	for some $\eps\ge 0$. Then there exists an unitary $U$ such that $\|U- \1\| \le 5\sqrt{n}\,\eps$ and
 \begin{align}
	    U\ket{v_i} = \ket{w_i}
	\end{align}
 for all $i\in[n].$
\end{lemma}
\begin{proof}
Define the orthogonal projections $E =\sum_{i=1}^n\kb{v_i}$ and $F=\sum_{i=1}^n\kb{w_i}$
and note that they satisfy
\begin{align}
\label{eq:MaaanNochMehrTurningLemmas}
    \left\|E-F\right\| \le \left\|\sum_{i=1}^n\left(\ket{v_i}-\ket{w_i}\right)\!\bra{v_i}\right\| + \left\|\sum_{i=1}^n\left(\ket{v_i}-\ket{w_i}\right)\!\bra{w_i}\right\| \le 2\eps\sqrt{n}, 
\end{align}
where we used the triangle inequality, the $*$-invariance of the operator norm and for the last inequality
\begin{align*}
    \left\|\sum_{i=1}^n\left(\ket{v_i}-\ket{w_i}\right)\!\bra{v_i}\right\| \le \sup_{\|\psi\|=1}\sum_{i=1}^n\left\|\ket{v_i}-\ket{w_i}\right\||\langle v_i,\psi\rangle| \le \eps\sqrt{n} \sup_{\|\psi\|=1}\sqrt{\sum_{i=1}^n|\langle v_i,\psi\rangle|^2} \le \eps\sqrt{n},
\end{align*}
which follows by the Cauchy-Schwarz inequality and similarly for the second term in \eqref{eq:MaaanNochMehrTurningLemmas}.
Hence, by Lemma~\ref{lem:Turning1Subspace} there exists an unitary $U_1$ such that $\|U_1-\1\|\le4\eps\sqrt{n}$ and
\begin{align*}
U_1EU_1^* = F.
\end{align*}
Furthermore, we define
the partial isometry 
\begin{align*}
    U_2=\sum_{i=1}^{n}\ket{w_i}\!\bra{v_i}
\end{align*} 
with support $\ran(E)$ and image $\ran(F).$ Clearly $U_2\ket{v_i} = \ket{w_i}$. Moreover, 
	\begin{align*}
	\|U_2-E\| &= \sup_{\|\psi\|=1} \|(U_2-E)\ket{\psi}\|  \le \sup_{\|\psi\|=1}\sum_{i=1}^n \left\| \ket{w_i}-\ket{v_i}\right\||\langle v_i,\psi\rangle|\le \eps \sqrt{n}.
	\end{align*}
	We can now define 
 \begin{align*}
     U = U_2 \oplus U_1(\1-E),
 \end{align*}
 which is unitary since $U_1(\1-E)$ is a partial isometry with support $\ran(\1-E)$ and image $\ran(\1-F).$ 
 Furthermore, it satisfies $U\ket{v_i}=\ket{w_i}$ and
 \begin{align*}
     \left\|U-\1\right| \le \left\|U_2-E\right\|+ \left\|U_1-\1\right\| \le 5\eps\sqrt{n}.
 \end{align*}
\end{proof}

\begin{lemma}
\label{lem:TurnVectorsSubspace}
Let $\left(\ket{\psi_i}\right)_{i=1}^n$ be an orthonormal system and $F$ be some orthogonal projection on some finite dimensional Hilbert space $\cH$ satisfying 
\begin{align}
\label{eq:TurnVectorsSquareAss}
  \sqrt{\sum_{l=1}^n  \left\|(\1-F)\ket{\psi_i}\right\|^2} \le \eps.
\end{align}
 for some $0\le \eps< 1$ and all $i\in[n].$ Then there exists an unitary $U$ such that $\|U-\1\| \le 2\, \eps$ and 
\begin{align}
\label{eq:TurnVectorsInSubspace}
FU \ket{\psi_i} = U \ket{\psi_i}
\end{align}
for all $i\in[n].$
In particular if 
\begin{align}
\label{eq:TurnVectorsNonSquared}
    F\ket{\psi_i} \approx^{\eps}\ket{\psi_i}
\end{align}
for some $0\le \eps< 1/\sqrt{n},$ then there exists an unitary $U$ such that $\|U-\1\| \le 2\sqrt{n}\, \eps$ satisfying \eqref{eq:TurnVectorsInSubspace}.
\end{lemma}
\begin{proof}
 We define the orthogonal projection $E = \sum_{i=1}^n \kb{\psi_i}$.  Note that we have by the Cauchy-Schwarz inequality and assumption \eqref{eq:TurnVectorsSquareAss}
\begin{align*}
\|E-FE\| &= \|(\1-F)E\| \le \sup_{ \|\psi\| =1}\sum_{i=1}^n\|(\1-F)\ket{\psi_i}\|\,|\langle \psi_i,\psi\rangle| \\&\le \eps\sup_{ \|\psi\| =1}\sqrt{\sum_{i=1}^n|\langle \psi_i,\psi\rangle|^2} \le \eps<1.
\end{align*}
Using now Lemma~\ref{lem:CloseProjSameDim} we see $\dim(\ran(FE))\ge \dim(E)$ and since the opposite inequality holds trivially we have $\dim(\ran(FE))= \dim(E).$
Let now $F'$ be the orthogonal projection onto $\text{ran}(FE)$ and note that it satisfies $F'=F'F$ and hence $F'E =F'FE=FE$. Since, as established $\dim(\ran(F')) = \dim(\ran(FE)) = \dim(\ran(E))$, $F'$ and $E$ are unitarily equivalent and hence also 
\begin{align*}
 \|F'(\1-E)\| =\|(\1-E)F'\| = \|(\1-F')E\| =\|(\1-F)E\| \le \eps.
\end{align*} 
Therefore, using orthogonality, we have
\begin{align*}
\|E-F'\| = \|(\1-F')E + F'(E-\1)\| = \max\Big\{\|F'(\1-E)\|,\|(\1-F')E\|\Big\} \le \eps.
\end{align*}
Hence, using now Lemma~\ref{lem:Turning1Subspace} again, there exists an unitary $U$ satisfying $\|U-\1\|\le 2 \eps$ and $UEU^*=F'$, which gives
\begin{align*}
FU\ket{\psi_i} =FUE\ket{\psi_i} = FF'U\ket{\psi_i} = F'U\ket{\psi_i}= UE\ket{\psi_i}=U\ket{\psi_i}.
\end{align*}

For the statement under the assumption \eqref{eq:TurnVectorsNonSquared}, we simply use that in this case
\begin{align*}
    \sqrt{\sum_{l=1}^n \left\|(\1-F)\ket{\psi_i}\right\|^2} \le\sqrt{\sum_{l=1}^n \eps^2} = \sqrt{n} \eps<1
\end{align*}
and apply the first part of the lemma.
 \end{proof}

\begin{lemma}
	\label{lem:TurningSubspaces}
Let $\left(E_l\right)_{l=1}^n$ and $\left(F_l\right)_{l=1}^n$ be families of orthogonal projections on some $d$-dimensional Hilbert space $\cH$ with $n\in[d]$  satisfying $\sum_{l=1}^n E_l\le \1$, $\sum_{l=1}^n F_l\le \1$ and
\begin{align*}
\|E_l-F_l\| \le \eps
\end{align*}
for some $0\le\eps< 1$ and all $l\in[n]$. Then there exists an unitary $U$ with $\|U-\1\| \le 6\eps\sqrt{n}. $ and 
\begin{align}
\label{eq:TurnedProj}
UE_lU^* = F_l
\end{align}
for all $l\in[n].$ 
\end{lemma}
\begin{proof} 
    Using for each $l\in[n]$ the Lemma~\ref{lem:Turning1Subspace} individually, we find unitaries $U_l$ with $\|U_l-\1\|\le2\eps$ and
    \begin{align*}
        U_lE_lU_l^* = F_l.
    \end{align*}
    We can now define the partial isometries $V_l = U_lE_l$ with support $\ran(E_l)$ and image $\ran(F_l)$, which by the above satisfy $\|V_l-E_l\|\le2\eps.$ We denote $E =\sum_{l=1}^nE_l$ and  $F =\sum_{l=1}^nF_l$ and define
    \begin{align*}
        V = \bigoplus_{l=1}^nU_l,
    \end{align*}
    which by construction is also a partial isometry with support $\ran(E)$ and image $\ran(F)$. Furthermore, note that we have
    \begin{align*}
        \|V-E\| &\le \sup_{\|\psi\|=1}\sum_{l=1}^n\left\|  (U_l-E_l)\psi\right\| \le 2\eps \sup_{\|\psi\|=1}\sum_{l=1}^n\left\|  E_l\psi\right\| \\&\le 2\eps \sqrt{n} \sup_{\|\psi\|=1}\sqrt{\sum_{l=1}^n\left\|  E_l\psi\right\|^2} \le 2\eps \sqrt{n},
    \end{align*}
    where for the second line we used the Cauchy-Schwarz inequality and then Pythagorean identity. This gives
    \begin{align*}
\left\|(\1-F) -(\1-E)\right\|=\left\|F -E\right\| = \left\|VEV^* -E\right\| \le 2\|V-E\| \le 4\eps\sqrt{n}.
    \end{align*}
    Hence, applying Lemma~\ref{lem:Turning1Subspace} again for $\1-E$ and $\1-F,$ we find another unitary $U_\perp$ which satisfies \begin{align*}
        \|U_\perp-\1\| \le4\eps\sqrt{n}
    \end{align*} 
    and $U_\perp (\1-E)U_\perp^* = \1 -F.$ Analogously to the above we define the partial isometry $V_\perp =U_\perp(\1-E)$, which has support $\ran(\1-E)$ and image $\ran(\1-F)$, and then finally the unitary
    \begin{align*}
        U = V \oplus V_\perp.
    \end{align*}
    Note that by construction, $U$ satisfies $UE_lU^* = F_l$ for all $l\in[n]$ and furthermore
    \begin{align*}
        \left\|U-\1\right\| \le \left\|V-E\right\| + \|(U_\perp-\1)(\1-E)\| \le 6\eps\sqrt{n}.
    \end{align*}
 \end{proof}
 
Another useful observation which is used in the proofs of the results below is that two channels with Stinespring isometries being close to each other in operator norm, are close to each other in diamond norm. This fact is stated and proved in the following lemma for later convenience.

\begin{lemma}
\label{lem:CloseStineCloseChannel}
Let  $V :\cH_A\to\cH_B\otimes\cH_E$ and $W:\cH_A\to\cH_B\otimes\cH_E$ be isometries on Hilbert spaces satisfying 
\begin{align}
    \|V-W\| \le \eps
\end{align} 
for $\eps\ge0$. Then, defining the quantum channels $\cN= \Tr_{E}(V(\,\cdot\,)V^*)$ and $\cM  = \Tr_{E}(W(\,\cdot\,)W^*)$, we have
\begin{align}
\frac{1}{2}\left\|\cN -\cM \right\|_\diamond \le \eps.
\end{align}

\end{lemma}

\begin{proof}
Using the definition of the diamond norm, monotonicity of the trace norm under under the partial trace and Hölder's inequality we find
\begin{align*}
\left\|\cN -\cM\right\|_\diamond &= \sup_{d\in\N,x\in\cT(\cH\otimes\C^d)\,\|x\|_1=1}\left\|\Tr_E(VxV^*) -\Tr_E(W x W^*)\right\|_1\\& \le \sup_{d\in\N,x\in\cT(\cH\otimes\C^d)\,\|x\|_1=1}\left\|VxV^* -W xW^*\right\|_1\\& \le \sup_{d\in\N,x\in\cT(\cH\otimes\C^d)\,\|x\|_1=1}\left\|(V-W)xV^*\right\|_1  +\sup_{d\in\N,x\in\cT(\cH\otimes\C^d)\,\|x\|_1=1}\left\|Wx(V^*-W^*)\right\|_1 \\&\le 2\|V-W\| \le 2\eps,
\end{align*}
where we dropped the identities $\id_d$ for notational simplicity.
\end{proof}

\subsection{Generalised depolarising trick}

In this section we establish the \emph{generalised depolarising trick} (Lemma~\ref{lem:GenDepLemma} below), which is a useful tool for fixing approximate fixed point equations in the sense of Definitions~\ref{def:FixingApproxFixPoints} and~\ref{def:RapidFixingApproxFixedPoint} and is employed in the proof of Theorem~\ref{thm:FixingUnitalChannels}. 

\begin{figure}[h]
\centering
   \begin{subfigure}{0.47\textwidth}
\includegraphics[width=\textwidth]{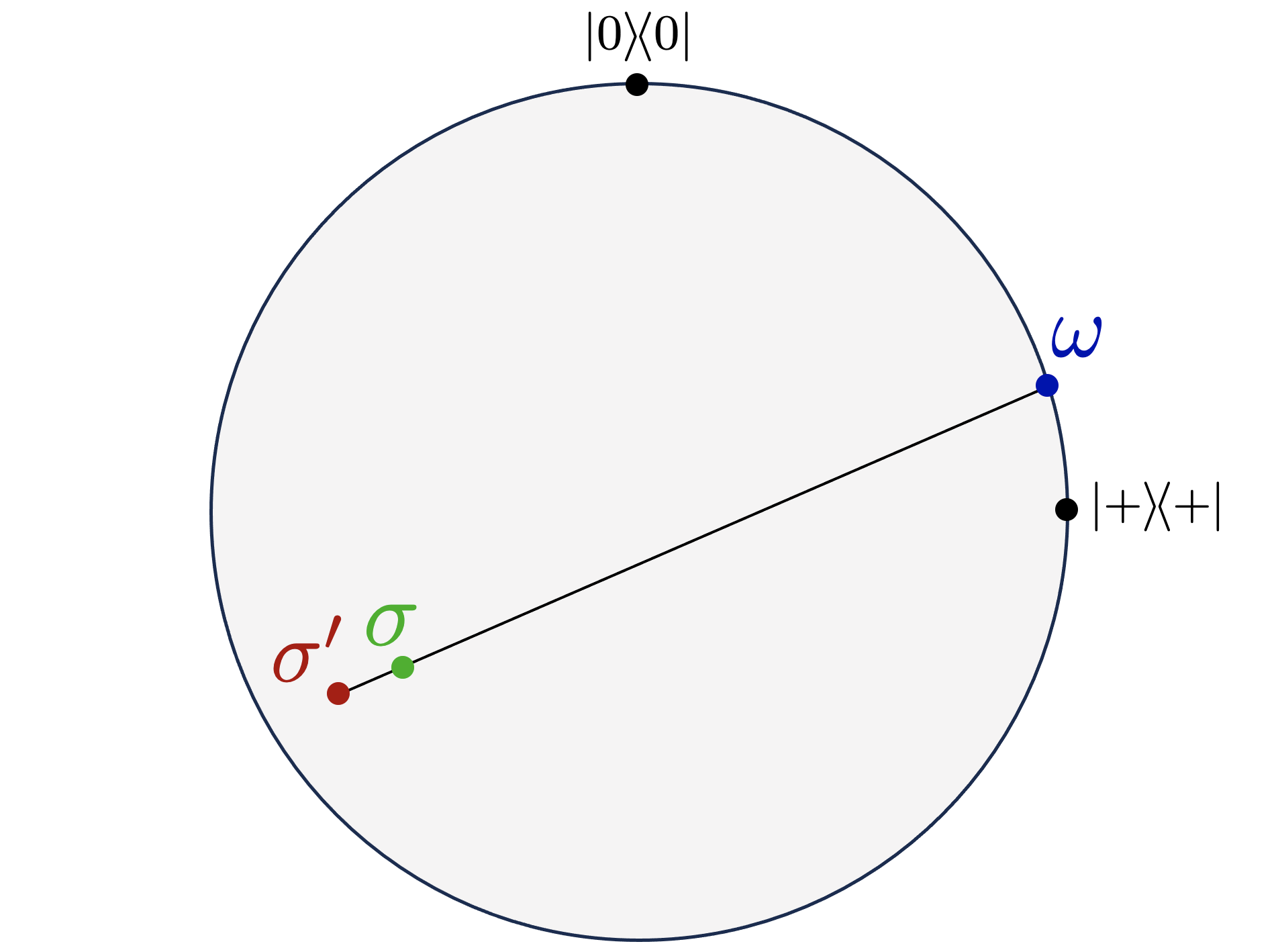}
\caption{Example 1}
\label{fig:GenSub1}
\end{subfigure}
  \begin{subfigure}{0.47\textwidth}
\includegraphics[width=\textwidth]{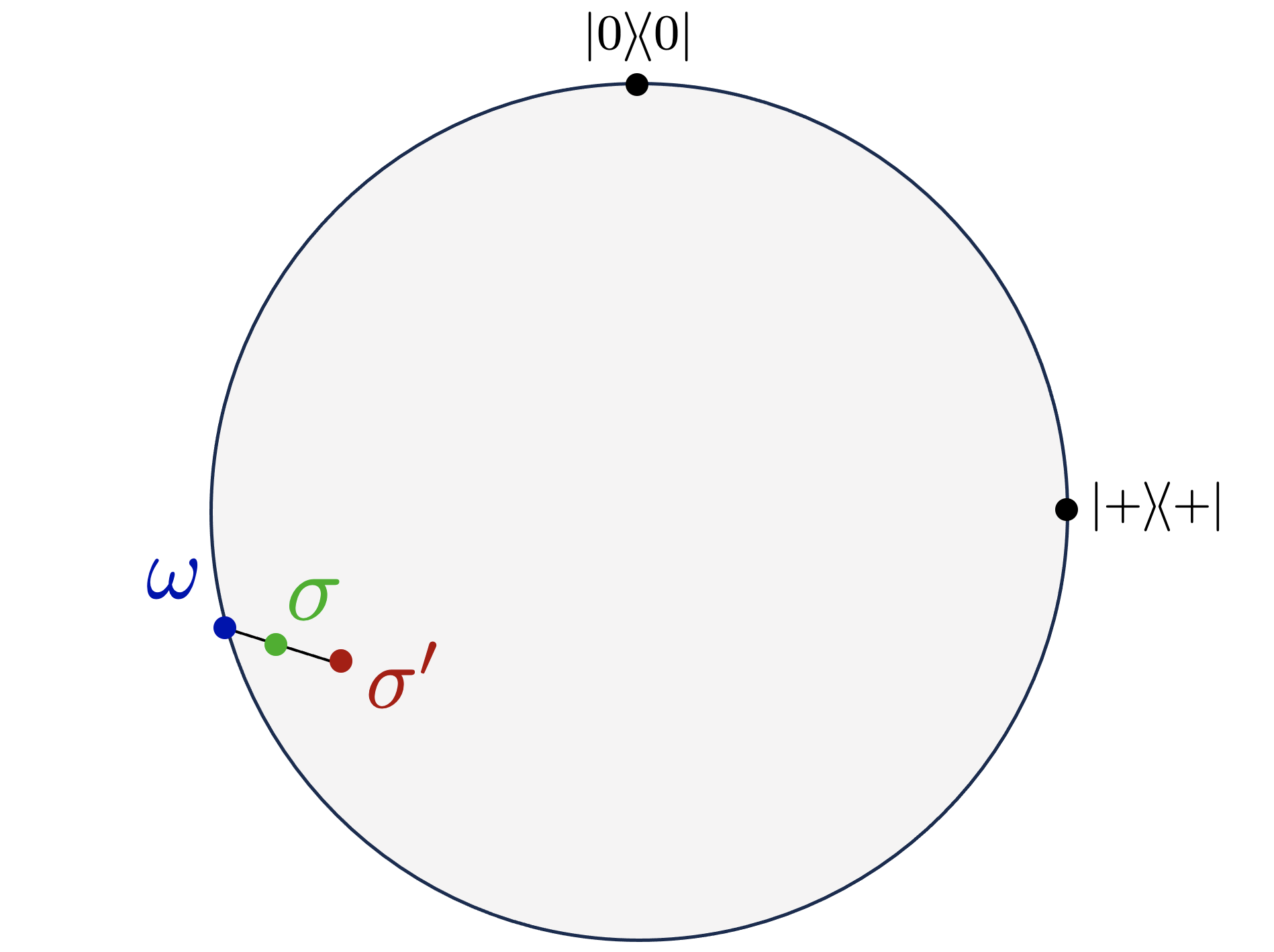}
\caption{Example 2}
\label{fig:GenSub2}
\end{subfigure}
    \caption{Consider the Bloch sphere representation of a qubit, i.e.~the isomorphism between real vectors with Euclidean norm less than or equal to 1 and states on $\C^2$ given by $r\mapsto\frac{1}{2}(\1 +r_x\sigma_x + r_y\sigma_y+r_z\sigma_z).$  In the figures only the slice of the Bloch sphere with $r_y=0$ is pictured. Two examples of states $\sigma$ and $\sigma'$ are illustrated, together with a third state $\omega$ lying on the crossing point of the continuation of the line from $\sigma'$ to $\sigma$ and the boundary of the sphere. Note, that $\omega$ is given by $(\sigma-(1-p_*)\sigma')/p_*$ with $p_*\in[0,1]$ being the smallest number such that $\sigma-(1-p)\sigma'\ge0.$ As $\sigma = (1-p_*)\sigma' + p_*\omega_{p_*},$ the value $p_*$ can pictorially be identified as the ratio between the distances of $\sigma'$ and $\sigma$ and $\sigma'$ and $\omega.$ The idea of the generalised deopolarising trick is to define the channel $\Phi = (1-p_*)\id + p_*\Tr(\placeholder)\omega$ which satisfies $\Phi\approx^{p_*}\id$ and 'pulls' $\sigma'$ onto $\sigma,$ i.e.~$\Phi(\sigma') =\sigma.$ Figure~\ref{fig:GenSub1} presents an example where $p_*$ is rather small and Figure~\ref{fig:GenSub2} an example where $p_*$ is rather big and hence the whole Bloch sphere must be significantly deformed by $\Phi$ in this case. Note, as shown in the proof of Lemma~\ref{lem:GenDepLemma}, we have $p_*\le p$ if $\sigma'\le(1+p)\sigma.$}
    \label{fig:GenDep}
\end{figure}

We consider two states $\sigma$ and $\sigma'$ which satisfy the operator inequality
\begin{align}
\label{eq:GenDepEqu}
    \sigma'\le (1+p)\sigma
\end{align}
for some $p\in[0,1].$ The main observation is then that we can find a generalised depolarising channel of the form
\begin{align}
\label{eq:genDepoDef}
    \Phi = (1-p)\id + p\Tr(\,\cdot\,)\,\omega
\end{align}
for some state $\omega,$ which `pulls back' $\sigma'$ onto $\sigma,$ i.e. $\Phi(\sigma')=\sigma.$ Furthermore, by the explicit form of \eqref{eq:genDepoDef} it is immediate to see that $\Phi\approx^{p}\id.$ In the context of an approximate fixed point equation $\cN(\sigma)\approx\sigma$ we can take $\sigma'=\cN(\sigma).$ Hence, if \eqref{eq:GenDepEqu} is satisfied we can define $\cM = \Phi\circ\cN$ which has $\sigma$ as an exact fixed point and is close to $\cN,$ i.e. $\cM\approx^{p}\cN.$ For an illustration of the generalised depolarising trick consider Figure~\ref{fig:GenDep}.

We now state and prove the generalised depolarising trick in the following lemma. In fact, we not just only prove it for single states $\sigma'$ and $\sigma$ but for families, as this the context in which it is employed in the proof of Theorem~\ref{thm:FixingUnitalChannels} below.

\begin{lemma}[Generalised depolarising trick]
\label{lem:GenDepLemma}
Let $\left(\sigma_i\right)_{i=1}^n$ and $\left(\sigma'_i\right)_{i=1}^n$ be families of states on a separable Hilbert space $\cH$ satisfying
\begin{align}
    \sigma'_i \le (1+p)\sigma_i
\end{align}
for some $p\in[0,1]$ and all $i\in[n].$ Furthermore, assume that $ \sigma'_i$ and  $\sigma'_j$ have orthogonal support for $i\neq j.$ Then there exists a  channel 
\begin{align}
\Phi\approx^{p} {\rm{id}}
\end{align}
such that
\begin{align}
    \Phi(\sigma'_i) = \sigma_i,
\end{align}
for all $i\in[n].$
\end{lemma}
\begin{proof}
Without loss of generality assume $p>0$ as otherwise $\sigma'_i=\sigma_i$ holds true immediately. Note that by assumption $\sigma_i - (1-p)\sigma'_i \ge 0$, which follows trivially for $p=1$ and otherwise by
\begin{align*}
\sigma_i - (1-p)\sigma_i' = (1-p)\left(\frac{1}{1-p}\sigma_i -\sigma_i'\right) \ge (1-p)\Big((1+p)\sigma_i -\sigma_i'\Big)\ge 0.
\end{align*}
Hence, we can define the normalised state $$\omega_i \coloneqq \frac{\sigma_i-(1-p)\sigma_i'}{p}.$$ Furthermore, since $\sigma'_i$ and $\sigma'_j$ have orthogonal support for $i\neq j$, there exists a family of orthogonal projections $\left(\pi_i\right)_{i=1}^n$ such that $\sum_{i=1}^n\pi_i=\1$ and 
\begin{align*}
    \pi_i\sigma_j'\pi_i = \delta_{ij}\sigma'_j,
\end{align*} 
with $\delta_{ij}$ denoting the Kronecker delta. With that we can now define 
the quantum channel 
\begin{align*}
\Phi(\,\cdot\,) \coloneqq (1-p){\rm{id}} + p
\sum_{i=1}^n\,\Tr(\pi_i(\,\cdot\,)\pi_i)\,\omega_i.
\end{align*}
Note, by the triangle inequality and the fact that every quantum channel has diamond norm equal to 1, we see
\begin{align*}
   \frac{1}{2} \left\|\Phi - {\rm{id}}\right\|_\diamond \le \frac{p}{2}  \left( \left\|{\rm{id}}\right\|_\diamond +  \left\|\sum_{i=1}^n\Tr(\pi_i(\,\cdot\,)\pi_i)\,\omega_i \right\|_\diamond\right) =p.
\end{align*}
Moreover, by definition we have for all $i\in[n]$
\begin{align*}
\Phi(\sigma_i')= (1-p)\sigma_i' + p\sum_{j=1}^n \Tr\left(\pi_j\sigma'_i\pi_j\right) \omega_j = (1-p)\sigma_i' + \sigma_i - (1-p)\sigma_i' = \sigma_i,
\end{align*}
which finishes the proof.
\end{proof}

\subsection{Unitary channels}
Theorem~\ref{thm:FixUnitary} below shows rapid fixiability of approximate fixed point equations in the sense of Definition~\ref{def:RapidFixingApproxFixedPoint} for $\cS=\mathfrak{S}(\cH)$ and $\cC$ being the set of unitary channels on $\cH:$

\begin{theorem}[Fixability for unitary channels]
\label{thm:FixUnitary}
Let $\rho$ be a state and $U$ an unitary on a $d$-dimensional Hilbert space $
\cH$ such that for $\eps\ge 0$ we have
\begin{align}
\label{eq:UNITHMAPRFIX}
    U\rho U^* \approx^{\eps} \rho.
\end{align}
Then there exist a state $\sigma$ and an unitary $V$ satisfying
\begin{align}
\sigma\approx^{4\,d^{5/4}\sqrt{\eps}}\rho,\quad\quad V\approx^{4\,d^{5/4}\sqrt{\eps}} U
\end{align} and
\begin{align}
V\sigma V^* = \sigma.
\end{align}
Furthermore, the state $\sigma$ is independent of $U.$

\end{theorem}
The proof of Theorem~\ref{thm:FixUnitary} can be outlined as follows: First we use the technique discussed in Section~\ref{sec:SpectralClustering} to find spectral clusters of the state $\rho$ separated by spectral gaps. Using Lemma~\ref{lem:BhathiaSpectralProj}, we know that the spectral projections of each of these clusters and their rotated versions under the unitary $U$ are close to each other due to the spectral gaps and the approximate fixed point equation \eqref{eq:UNITHMAPRFIX}. Therefore, by the rotation lemma~\ref{lem:TurningSubspaces}, we can find an unitary close to the identity which rotates them back. The composition of this new unitary and the original unitary then defines the new unitary channel. Furthermore, we take as the new state the one defined through spectral clustering in \eqref{eq:SpecClusterState}. Both new state and channel are close to the original ones and furthermore satisfy an exact fixed point equation. 

\smallskip

\begin{proof}[Proof of Theorem~\ref{thm:FixUnitary}]
We employ the spectral clustering technique outlined in Section~\ref{sec:SpectralClustering} for $\rho$ and $\delta\ge0$ to be determined later. In particular the spectrum of $\rho$ is separated into $n$ clusters such that spectral points from different clusters are at least $\delta$ far from each other.
\\ \indent We denote the spectral projections of the $l$th spectral cluster defined in \eqref{eq:ClusterProjection} by $E_l$ and the rotated projection by
\begin{align}
     F_l= UE_lU^*.
 \end{align}
By construction, in particular the seperation of spectral points established in \eqref{eq:Clustering}, together with Lemma~\ref{lem:BhathiaSpectralProj} we have for all $l\in[n]$
\begin{align*}
    \left\|E_l - F_l\right\|  \le \frac{2\|\rho -U\rho U^*\|}{\delta} \le \frac{2\|\rho -U\rho U^*\|_1}{\delta} \le \frac{4\eps}{\delta}.
\end{align*}
Using now Lemma~\ref{lem:TurningSubspaces} we see that there exists an unitary $U'$ such that $\|U'-\1\|\le 24\eps\sqrt{n}/\delta$ and for all $l\in[n]$
\begin{align*}
 U'F_l U'^* = E_l.
\end{align*}
Moreover, define $V =  U'U$ which satisfies 
\begin{align*}
\left\|V-U\right\| = \|U'-\1\|\le \frac{24\eps\sqrt{n}}{\delta} \le \frac{24\eps\sqrt{d}}{\delta}.
\end{align*}
Considering the state defined in \eqref{eq:SpecClusterState}, i.e.
\begin{align*}
    \sigma = \sum_{l=1}^n\mu_l E_l
\end{align*}
where $\mu_l$ is the average of the spectral point of the $l$th cluster defined in \eqref{eq:AverageSpecCluster}, we see that $\sigma$ is a fixed point of the unitary channel associated with $V$: 
\begin{align*}
    V\sigma V^* = \sum_{l=1}^n \mu_l \,U'UE_lU^*U'^* = \sum_{l=1}^n \mu_l \, U'F_l U'^* = \sum_{l=1}^n \mu_l E_l = \sigma.
\end{align*}
Furthermore, by Lemma~\ref{lem:ApproxClusterState}
 we know
\begin{align*}
    \frac{1}{2}\left\|\rho-\sigma\right\|_1 \le \frac{d^2\delta}{2}.
\end{align*}
Picking now \begin{align*}
    \delta =  \frac{\sqrt{48\,\eps}}{d^{3/4}}
\end{align*}
gives
\begin{align*}
     \frac{1}{2}\left\|\rho -\sigma\right\|_1 &\le \sqrt{\frac{48}{4}}d^{5/4}\sqrt{\eps} \le 4\,d^{5/4}\sqrt{\eps},\\\left\|U-V\right\| &\le 4\,d^{5/4}\sqrt{\eps},
\end{align*}
which finishes the proof. Furthermore, note that the state $\sigma$ solely depends on $\rho$ as outlined in Section~\ref{sec:SpectralClustering} but not on the unitary $U.$

\end{proof}
\subsection{Mixed-unitary channels}

In the following we consider $\cC$ to be the set of mixed-unitary channels on some $d$-dimensional Hilbert space $\cH$. More precisely, this means
\begin{align*}
    \cC = {\rm{conv}}\!\left\{\cU\right\}\coloneqq {\rm{conv}}\left\{\cU\,\Big|\, \cU(\,\cdot\,)=  U(\,\cdot\,)U^*\ \text{with $U$ unitary}\right\},
\end{align*}
where ${\rm{conv}}$ denotes taking the convex hull.\comment{It is well know that the unitary group on a $d$-dimensional Hilbert space is a real Lie group of dimension $d^2$ (see e.g.~\cite[Chapter 2.2]{Ziller_LieGroup_2010}). Therefore, by Whitney's embedding theorem \cite[Theorem~6.19]{lee_introduction_2012} it, and by that also the set of unitary channels, can be smoothly embedded in $\R^{2d^2}.$ } 
From Proposition~4.9 in \cite{watrous_2018}, which is a consequence of Caratheodory's theorem \cite{caratheodory_uber_1911}, we 
see that for every $\cN \in {\rm{conv}}\!\left\{\cU\right\}$ there exists a probability vector $\left(p_i\right)_{i=1}^{d^4}\in\R^{d^4}$ and unitaries $(U_i)_{i=1}^{d^4}$ on $\cH$ such that
\begin{align}
\label{eq:MixtureUnitary}
    \cN(\rho) = \sum_{i=1}^{d^4} p_i \,U_i\rho U_i^* 
\end{align}
for all states $\rho.$

Theorem~\ref{thm:FixingMixtureUnitaries} shows rapid fixiability of approximate fixed point equations in the sense of Definition~\ref{def:RapidFixingApproxFixedPoint} for $\cS=\mathfrak{S}(\cH)$ and $\cC={\rm{conv}}\!\left\{\cU\right\}:$  

\begin{theorem}[Fixability for mixed-unitary channels]
\label{thm:FixingMixtureUnitaries}
Let $\rho$ be a state and \ $\cN\in{\rm{conv}}\!\left\{\cU\right\}$ on a $d$-dimensional Hilbert space $\cH$ such that for $\eps\ge 0$ we have
\begin{align}
\cN(\rho) \approx^{\eps} \rho.
\end{align}
Then there exist a state $\sigma$ and a channel $\cM\in{\rm{conv}}\!\left\{\cU\right\}$ satisfying
\begin{align}\sigma\approx^{4d^2\,\eps^{1/5}}\rho,\quad\quad \cM\approx^{7d^2\,\eps^{1/5}}\cN\end{align} and
\begin{align}
\cM(\sigma) =\sigma.
\end{align}
\end{theorem}

The following lemma shows that if $\rho$ is an approximate fixed point of $\cN\in{\rm{conv}}\!\left\{\cU\right\}$ of the form \eqref{eq:MixtureUnitary}, then it is also an approximate fixed point of all unitary channels $U_i(\,\cdot\,)U_i^*$ provided that the corresponding probability is not too small, i.e.~$p_i\gg d\eps.$ We use this fact in the proof of Theorem~\ref{thm:FixingMixtureUnitaries} to apply Theorem~\ref{thm:FixUnitary} for each of those unitary channels individually. Importantly, as the state obtained in Theorem~\ref{thm:FixUnitary} is independent of the unitary, we can pick the same state $\sigma$ working for all of those unitaries. 

We now first state and prove Lemma~\ref{lem:MixturUnitarAssist} and then give the detailed proof of Theorem~\ref{thm:FixingMixtureUnitaries}.
\begin{lemma}
\label{lem:MixturUnitarAssist}
Let $\rho$ be quantum state and $\cN\in {\rm{conv}}\!\left\{\cU\right\}$ be a mixed-unitary channel of the form \eqref{eq:MixtureUnitary} on a $d$-dimensional Hilbert space $\cH$ such that for $\eps\ge0$ we have
\begin{align}
\cN(\rho) \approx^\eps \rho.
\end{align} 
Then for all $i\in[d^4]$ we have
\begin{align}
    \left\|U_i\rho U^*_i -\rho \right\|_2 \le \sqrt{\frac{4\eps}{p_i}}.
\end{align}
Consequently, we have 
\begin{align}
\label{eq:MixUniApproxSinglU}
    U_i\rho U_i^* \approx^{\sqrt{d\eps/p_i}}\rho.
\end{align}
\end{lemma}
\begin{proof}
We use the fact that the Hilbert-Schmidt norm is dominated by the trace-norm and the dual representation of the Hilbert-Schmidt norm to see for every $j\in[d^4]$
\begin{align*}
    2\eps &\ge \left\|\rho -\sum_{i=1}^{d^4}p_i U_i\rho U_i^*\right\|_1 \ge \left\|\rho -\sum_{i=1}^{d^4}p_i U_i\rho U_i^*\right\|_2 = \sup_{\|\omega\|_2=1}\left|\Tr\left(\omega^*\left(\rho -\sum_{i=1}^{d^4} U_i\rho U_i^*\right)\right)\right|\\& \ge \Tr\left(\frac{\rho}{\|\rho\|_2}\left(\rho -\sum_{i=1}^{d^4} p_iU_i\rho U_i^*\right)\right) \ge \|\rho\|_2 - \frac{p_j}{\|\rho\|_2}\Tr\left(\rho U_j\rho U_j^*\right) -\sum_{i\neq j}p_i\|\rho\|_2 \\&= p_j\left(\|\rho\|_2 - \frac{1}{\|\rho\|_2}\Tr\left(\rho U_j\rho U_j^*\right)\right).
\end{align*}
From that we get
\begin{align*}
\Tr\left(\rho U_j\rho U_j^*\right) \ge \|\rho\|^2_2 - \frac{2\eps\|\rho\|_2}{p_j}
\end{align*}
and therefore 
\begin{align*}
    \left\|\rho - U_j\rho U_j^*\right\|^2_2 = 2\left(\|\rho\|^2_2 - \Tr\left(\rho U_j\rho U_j^*\right)\right) \le \frac{4\eps\|\rho\|_2}{p_i} \le \frac{4\eps}{p_i}.
\end{align*}
For \eqref{eq:MixUniApproxSinglU} we use the fact that the trace norm of an operator is upper bounded by $\sqrt{d}$  times its Hilbert-Schmidt norm, which follows by Hölder's inequality, and hence
\begin{align*}
    \frac{1}{2}\left\|\rho - U_j\rho U_j^*\right\|_1 \le \frac{1}{2}\sqrt{d}\left\|\rho - U_j\rho U_j^*\right\|_2 \le \sqrt{\frac{d\eps}{p_j}}.
\end{align*}
\end{proof}

\begin{proof}[Proof of Theorem~\ref{thm:FixingMixtureUnitaries}]
We write $\cN$ in the form \eqref{eq:MixtureUnitary}, i.e.~
\begin{align*}
    \cN(\rho) = \sum_{i=1}^{d^4} p_i \,U_i\rho U_i^*
\end{align*}
for some probability vector $\left(p_i\right)_{i=1}^{d^4}$ and unitaries $\left(U_i\right)_{i=1}^{d^4}.$ \ Let now $\delta\ge 0$ to be determined later.
Let $I\subset[d^4]$ be the subset of indices such that
\begin{align*}
    p_i\ge \delta,\quad\quad\text{for all $i\in I$.}
\end{align*}
We know by Lemma~\ref{lem:MixturUnitarAssist} that for all $i\in I$ we have
\begin{align*}
    U_i\rho U_i^* \approx^{\sqrt{d\eps/\delta}} \rho.
\end{align*}
Using now Theorem~\ref{thm:FixUnitary}, we see that there exists state a $\sigma$ and for all $i\in I$ an unitary $V_i$
satisfying
\begin{align}
\label{eq:MixUnitCLosUni}
\frac{1}{2}\left\|\sigma-\rho\right\|_1 \le \frac{4\,d^{3/2}\eps^{1/4}}{\delta^{1/4}},\quad\quad \left\|V_i -U_i\right\| \le \frac{4\,d^{3/2}\eps^{1/4}}{\delta^{1/4}}
\end{align} and
\begin{align}
V_i\sigma V_i^* = \sigma.
\end{align}
Note, that we have used explicitly that the $\sigma$ constructed in Theorem~\ref{thm:FixUnitary} is independent of the unitary $U_i.$ Define now the the quantum channel
\begin{align*}
 \cM(\,\cdot\,) = \sum_{i\in I}p_i V_i(\,\cdot\,) V_i^* + \, \sum_{i\in[d^4]\setminus I}p_i\,\id,
\end{align*}
which by definition is a mixed-unitary channels. Note that by construction we have
\begin{align*}
 \cM(\sigma) = \sum_{i\in I}p_i V_i\sigma V_i^* + \sum_{i\in[d^4]\setminus I}p_i\,\sigma = \sum_{i=1}^{d^4}p_i\,\sigma=\sigma.
\end{align*}
Moreover, using \eqref{eq:MixUnitCLosUni}, we have
\begin{align*}
    \frac{1}{2}\left\|\cN-\cM\right\|_\diamond \le \sum_{i\in I} p_i\|V_i-U_i\| + \sum_{i\in[d^4]\setminus I}p_i,\le \frac{4\,d^{3/2}\eps^{1/4}}{\delta^{1/4}} + \delta d^4,
\end{align*}
where we used in the first inequality Lemma~\ref{lem:CloseStineCloseChannel} for all $i\in I$ and for all $i\in[d^4]\setminus I$ the fact that $\frac{1}{2}\|U_i(\,\cdot\,) U_i^* - \id\|_\diamond\le1.$ Optimising now over $\delta\ge 0$
 and picking 
 \begin{align*}
     \delta = \frac{4^{4/5}\,\eps^{1/5}}{d^2}
 \end{align*}
 gives
 \begin{align*}
\frac{1}{2}\left\|\cN-\cM\right\|_\diamond &\le 2 \left(4^{4/5} d^2\,\eps^{1/5}\right) \le 7 d^2\,\eps^{1/5}, \\
\frac{1}{2}\left\|\sigma-\rho\right\|_1 &\le 4^{4/5} d^2\,\eps^{1/5}\le 4 d^2\,\eps^{1/5},
 \end{align*}
 which finishes the proof.
\end{proof}
\subsection{Unital channels}
Theorem~\ref{thm:FixingUnitalChannels} below shows rapid fixiability of approximate fixed point equations in the sense of Definition~\ref{def:RapidFixingApproxFixedPoint} for $\cS=\mathfrak{S}(\cH)$ and $\cC$ being the set of unital channels, i.e.~channels $\cN$ satisfying $\cN(\1)=\1.$  
\begin{theorem}[Fixability for unital channels]
\label{thm:FixingUnitalChannels}
Let $\rho$ be a state and $\cN$ a unital channel on some $d$-dimensional Hilbert space $\cH$ such that for $\eps\ge 0$ we have
\begin{align}
\cN(\rho) \approx^{\eps} \rho.
\end{align}
Then there exist a state $\sigma$ and an unital channel $\cM$ satisfying
\begin{align}\sigma\approx^{7d^{5/3}\eps^{1/6}}\rho,\quad\quad \cM\approx^{7d^{5/3}\eps^{1/6}}\cN\end{align} and
\begin{align}
\cM(\sigma) =\sigma.
\end{align}
\end{theorem}
On of the main ingredients of the proof of Theorem~\ref{thm:FixingUnitalChannels} is Lemma~\ref{lem:ApproxBau} below, which can be seen as an approximate version of a well-known result about fixed points of quantum channels (see e.g.~\cite[Proposition 6.8]{Wolf12} and \cite[Lemma 1]{Burgarth_2013}). In particular, Lemma~\ref{lem:ApproxBau} shows that if a linear operator $A$ is an approximate fixed point of some channel $\cN,$ then so is its real- and imaginary- and their respective positive- and negative parts. Since in the case of Theorem~\ref{thm:FixingUnitalChannels} the channel $\cN$ is unital and has therefore additionally to the approximate fixed point $\rho$ also the exact one $\1,$ we can employ Lemma~\ref{lem:ApproxBau} to find multiple additional approximate fixed points of $\cN.$ More precisely, in Lemma~\ref{lem:ManyApproxFixUnital} we find by taking weighted differences of $\rho$ and $\1$ and then their respective positive parts, that all spectral projections of $\rho$ are approximate fixed points of $\cN$ as well. In the proof of Theorem~\ref{thm:FixingUnitalChannels} we then again employ the spectral clustering technique outlined in Section~\ref{sec:SpectralClustering} and focus on the spectral projections of the corresponding spectral clusters. We can then fix the approximate fixed point equation for each of those projections using first the rotation lemma~\ref{lem:TurnVectorsSubspace}, but on Stinespring level involving the environment of the channel, and then the, what we call, \emph{generalised depolarising trick} in Lemma~\ref{lem:GenDepLemma}. By that we obtain a new unital channel $\cM$ which is close to $\cN$ and which has the spectral cluster state $\sigma$ defined in \eqref{eq:SpecClusterState} as an exact fixed point.

We now state and prove the mentioned Lemmas~\ref{lem:ApproxBau} and~\ref{lem:ManyApproxFixUnital} and then give the proof of Theorem~\ref{thm:FixingUnitalChannels}.

\begin{lemma}
\label{lem:ApproxBau}
Let $A$ be a linear operator on a separable Hilbert space and $\cN$ a channel\footnote{For the lemma it suffices to assume that $\cN$ is linear, positive and trace-preserving.} such that for $\eps\ge0$ we have
\begin{align}
\label{eq:AApprox}
    \frac{1}{2}\left\|\cN(A) -A\right\|_1\le\eps
\end{align}
Then also we also have for the corresponding real and imaginary parts $\Re(A) = (A+A^*)/2$ and $\Im(A)= (A-A^*)/(2i)$
\begin{align}
\label{eq:ApproxFixRealIm}
    \frac{1}{2}\left\|\cN(\Re(A)) - \Re(A)\right\|_1\le \eps,\quad\quad\frac{1}{2}\left\|\cN(\Im(A)) - \Im(A)\right\|_1\le \eps,
\end{align}
and furthermore for their positive- and negative parts
\begin{align}
\label{eq:ApproxPosNeg}
   \nn \frac{1}{2}\left\|\cN([\Re(A)]_+) - [\Re(A)]_+\right\|_1&\le \frac{3}{2}\eps+\sqrt{\eps\Tr([\Re(A)]_+)},\\\nn\frac{1}{2}\left\|\cN([\Re(A)]_-) - [\Re(A)]_-\right\|_1&\le \frac{3}{2}\eps+\sqrt{\eps\Tr([\Re(A)]_-)}\\  \nn\frac{1}{2}\left\|\cN([\Im(A)]_+) - [\Im(A)]_+\right\|_1&\le \frac{3}{2}\eps+\sqrt{\eps\Tr([\Im(A)]_+)},\\\frac{1}{2}\left\|\cN([\Im(A)]_-) - [\Im(A)]_-\right\|_1&\le \frac{3}{2}\eps+\sqrt{\eps\Tr([\Im(A)]_-)}.
\end{align}
\end{lemma}
\begin{proof}
The proof uses similar arguments to the ones in \cite[Proposition 6.8]{Wolf12} and \cite[Lemma 1]{Burgarth_2013}, which proved the result in the exact case $\eps=0.$

Since $\cN$ is positive and in particular hermiticity preserving, i.e.~$\cN(A^*) = \cN(A)^*$, \eqref{eq:ApproxFixRealIm} immediately follows from \eqref{eq:AApprox} and the triangle inequality.

We now focus on proving \eqref{eq:ApproxPosNeg}. For that let $X$ be a self-adjoint trace class operator satisfying
\begin{align}
\label{eq:XApprox}
    \frac{1}{2}\left\|\cN(X) -X\right\|_1\le\eps.
\end{align}

 Let $\pi_+$ be the projection onto the orthogonal sum of the support of $[X]_+$ and the kernel of $X$ and $\pi_-$ the projection onto the support of $[X]_-$, such that we have $\pi_+ +\pi_-=\1$. Using \eqref{eq:XApprox} and the variational expression of the trace distance \eqref{eq:VarExpTraceDistance} we have
 \begin{align*}
 \Tr\big(\pi_{+}\left(X-\cN(X)\right)\big) \le \frac{1}{2} \big\|\cN(X)-X\big\|_1 \le \eps.
 \end{align*}
 Therefore, we have
 \begin{align*}
     \Tr\big([X]_+\big) &= \Tr\big(\pi_+X\big) \le\Tr\big(\pi_+\,\cN(X)\big) +\eps = \Tr\big(\pi_+\,\cN([X]_+)\big) - \Tr\big(\pi_+\,\cN([X]_-)\big)+\eps \\&\le \Tr\big(\pi_+\,\cN([X]_+)\big) +\eps\le     \Tr\big([X]_+\big) +\eps.
 \end{align*}
 This shows $ \Tr\big(\pi_+\,\cN([X]_-)) \le \eps$ and by the same argument with positive- and negative parts interchanged also  $\Tr\big(\pi_-\,\cN([X]_+) \le \eps.$ By Hölder's inequality and positive semi-definiteness of $\cN([X]_+)$ and $\cN([X]_-)$ we see
 \begin{align*}
    \left\|\pi_-\cN([X]_+)\right\|_1 &\le \sqrt{\Tr\big(\pi_-\,\cN([X]_+)\big)}\sqrt{\Tr\big([X]_+\big)} \le \sqrt{\eps\Tr\big([X]_+\big)}\\
     \left\|\pi_+\cN([X]_-)\right\|_1 &\le\sqrt{\eps\Tr([X]_-)}
 \end{align*}
 This gives
 \begin{align*}
    \left\|\cN([X]_+) - [X]_+\right\|_1 &\le \left\|\pi_+(\cN([X]_+) - X)\pi_+\right\|_1 + \left\|\pi_-\cN([X]_+)\right\|_1 + \left\|\pi_-\cN([X]_+)\pi_+\right\|_1 \\&\le \left\|\pi_+(\cN(X) - X)\pi_+\right\|_1 +\left\|\pi_+(\cN([X]_-)\pi_+\right\|_1 +2\sqrt{\eps\Tr\big([X]_+\big)}\\& \le 2\eps + \Tr(\pi_+\cN([X]_-)) +2\sqrt{\eps\Tr\big([X]_+\big)} \le 3\eps + 2\sqrt{\eps\Tr\big([X]_+\big)}
 \end{align*}
 By symmetry we also get 
\begin{align*}
   \frac{1}{2} \left\|\cN([X]_-) - [X]_-\right\|_1 \le\frac{3}{2}\eps + \sqrt{\eps\Tr\big([X]_-\big)},
\end{align*}
which finishes the proof
\end{proof}
\begin{lemma}
\label{lem:ManyApproxFixUnital}
Let $\rho$ be a state and $\cN$ an unital channel on some $d$-dimensional Hilbert space $\cH$ such that for $0\le\eps\le 1$ we have
\begin{align}
\cN(\rho) \approx^{\eps} \rho.
\end{align}
Write $\rho$ in spectral decomposition
\begin{align}
\label{eq:SpectDecompUnital}
    \rho = \sum_{i=1}^m\lambda_i \pi_i
\end{align}
with $m\in[d],$ spectral points $\lambda_1 >\lambda_2>\cdots>\lambda_m$ and non-zero orthogonal projections $\left(\pi_i\right)_{i=1}^m$ summing to $\1$. Then for all $j\in[m-1]$ we have
\begin{align}
\label{eq:SumProjFix}
    \frac{1}{2}\left\|\cN\left(\sum_{i=1}^{j}\pi_i\right) - \sum_{i=1}^{j}\pi_i\right\|_1 \le \frac{5\sqrt{\eps}}{\delta_j}
\end{align}
where $\delta_j = \lambda_j-\lambda_{j+1}>0.$
\end{lemma}
\begin{remark}
Lemmas~\ref{lem:ApproxBau} and~\ref{lem:ManyApproxFixUnital} provide a way to construct many new (approximate) fixed points of a quantum channel $\cN$ from a few. By that, having only knowledge of the channel's output on some input states, we might be able to determine the whole channel fully or at least partially. Let us illustrate this by considering the example of an unital channel $\cN,$ which we know to have an additional exact fixed point state $\rho.$ Assume furthermore that the state $\rho$ has simple spectrum and hence can be written in spectral decomposition
\begin{align*}
    \rho = \sum_{i=1}^d\lambda_i\kb{\psi_i}
\end{align*}
with spectral points $\lambda_1>\lambda_2>\cdots>\lambda_d$ and an orthonormal basis $\left(\psi_i\right)_{i=1}^d.$ From Lemma~\ref{lem:ManyApproxFixUnital} we can deduce that also all of the spectral projections of $\rho$ are fixed points of $\cN,$ i.e.
\begin{align*}
\cN(\kb{\psi_i}) = \kb{\psi_i}
\end{align*}
for all $i\in[d].$ This gives for $V:\cH\to\cH\otimes\cH_E$ being a Stinespring isometry of $\cN$ that
\begin{align*}
    V\ket{\psi_i}=\ket{\psi_i}\ket{\phi_i}
\end{align*}
for some normalised vectors $\left(\ket{\phi_i}\right)_{i=1}^d$ on the environment Hilbert space $\cH_E.$ Therefore, for any linear operator $A$ which can be written in matrix components as $A=\sum_{i,j=1}^da_{i,j}\ket{\psi_i}\!\bra{\psi_j}$ we see that the channel's output is given by
\begin{align*}
    \cN(A) = \sum_{i,j=1}^da_{i,j}\langle \phi_j,\phi_i\rangle\, \ket{\psi_i}\!\bra{\psi_j}.
    \end{align*}
    Therefore, only from the two fixed points $\1$ and $\rho,$ we are able to determine the channel $\cN$ uniquely up to knowledge of the complex numbers $\langle \phi_j,\phi_i\rangle.$
\end{remark}
\begin{proof}[Proof of Lemma~\ref{lem:ManyApproxFixUnital}]
 Define
\begin{align*}
    A_1 = \rho - \lambda_2\1
\end{align*}
and note that since $\cN$ is unital we have
\begin{align*}
   \frac{1}{2}\left\| \cN(A_1) - A_1\right\|_1 = \frac{1}{2}\left\|\cN(\rho) - \rho \right\|_1 \le \eps.
\end{align*}
Therefore, using Lemma~\ref{lem:ApproxBau}, we find for the positive part \begin{align*}
    \frac{1}{2}\|\cN([A_1]_+)-[A_1]_+\|_1\le\frac{3}{2}\eps  + \sqrt{\eps} \le \frac{5}{2}\sqrt{\eps},
\end{align*}
where we have used $\Tr([A_1]_+)\le \Tr(\rho)=1$ and $\eps\le 1.$
Using the spectral decomposition \eqref{eq:SpectDecompUnital}, we can read off $$[A_1]_+ = (\lambda_1-\lambda_2)\pi_1=\delta_1 \pi_1$$ and hence
\begin{align*}
    \frac{1}{2}\left\|\cN(\pi_1) - \pi_1\right\|_1 \le\frac{5\sqrt{\eps}}{2\delta_1}.
\end{align*}
Iterating the above procedure, we define for all $j\in[m-1]$
\begin{align*}
    A_j = \rho - \lambda_{j+1}\1 = \sum_{i=1}^m (\lambda_i-\lambda_{j+1})\pi_i,
\end{align*}
which also satisfies $\frac{1}{2}\|\cN(A_j) - A_j\|\le\eps$ and by Lemma~\ref{lem:ApproxBau} also
\begin{align}
\label{eq:PosIteratingUnital}
    \frac{1}{2}\|\cN([A_j]_+) - [A_j]_+\|_1\le \frac{5}{2}\sqrt{\eps} ,
\end{align}
where as above we have used $\Tr([A_j]_+)\le1$ and $\eps\le 1.$
Noting now that 
\begin{align*}
    [A_j]_+ &= \sum_{i=1}^{j}(\lambda_i-\lambda_{j+1})\pi_i =  \sum_{i=1}^{j-1}(\lambda_i-\lambda_{j})\pi_i + \sum_{i=1}^{j}(\lambda_{j}-\lambda_{j+1})\pi_i \\&= [A_{j-1}]_+ + \delta_j\sum_{i=1}^{j}\pi_j.
\end{align*}
we see from \eqref{eq:PosIteratingUnital} and the triangle inequality that
\begin{align}
    \frac{1}{2}\left\|\cN\left(\sum_{i=1}^{j}\pi_j\right) - \sum_{i=1}^{j}\pi_j\right\|_1 \le \frac{5\sqrt{\eps}}{\delta_j}.
\end{align}
for all $j\in[m-1].$
\end{proof}

\begin{proof}[Proof of Theorem~\ref{thm:FixingUnitalChannels}]
We can take without loss of $\eps\le 1$ as other wise the statement is trivially true, e.g by taking the maximally mixed state $\sigma=\1/d$ which is a fixed point of $\cN$ by unitality and furthermore clearly $\frac{1}{2}\|\rho-\sigma\|_1\le1\le7d^{5/3}.$
\\ \indent We use the spectral clustering technique discussed in Section~\ref{sec:SpectralClustering} for $\rho$ and $\delta\ge\eps\ge0$ to be determined later. In particular for that write $\rho$ in spectral decomposition
\begin{align}
\label{eq:SpectDecompUnital2}
    \rho = \sum_{i=1}^m\lambda_i \pi_i
\end{align}
with $m\in[d],$ spectral points $\lambda_1 >\lambda_2>\cdots>\lambda_m$ and non-zero orthogonal projections $\left(\pi_i\right)_{i=1}^m$ summing to $\1.$ As explained in Section~\ref{sec:SpectralClustering}, $k_l\in[m]$ is the $l$th largest number such that
\begin{align*}
    \delta_{k_l} = \lambda_{k_l} - \lambda_{k_l+1} > \delta.
\end{align*}
 The spectral projections corresponding $l$th spectral cluster defined in \eqref{eq:ClusterProjection} is  then explicitly given by
 \begin{align*}
     E_l = \sum_{i=k_{l-1}+1}^{k_l} \pi_i
 \end{align*}
 Furthermore, $n\in[m]$ denotes the number of spectral clusters constructed in that way. \comment{We define the set \begin{align}
    I = \left\{i\in[n-1]\,\Big|\,\delta_i\ge \delta\right\}
\end{align}
For every $i\in I$ we now define orthogonal projection
\begin{align*}
    E_i = \sum_{j=i_{(<)}+1}^i\pi_j 
\end{align*}
where $i_{(<)} = 0$ if $i$ is the smallest element in $I$ and otherwise we take $i_{(<)}$ to be the largest element in $I$ such that $i_{(<)} < i.$  Define for $i\in I$ the average \begin{align*}
    \mu_i = \frac{1}{i-i_{(<)}}\sum_{j=i_{(<)}+1}^i \lambda_j\ge 0
\end{align*} and with that 
\begin{align*}
    \sigma = \sum_{i\in I} \mu_iE_i,
\end{align*}
which is a state as clearly $\sigma\ge 0$ and $\Tr(\sigma) = \Tr(\rho) =1.$
Furthermore, note that since for all $j= i_{(<)} +1,\dots,i-1$ we have $\lambda_j -\lambda_{j+1} = \delta_j \le\delta$ and therefore 
\begin{align*}
   \sum_{j} |\mu_{i} - \lambda_j|            
\end{align*}
\begin{align}
\label{eq:StatesCloseUnital}
    \frac{1}{2}\left\|\rho- \sigma\right\|_1 \le \frac{d^2\delta}{2}.
\end{align}}
Using Lemma~\ref{lem:ManyApproxFixUnital} we see for all $l\in [n]$
\begin{align}
 \label{eq:UnitalEiFix}  \nn \frac{1}{2}\left\|\cN(E_l) - E_l\right\|_1  &\le   \frac{1}{2}\left\|\,\cN\left(\sum_{i=1}^{k_l}\pi_i\right) - \sum_{i=1}^{k_l}\pi_j \right\|_1  +  \frac{1}{2}\left\|\,\cN\left(\sum_{i=1}^{k_{l-1}}\pi_i\right) - \sum_{i=1}^{k_{l-1}}\pi_i\right\|_1 \\& \le\frac{5
 \sqrt{\eps}}{\delta_{k_l}} + \frac{5\sqrt{\eps}}{\delta_{k_{l-1}+1}} \le \frac{10\sqrt{\eps}}{\delta}.
\end{align}
Now we denote for $l\in [n]$ the dimension $d_l= \dim(\ran(E_l))$ and by $\left(\psi_{l,j}\right)_{j=1}^{d_l}\subseteq\cH$ an orthonormal basis of $\ran(E_l)$ and hence in particular
\begin{align*}
E_l = \sum_{j=1}^{d_l} \kb{\psi_{l,j}}.
\end{align*}
Let $V :\cH\to\cH\otimes\cH_E$ be a Stinespring isometry for the quantum channel $\cN,$ i.e.~$\cN(\,\cdot\,) = \Tr_E\left(V(\,\cdot\,)V^*\right).$ Now note that \eqref{eq:UnitalEiFix} gives for all $l\in [n]$
\begin{align*}
  \sum_{j=1}^{d_l}\, \Tr&\big(((\1-E_l)\otimes\1_E) V\kb{\psi_{l,j}}V^*\big) = \sum_{j=1}^{d_l}\Tr\left((\1-E_l)\cN(\kb{\psi_{l,j}})\right) \\&\le \max_{0\le\Lambda\le\1}\Tr\left(\Lambda(\cN(E_l) - E_l)\right) = \frac{1}{2}\left\|\cN(E_l) - E_l\right\|_1 \le \frac{10\sqrt{\eps}}{\delta} ,
\end{align*}
where we have used the variational expression of the trace distance \eqref{eq:VarExpTraceDistance}. From that we see
\begin{align*}
    \sqrt{\sum_{j=1}^{d_i}\big\|((\1-E_l)\otimes\1_E )V\ket{\psi_{l,j}}\big\|^2} = \sqrt{\sum_{j=1}^{d_l}\, \Tr\big(((\1-E_l)\otimes\1_E) V\kb{\psi_{l,j}}V^*\big)} \le \sqrt{\frac{10}{\delta}}\,\eps^{1/4}.
\end{align*}
Therefore, applying Lemma~\ref{lem:TurnVectorsSubspace}, we can find for all $l\in [n]$ an unitary $U_l$ on $\cH\otimes\cH_E$ with
\begin{align}
\label{eq:Ui1UnitalProof}
\|U_l - \1\|\le 2\sqrt{\frac{10}{\delta}} \,\eps^{1/4}\le 7\frac{\eps^{1/4}}{\sqrt{\delta}}
\end{align}
and 
\begin{align}
\label{eq:Ui2UnitalProof}
(E_l\otimes\1_E)U_lV\ket{\psi_{l,j}} = U_lV\ket{\psi_{l,j}}.
\end{align}
In particular $(E_l\otimes\1_E)U_lV E_l = U_lV E_l$ and since the projections $E_l,E_{l'}$ have orthogonal images for $l\neq l'$, we can combine these unitaries to define the isometry $V':\cH\to\cH\otimes\cH_E$ given by
\begin{align*}
    V' \coloneqq \bigoplus_{l=1}^{n} U_lVE_l.
\end{align*}
By construction, using \eqref{eq:Ui1UnitalProof} and \eqref{eq:Ui2UnitalProof}, the isometry $V'$ satisfies
\begin{align}
    \label{eq:UnitalIsomClose}
    \left\|V'-V\right\| \le \sup_{\|\psi\|=1}\sum_{l=1}^{n} \left\|(U_l-\1)VE_l\ket{\psi}\right\|  \le 7\sqrt{\frac{n}{\delta}} \,\eps^{1/4}\sup_{\|\psi\|=1}\sqrt{\sum_{i=1}^{n}\|E_i\ket{\psi}\|^2} =7\sqrt{\frac{n}{\delta}} \,\eps^{1/4}
\end{align}
and 
\begin{align}
\label{eq:IsometrySupportUnital}
   (E_l\otimes\1_E) V'E_lV'^*(E_l\otimes\1_E) = V'E_lV'^*.
\end{align}
Define the quantum channel $\cN'(\,\cdot\,) \coloneqq \Tr_E(V'(\,\cdot\,)V'^*)$ which by \eqref{eq:UnitalIsomClose} and Lemma~\ref{lem:CloseStineCloseChannel} satisfies
\begin{align}
\label{eq:N'ApproxUni}
    \frac{1}{2}\left\|\cN' -\cN\right\|_\diamond \le 7\sqrt{\frac{n}{\delta}} \,\eps^{1/4} \le 7
\sqrt{\frac{d}{\delta}}\eps^{1/4}
\end{align}
Since $\cN'$ is close to $\cN$ and the projections $E_l$ are approximate fixed points of $\cN,$ cf. \eqref{eq:UnitalEiFix}, they are also approximate fixed points of the new channel $\cN':$
\begin{align}
\label{eq:UnitalN'Ei}
 \nn\frac{1}{2}\|\cN'(E_l)-E_l\|_1&\le\frac{1}{2}\left\|\cN'(E_l)-E_l\right\|_1 \le\frac{d_l}{2}\left\|\cN' -\cN\right\|_\diamond + \frac{1}{2}\left\|\cN(E_l)-E_l\right\|_1\\& \le 7d_l\sqrt{\frac{d}{\delta}} \,\eps^{1/4} + \frac{10\sqrt{\eps}}{\delta} \le 17d^{3/2}\frac{\eps^{1/4}}{\sqrt{\delta}},
\end{align}
where for the last inequality we have used $\delta\ge\eps,$ and $d_l\le d.$ 
Since, by \eqref{eq:IsometrySupportUnital} we also have 
\begin{align}
\label{eq:UnitalOrthSupport}
E_l \,\cN'(E_l)E_l = \cN'(E_l)
\end{align}
and hence $\cN'(E_l) - E_l$ has support in $\ran(E_l),$ we get from \eqref{eq:UnitalN'Ei} the operator inequality 
\begin{align*}
  \cN'(E_l) = E_l + \cN'(E_l) - E_l \le E_l + \|\cN'(E_l)-E_l\|E_l\le \left(1+  17d^{3/2}\frac{\eps^{1/4}}{\sqrt{\delta}}\right) E_l.
\end{align*}
Using again that by \eqref{eq:UnitalOrthSupport} the operators $\cN'(E_l)$ and $\cN'(E_{l'})$ have orthogonal support for $l\neq l'$, we can use the generalised depolarising trick in Lemma~\ref{lem:GenDepLemma} to see that there exists a quantum channel $\Phi$ with \begin{align}
\label{eq:PhiApproxUni}
    \frac{1}{2}\|\Phi-\id\|_\diamond\le 17d^{3/2}\frac{\eps^{1/4}}{\sqrt{\delta}}
\end{align} and
\begin{align*}
    \Phi(\cN'(E_l)) = E_l.
\end{align*}
for all $l\in[n].$
Define now finally the map 
\begin{align*}
    \cM \coloneqq \Phi\circ\cN'
\end{align*}
which by definition is a quantum channel 
and satisfies
\begin{align}
\label{eq:MElFix}
    \cM(E_l) &= E_l\quad\quad\text{for all $l\in[n]$ }
\end{align}
From that we already see $\cM(
\1)=\1$ and hence $\cM$ is a unital channel.

We take $\sigma$ to be the state defined in \eqref{eq:SpecClusterState}, i.e.
\begin{align*}
    \sigma = \sum_{l=1}^n \mu_l E_l,
\end{align*}
with $\mu_l$ being the average of the spectral point in the $l^{th}$ cluster defined in \eqref{eq:AverageSpecCluster}. From \eqref{eq:MElFix} it is now immediate that
    \begin{align*}
    \cM(\sigma) &= \sigma.
\end{align*}
Furthermore, from \eqref{eq:N'ApproxUni} and \eqref{eq:PhiApproxUni} we see
\begin{align}
\label{eq:FinalChannelCloseUnital}
    \frac{1}{2}\left\|\cM - \cN \right\|_\diamond \le \frac{1}{2}\left\|\Phi - \id \right\|_\diamond + \frac{1}{2}\left\|\cN' -\cN\right\|_\diamond \le 24d^{3/2}\frac{\eps^{1/4}}{\sqrt{\delta}}.
\end{align}
From Lemma~\ref{lem:ApproxClusterState} we know
\begin{align}
\label{eq:StatesCloseUnital}
    \frac{1}{2}\left\|\rho-\sigma\right\|_1\le \frac{d^2\delta}{2}.
\end{align}
Optimising now over $\delta$ and choosing
\begin{align*}
    \delta = 48^{2/3} \frac{\eps^{1/6}}{d^{1/3}},
\end{align*}
we get by plugging this into \eqref{eq:StatesCloseUnital} and \eqref{eq:FinalChannelCloseUnital} 
\begin{align*}
\frac{1}{2}\left\|\rho - \sigma \right\|_1&\le \frac{48^{2/3}}{2} d^{5/3}\eps^{1/6} \le 7 d^{5/3}\eps^{1/6},
\\\frac{1}{2}\left\|\cM - \cN \right\|_\diamond&\le 7d^{5/3}\eps^{1/6}.
\end{align*}
\end{proof}
\subsection{Local channels and pure states}
In the following we consider a bipartiate Hilbert space $\cH_{AB}=\cH_A\otimes\cH_B$ with $\cH_A$ and $\cH_B$ being Hilbert spaces of finite dimension $d_A$ and $d_B$ respectively. Theorem~\ref{thm:ApproxLocalFixPure} below shows rapid fixability of approximate fixed points in the sense of Defintion~\ref{def:RapidFixingApproxFixedPoint} for  $\cS$ being the set of pure states on $\cH_{AB}$ and $\cC=\id_A\otimes{\rm{CPTP}}(\cH_B).$
\begin{theorem}[Fixability for local channels and pure states]
	\label{thm:ApproxLocalFixPure}
	Let $\cH_{AB}=\cH_A\otimes\cH_B$ be a finite-dimensional, bipartite Hilbert space with $d_A \coloneqq\dim(\cH_A)$ and $d_B\coloneqq\dim(\cH_B)$. Let $\rho_{AB} = \kb{\Psi}_{AB}$ be a pure state on $\cH_{AB}$ and $\cN_B\in{\rm{CPTP}}(\cH_B)$ such that for $\eps\ge 0$ we have
	\begin{align}
	\label{eq:ApproxLocalFixPure}
	(\id_A\otimes\cN_{B})(\rho_{AB}) \approx^{\eps} \rho_{AB}.
	\end{align}
	Then there exist a pure state $\sigma_{AB} = \kb{\Phi}_{AB}$ and a channel $\cM_B\in{\rm{CPTP}}(\cH_B)$ such that 
	\begin{align}
    \label{eq:LocalRapidFixability}
	\rho_{AB} \approx^{7\sqrt{d_*}\,\eps^{1/3}} \sigma_{AB},\quad\quad \cM_B \approx^{7\sqrt{d_*}\,\eps^{1/3}} \cN_B,
	\end{align}
	where $d_*= \min\{d_A,d_B\},$ and 
	\begin{align}
	(\id_A\otimes\cM_{B}) (\sigma_{AB}) = \sigma_{AB}.
	\end{align}
Furthermore, the state $\sigma_{AB}$ is independent of $\cN_B.$
\end{theorem}

For the proof of Theorem~\ref{thm:ApproxLocalFixPure} we express the state vector $\ket{\Psi}_{AB}$ in Schmidt decomposition as
\begin{align}
\label{eq:SChmidtLocalOutlineFix}
\ket{\Psi}_{AB} = \sum_{i=1}^{d_*} \sqrt{\mu_i}\ket{e_i}_A\ket{f_i}_B.
 \end{align}
 The idea is then to define the new state vector $\ket{\Phi}_{AB}$ approximating $\ket{\Psi}_{AB}$ by setting Schmidt coefficients in \eqref{eq:SChmidtLocalOutlineFix} below a certain threshold $\delta$ equal to zero. Denoting by $V:\cH_B\to\cH_B\otimes\cH_E$ a Stinespring isometry of the channel $\cN_B,$ we show for remainder of the Schmidt vectors $\ket{f_i}_B$ with large Schmidt coefficients that 
 \begin{align}
 \label{eq:approxSchmidtVectorOutline}
     V\ket{f_i}_B \approx \ket{f_i}_B\ket{\beta}_E
 \end{align}
 for some fixed state vector $\ket{\beta}_E$ on the environment Hilbert space $\cH_E$ independent of the index $i.$ Hence, using the rotation lemma~\ref{lem:UnitaryCloseVectors} we can find an unitary $U$ close to the identity on $\cH_B\otimes\cH_E$ such that its application onto the left hand side of \eqref{eq:approxSchmidtVectorOutline} makes this an exact equality. Therefore, defining the channel $\cM_B$ with Stinespring isometry $UV,$ we see that when applied onto the state $\kb{\Phi}_{AB}$ no correlations between the system and environment are produced and the state is left invariant. Similarly as in the proof of Theorems~\ref{thm:FixUnitary}, \ref{thm:FixingMixtureUnitaries} and~\ref{thm:FixingUnitalChannels} we then optimise over the parameter $\delta$ to obtain good approximation error for both state and channel. 

\begin{proof}[Proof of Theorem~\ref{thm:ApproxLocalFixPure}]
Without loss of generality assume $\eps\le 1/2$ as otherwise the statement is trivially true by simply taking $\cM_B =\id_B$ and $\sigma_{AB}=\rho_{AB}$ which satisfies \eqref{eq:LocalRapidFixability}.
\\ \indent
	Consider the Schmidt decomposition \begin{align}
 \label{eq:Schmidt1LocalFix}\ket{\Psi}_{AB} = \sum_{i=1}^{d_*} \sqrt{\mu_i}\ket{e_i}_A\ket{f_i}_B
 \end{align}
 with $d_* =\min\{d_A,d_B\},$ ordered Schmidt coefficients $\mu_1 \ge \mu_2 \ge \cdots \ge \mu_{d_*} \ge 0$ summing to 1 and $\left(\ket{e_i}_A\right)_{i=1}^{d_*}$ and $\left(\ket{f_i}_B\right)_{i=1}^{d_*}$ being orthonormal systems on $\cH_A$ and $\cH_B$ respectively. Let $V :\cH_{B}\to\cH_{B}\otimes\cH_E$ be a Stinespring isometry of $\cN_{B}$. 
	Denoting the identity on $\cH_A$ by $\1_A$, we write in Schmidt decomposition over the bipartition $AB$ and $E$
	\begin{align}\label{eq:Schmidt2LocalFix} \left(\1_{A}\otimes V\right)\ket{\Psi}_{AB} = \sum_{i= 1}^{d_Ad_B} \sqrt{\lambda_i}\ket{\alpha_i}_{AB}\ket{\beta_i}_E,\end{align}
	again with ordered Schmidt coefficients $\lambda_1 \ge \lambda_2 \ge \cdots \ge \lambda_{d_A d_B} \ge 0$ summing to 1 and $\left(\ket{\alpha_i}_{AB}\right)_{i=1}^{d_Ad_B}$ and $\left(\ket{\beta_i}_{E}\right)_{i=1}^{d_Ad_B}$ being orthonormal systems on $\cH_{AB}$ and $\cH_E$ respectively.  By assumption \eqref{eq:ApproxLocalFixPure} we have
	\begin{align*}
	\kb{\Psi}_{AB} \approx^\eps \id_A\otimes\cN_{B}(\kb{\Psi}_{AB}) = \Tr_{E}\left(\1_{A}\otimes V\left(\kb{\Psi}_{AB}\right)\1_{A}\otimes V^* \right)= \sum_{i=1}^{d_A d_B} \lambda_i \kb{\alpha_i}_{AB}.
	\end{align*}
	Therefore, by \eqref{eq:ContOfSpectrum} we have
	\begin{align}
 \label{eq:LocalFixLeadingSchmidt}
	\lambda_1 \ge 1-\eps
	\end{align}
and 
	\begin{align*}
	&\frac{1}{2}\left\|\kb{\Psi}_{AB} - \kb{\alpha_1}_{AB}\right\|_1 \\&\le \frac{1}{2}\left\|\kb{\Psi}_{AB} - \sum_{i=1}^{d_Ad_B}\lambda_i\kb{\alpha_i}_{AB}\right\|_1 + \frac{1}{2}\left\|\kb{\alpha_1}_{AB} - \sum_{i=1}^{d_Ad_B}\lambda_i\kb{\alpha_i}_{AB}\right\|_1 \\&\le 2\eps. 
	\end{align*}
    Rewriting the expression of the trace distance for pure states in \eqref{eq:PureTraceDistance}, the above gives for $\theta\in[0,2\pi]$ such that $e^{i\theta}= \frac{\langle\alpha_1,\Psi\rangle_{AB}}{|\langle\alpha_1,\Psi\rangle_{AB}|}$ we have
    \begin{align*}
         e^{i\theta}\langle\Psi,\alpha_1\rangle_{AB} \ge \sqrt{1-4\eps^2}\ge 1-4\eps^2
    \end{align*}
    and therefore
	\begin{align*}
	\left\|\ket{\Psi}_{AB} - e^{i\theta} \ket{\alpha_1}_{AB}\right\| = \sqrt{2\left(1-e^{i\theta}\langle\Psi,\alpha_1\rangle_{AB}\right)} \le \sqrt{8}\,\eps\le 3\eps.
	\end{align*}
	Using this now together with \eqref{eq:LocalFixLeadingSchmidt} and the Schmidt decompositions \eqref{eq:Schmidt1LocalFix} and \eqref{eq:Schmidt2LocalFix}, we get for all $j\in\{1,\cdots, d_*\}$
	\begin{align*}
	e^{-i\theta}\sqrt{\mu_j}&\approx^{\eps}e^{-i\theta}\sqrt{\mu_j\lambda_1}=	e^{-i\theta}\sqrt{\lambda_1}\bra{e_j,f_j}\Psi\rangle\approx^{3\eps}\sqrt{\lambda_1} \bra{e_j,f_j} \alpha_1\rangle\\&=\bra{e_j,f_j,\beta_1}\1_{A}\otimes V \ket{\Psi} = \sqrt{\mu_j}\bra{f_j,\beta_1}V \ket{f_j},
	\end{align*}
 where we dropped the system labels for readability.
 Rewriting the above, we have by the triangle inequality
 \begin{align*}
     1- \Re(e^{i\theta}\bra{f_j,\beta_1}V \ket{f_j})\le \left|1- e^{i\theta}\bra{f_j,\beta_1}V \ket{f_j}\right| \le  \frac{4\eps}{\sqrt{\mu_j}}
 \end{align*}
	and therefore 
	\begin{align}
    \label{eq:LocalSchmidtVectorCloseness}
	\left\|e^{-i\theta}\ket{f_j}_B\ket{\beta_1}_E - V \ket{f_j}_{B} \right\|^2 = 2\left(1-\Re\left(e^{i\theta}\bra{f_j,\beta_1}V \ket{f_j}\right)\right) \le \frac{8\eps}{\sqrt{\mu_j}}.
	\end{align}
	Let now $0\le\delta< \frac{1}{\sqrt{d_*}}$ (whose exact value is to be determined later) and pick   $i_*\in\{1,\cdots, d_*\}$ such that 
 \begin{align*}\sqrt{\mu_{i_*}}\ge \delta \ge   \sqrt{\mu_{i_*+1}}.
 \end{align*} 
 Note that by definition of $i_*$ we have
 \begin{align*}
     1-\sum_{i=1}^{i_*}\mu_i = \sum_{i=i_*+1}^{d_*}\mu_i \le d_*\delta^2
 \end{align*}
 and hence, denoting the normalisation constant $c\coloneqq \sqrt{\sum_{i=1}^{i_*}\mu_i}\ge 1-d_*\delta^2 >0,$
 we can define the normalised state vector by truncating 
  the Schmidt-decomposition in \eqref{eq:Schmidt1LocalFix} as
 \begin{align*}
 \ket{\Phi}_{AB} \coloneqq \frac{1}{c} \sum_{i=1}^{i_*} \sqrt{\mu_i}\ket{e_i}_A\ket{f_i}_B\end{align*}
 and the pure state $\sigma_{AB} \coloneqq \kb{\Phi}_{AB}$. 
 Note, that by using the expression of the trace distance for pure states \eqref{eq:PureTraceDistance}, we have  
 \comment{\begin{align*}
     \left\|\ket{\Phi}_{AB} -\ket{\Psi}_{AB}\right\|&\le \left\|\ket{\Phi}_{AB}-\sum_{i=1}^{i_*} \sqrt{\mu_i}\ket{e_i}_A\ket{f_i}_B\right\| + \left\|\sum_{i=1}^{i_*} \sqrt{\mu_i}\ket{e_i}_A\ket{f_i}_B-\ket{\Psi}_{AB}\right\| \\&\le 1-c + \sqrt{\sum_{i=i_*+1}^{d_*} \mu_i} \le 2\sqrt{d_*}\,\delta,
 \end{align*}}
 \begin{align}
 \label{eq:LocalStateCloseness}
\frac{1}{2}\left\|\sigma_{AB}-\rho_{AB}\right\|_1 = \sqrt{1-|\langle\Phi_{AB},\Psi_{AB}\rangle|^2} = \sqrt{1-\sum_{i=1}^{i_*}\mu_i} \le \sqrt{d_*}\,\delta.
 \end{align}
 By \eqref{eq:LocalSchmidtVectorCloseness} we have for all $j \in\{ 1,\cdots,i_*\}$ 
	\begin{align*}
	\left\|e^{-i\theta}\ket{f_j}_B\ket{\beta_1}_E - V \ket{f_j}_{B} \right\| \le \sqrt{\frac{8\eps}{\delta}}
	\end{align*}
\comment{ and hence for the orthogonal projections $E_i = \kb{f_i}_B\otimes\kb{\beta_1}_E$ and $F_i=V\kb{f_i}_{B}V^*$ we have
 \begin{align*}
     \left\|E_i -F_i\right\| \le 2\left\|e^{-i\theta}\ket{f_i}_B\ket{\beta_1}_E -  V\ket{f_i}_{B} \right\| \le 2\sqrt{8} \sqrt{\frac{\eps}{\delta}} 
 \end{align*}}
	and hence by Lemma~\ref{lem:UnitaryCloseVectors}
 there exists an unitary operator $U:\cH_B\otimes\cH_E\to\cH_B\otimes\cH_E$ such that \begin{align*}
     \|U-\1\|\le 5\sqrt{8}\sqrt{\frac{i_*\eps} {\delta}}\le 15\,\sqrt{\frac{d_*\eps}{\delta},}
 \end{align*} and 
 \begin{align*}
     UV\ket{f_j}_B = e^{-i\theta}\ket{f_j}_B\ket{\beta_1}_E.
 \end{align*}
 Hence, we can define an isometry $V' \coloneqq UV:\cH_B\to\cH_B\otimes\cH_E$, which satisfies $\|V'-V\|\le\|U-\1\|\le15\sqrt{\frac{d_*\eps}{\delta}}$
 \comment{ and for all $i,j\in\{1,\cdots,i_*\}$ 
  \begin{align*}
     V'\ket{f_i}\!\bra{f_j}_BV'^* = \ket{f_i}\!\bra{f_j}_B\otimes\kb{\beta_1}_E.
 \end{align*}}
 and 
 \begin{align*}
     (\1_A\otimes V')\sigma_{AB}(\1_A\otimes V'^*) &= \sum_{i,j=1}^{i_*}\sqrt{\mu_i\mu_j}\ket{e_i}\!\bra{e_j}_B\otimes V'\ket{f_i}\!\bra{f_j}_BV'^* \\&= \sum_{i,j=1}^{i_*}\sqrt{\mu_i\mu_j}\ket{e_i}\!\bra{e_j}_B\otimes\ket{f_i}\!\bra{f_j}_B\otimes\kb{\beta_1}_E = \sigma_{AB}\otimes\kb{\beta_1}_E.
 \end{align*}
 We can now define the quantum channel on $\cH_B$ by
	$\cM_B(\,\cdot\,) \coloneqq \Tr_E(V'(\,\cdot\,) V'^*)$ which satisfies by Lemma~\ref{lem:CloseStineCloseChannel}
	\begin{align}
 \label{eq:LocalChannelCloseness}
	    \frac{1}{2}\left\|\cM_B-\cN_B\right\|_\diamond\le 15\,\sqrt{\frac{d_*\eps}{\delta}}
	\end{align} and furthermore the fixed point equation
	\begin{align*}
	(\id_A\otimes\cM_B)(\sigma_{AB}) = \sigma_{AB}
	\end{align*} 
 Next, optimising over 
 $\delta$ in \eqref{eq:LocalStateCloseness} and \eqref{eq:LocalChannelCloseness} and choosing\footnote{Note that without loss of generality $\eps$ is small enough such with the choice \eqref{eq:DeltaLocalChoice} we have $\delta<1/\sqrt{d_*}$ (as we demanded above) as otherwise the approximations \eqref{eq:ApproxLocalStuffLast} of $\rho_{AB}$ and $\cN_B$ by a fixed point pair $\sigma_{AB}$ and $\cM_B$ is trivially satisfied.}
 \begin{align}
 \label{eq:DeltaLocalChoice}
     \delta = 15^{2/3} \eps^{1/3},
 \end{align}
 yields
\begin{align}
\label{eq:ApproxLocalStuffLast}
    \nn\frac{1}{2}\left\|\sigma_{AB}-\rho_{AB}\right\|_1 &\le 15^{2/3} \sqrt{d_*}\eps^{1/3}\le 7\sqrt{\min\{d_A,d_B\}}\,\eps^{1/3},\\
    \frac{1}{2}\left\|\cM_B-\cN_B\right\|_\diamond&\le 7\sqrt{\min\{d_A,d_B\}}\,\eps^{1/3}.
\end{align}
\end{proof}

\section{Impossibility of rapid fixability for local channels}
\label{sec:MultipleImpossLocal}
In the previous sections we have shown that for a variety of natural choices of $\cS$ and $\cC$, approximate fixed point equations are rapidly fixable in the sense of Definition~\ref{def:RapidFixingApproxFixedPoint}. In particular,  Theorem~\ref{thm:ApproxLocalFixPure} shows that rapid fixability is possible in the bipartite regime $\cH_{AB}=\cH_A\otimes\cH_B$ with $\cS$ being the set of pure states on $\cH_{AB}$ and $\cC=\id_A\otimes{\rm{CPTP}}(\cH_B).$ In contrast,
we show in this section that if one considers instead the full set of general mixed states, $\cS=\mathfrak{S}(\cH_{AB})$, rapid fixability of approximate fixed points is no longer possible for $\cC=\id_A\otimes{\rm{CPTP}}(\cH_B).$  We do this by providing a concrete counterexample. As a matter of fact, we also see that named impossibility already occurs in the classical bipartite regime, i.e.~for $\cS={\rm{Prob}}(\cX\times\cY)$ and $\cC = \id_X\otimes{\rm{Stoch}}(\mathcal{Y})$. Both of these results are stated in Corollary~\ref{cor:ImpossiLocalFix} below. 

To prove Corollary~\ref{cor:ImpossiLocalFix}, we first focus on the question of fixability of multiple approximate fixed points at once. Similarly to Definition~\ref{def:FixingApproxFixPoints}, we say that for a $d$-dimensional Hilbert space $\cH,$ $n\in\N$ and $\cS\subseteq\mathfrak{S}(\cH)$ and $\cC\subseteq {\rm{CPTP}}(\cH)$, \emph{$n$ approximate fixed point equations are fixable at once} for the pair $\cS\subseteq\mathfrak{S}$ and $\cC\subseteq{\rm{CPTP}}(\cH)$ if the following holds: There exist approximation functions $f ,g:\N\times\R_+\to\R_+$ such that 
\begin{align}
    \lim_{\eps\to 0}f(d,\eps)=\lim_{\eps\to 0}g(d,\eps)=0,
\end{align}
and for all $\rho_1,\cdots,\rho_n\in\cS,\,\cN\in\cC$ and $\eps\ge 0$ such that $\cN(\rho_i)\approx^{\eps}\rho_i$ for all $i\in[n]$, we can find new states and a channel of the same structure:
\begin{align}
 \sigma_i\in\cS\quad\quad\text{and}\quad\quad\cM\in\cC,\quad
\end{align}
which are close to original ones, i.e.
\begin{align}
\sigma_i\approx^{f(d,\eps)} \rho_i \quad\quad\text{and}\quad\quad\cM\approx^{g(d,\eps)}\cN,
\end{align}
 and which satisfy the fixed point equations
\begin{align}
\cM(\sigma_i)=\sigma_i.
\end{align}
 Importantly, here we demand that the 
 new channel $\cM$ satisfies an exact fixed point equation for all states $\sigma_i$ at the same time.

Furthermore, analogously to Definition~\ref{def:RapidFixingApproxFixedPoint} we say $n$ approximate fixed point equations are \emph{rapidly fixable at once} if the functions $f$ and $g$ in the above satisfy
\begin{align}
f(d,\eps) = c\,d^{b}\eps^{a}\quad\quad\text{and}\quad\quad g(d,\eps) = c'd^{b'}\eps^{a'},
\end{align}
for some $a,a'>0$ (and typically $a,a'\le 1 $) and $b',b',c,c'\ge 0$ all independent of $d$ and $\eps.$

Theorem~\ref{thm:MultipleFixCounterEx} below shows (via a concrete counterexample) that already in the classical case, i.e.~for $\cS= {\rm{Prob}}(\cX)$ and $\cC= {\rm{Stoch}}(\cX)$, rapid fixability of two approximate fixed points at once is in general impossible.
Here, we find for every dimension $d$, two probability vectors and a stochastic matrix which satisfy approximate fixed point equations with error parameter vanishing exponentially as $d\to\infty$, i.e.~$\eps=\frac{1}{2^d}.$ However, for any two new probability vectors and a stochastic matrix, which fix named approximate fixed point equations at once, at least one needs to be at a constant distance from the corresponding original one. 
\begin{theorem}[No rapid fixing of 2 approximate fixed points in the classical case]
\label{thm:MultipleFixCounterEx}
For every $d\in\N$ there exist probability vectors $\vec P_1$ and $\vec P_2$ and a stochastic matrix $T$ on a classical alphabet $\cX$ with $|\cX|=d$
such that
\begin{align}
    \frac{1}{2}\left\|\vec P_1 -\vec P_2\right\|_1 = 1 
\end{align}
and
\begin{align}
    \frac{1}{2}\left\|T\vec P_1 - \vec P_1 \right\|_1 \le \frac{1}{2^d}, \quad\quad \frac{1}{2}\left\|T\vec P_2 - \vec P_2 \right\|_1 \le \frac{1}{2^d},
\end{align}
and which satisfy the following: For 
any pair of probability vectors $\vec Q_1$ and $\vec Q_2$ and any stochastic matrix $S$ such that \begin{align}
    S\vec Q_1 = \vec Q_1 \quad\quad\text{and}\quad\quad S\vec Q_2 = \vec Q_2,
\end{align} 
we either have
\begin{align}
\label{eq:FixedPointBadApproxCounter}
   \|S-T\| \ge \frac{1}{4}\quad\text{or}\quad\frac{1}{2}\left\|\vec Q_1 -\vec P_1\right\|_1\ge \frac{1}{2}\quad\text{or}\quad\frac{1}{2}\left\|\vec Q_2 -\vec P_2\right\|_1\ge \frac{1}{2}. 
\end{align}
This shows that two approximate fixed points are, in general,  not rapidly fixable at once for the pair $\cS= {\rm{Prob}}(\mathcal{\cX})$ and $\mathcal{\cC}=
\Stoch(\mathcal{\cX})$ for classical alphabets of size $|\cX| =d.$
\end{theorem}

We can embed this counterexample into the quantum setting using the natural constructions \eqref{eq:ProbClassicalState}
and \eqref{eq:StochMatrixClassicalChannel}. Hence, we find for every dimension $d$, two states and a channel which satisfy approximate fixed point equations with error parameter vanishing exponentially in the dimension. Importantly, in the quantum setting we have a bigger set of possible states and channels to fix named approximate fixed point equation. Therefore, the impossibility statement of Theorem~\ref{thm:MultipleFixCounterEx} does not immediately also give the corresponding impossibility statement in the quantum case. However, using a slight adjustment of the classical argument we are still able to show that for any new two states and a channel, which fix named approximate fixed point equations at once, at least one needs to be at a distance of order $\mathcal{O}(1/d)$ from the corresponding original one. Since, as mentioned, we have for the error parameter $\eps=\frac{1}{2^d}$, this shows that these two approximate fixed points are also not rapidly fixable in the quantum setting:

\begin{theorem}[No rapid fixing of 2 approximate fixed points in quantum case]
\label{thm:MultipleFixCounterExQuant}
For every $d\in\N$ there exist states $\rho_1$ and $\rho_2$ and a channel $\cN$ on a $d$-dimensional Hilbert space $\cH$ such that
\begin{align}
    \frac{1}{2}\left\|\rho_1 -\rho_2\right\|_1 = 1 
\end{align}
and
\begin{align}
\label{eq:MultipleApproxQuantCounter}
    \frac{1}{2}\left\|\cN(\rho_1) - \rho_1 \right\|_1 \le \frac{1}{2^d}, \quad\quad \frac{1}{2}\left\|\cN(\rho_2)- \rho_2 \right\|_1 \le \frac{1}{2^d},
\end{align}
and which satisfy the following: For all states $\sigma_1$ and $\sigma_2$ and channels $\cM$ such that \begin{align}
    \cM(\sigma_1) = \sigma_1 \quad\quad\text{and}\quad\quad \cM(\sigma_2) = \sigma_2
\end{align} 
we either have
\begin{align}
\label{eq:FixedPointBadApproxCounterQuantum}
   \frac{1}{2}\|\cM-\cN\|_\diamond \ge \frac{1}{16d}\quad\text{or}\quad\frac{1}{2}\left\|\sigma_1 -\rho_1\right\|_1\ge \frac{1}{2}\quad\text{or}\quad\frac{1}{2}\left\|\sigma_2 -\rho_2\right\|_1\ge \frac{1}{2}. 
\end{align}
This shows two approximate fixed points are, in general, not rapidly fixable at once for the pair $\cS= \mathfrak{S}(\cH)$ and $\cC = {\rm{CPTP}}(\cH).$
\end{theorem}

We can encode the counterexamples of Theorem~\ref{thm:MultipleFixCounterEx} and Theorem~\ref{thm:MultipleFixCounterExQuant} into the language of bipartite states. By that, we see the impossibility of rapidly fixing approximate fixed point equations in the bipartite setting with channels acting trivially on the first of the systems. We present this statement in the following corollary and prove it using the results from Theorems~\ref{thm:MultipleFixCounterEx} and~\ref{thm:MultipleFixCounterExQuant}. We then continue to provide the proofs for Theorems~\ref{thm:MultipleFixCounterEx} and~\ref{thm:MultipleFixCounterExQuant}.

\begin{corollary}
\label{cor:ImpossiLocalFix}
  Approximate fixed point equations are in general not rapidly fixable in the sense of Defintion~\ref{def:RapidFixingApproxFixedPoint} for the following cases:
  \begin{enumerate}
      \item \label{it:QuantImposs} $\cS\in\mathfrak{S}(\cH_{AB})$ and $\cC = \id_A\otimes{\rm{CPTP}}(\cH_B)$, with $\cH_{AB}=\cH_A\otimes\cH_B$ being a bipartite Hilbert space, with $\dim(\cH_A)\ge 2$  
    \item \label{it:ClassImposs} $\cS= {\rm{Prob}}(\mathcal{X}\times\mathcal{Y})$ and $\cC = \id_X\otimes{\rm{Stoch}}(\mathcal{Y})$ for $|\mathcal{X}|\ge 2.$
  \end{enumerate}
\end{corollary}
\begin{proof}[Proof of Corollary~\ref{cor:ImpossiLocalFix} using Theorems~\ref{thm:MultipleFixCounterEx} and~\ref{thm:MultipleFixCounterExQuant}]
 We focus on the quantum setting \ref{it:QuantImposs} above, as the statement in the classical setting~\ref{it:ClassImposs} follows exactly the same way using Theorem~\ref{thm:MultipleFixCounterEx}.
Consider for $d\in\N$, a $d$-dimensional Hilbert space, $\cH_B$. Furthermore, let $\cH_A$ be a Hilbert space with $d_A\coloneqq\dim(\cH_A)\ge 2$, and we denote by $\left\{\ket{i}\right\}_{i=1}^{d_A}$ an orthonormal basis on $\cH_A.$ Consider the bipartite state on $\cH_{AB} =\cH_A\otimes\cH_B$ defined by
\begin{align}
    \rho_{AB} \coloneqq \frac{1}{2}\kb{1}\otimes \rho_1 + \frac{1}{2}\kb{2}\otimes \rho_2,
\end{align}
where $\rho_1$ and $\rho_2$ are the states given in Theorem~\ref{thm:MultipleFixCounterExQuant}. Furthermore, by $\cN_B$ we denote the channel from Theorem~\ref{thm:MultipleFixCounterExQuant}. Using \eqref{eq:MultipleApproxQuantCounter}, we see by block diagonality
\begin{align*}
    \frac{1}{2}\left\|(\id_A\otimes\cN_B)(\rho_{AB}) - \rho_{AB}\right\|_1 = \frac{1}{4}\left\|\cN_B(\rho_1)-\rho_1\right\|_1 +\frac{1}{4}\left\|\cN_B(\rho_2)-\rho_2\right\|_1 \le \frac{1}{2^d} =:\eps_d.
\end{align*}
Let $\sigma_{AB}\in\mathfrak{S}(\cH_{AB})$ be a state and $\cM_B\in{\rm{CPTP}}(\cH_B)$ be a channel such that
\begin{align}
\label{eq:FIXEDPOINTCOUNTER}
    (\id_A\otimes\cM_{B})(\sigma_{AB}) =\sigma_{AB}.
\end{align}
without loss of generality 
\begin{align}
\label{eq:CQFORMCOUNTER}
    \sigma_{AB} = \sum_{i=1}^{d_A} q_i\kb{i}\otimes \sigma_i 
\end{align}
for some probability distribution $\left(q_i\right)_{i=1}^{d_A}$ and states $\left(\sigma_i\right)_{i=1}^d\subseteq\mathfrak{S}(\cH_B)$ as otherwise consider for the dephasing channel
\begin{align*}
    \cE_A(\,\cdot\,) \coloneqq \sum_{i=1}^{d_A} \kb{i}\ \bra{i}(\,\cdot\,)\ket{i}
\end{align*}
the state $(\cE_{A}\otimes\id_B)(\sigma_{AB}),$ which is of the form \eqref{eq:CQFORMCOUNTER} and furthermore is a fixed point of $\id_A\otimes\cM_B$ by \eqref{eq:FIXEDPOINTCOUNTER}. Assume furthermore 
\begin{align}
\label{eq:CQStateClosenessCounter}
    \frac{1}{2}\left\|\sigma_{AB}-\rho_{AB}\right\|_1 < \frac{1}{8}.
\end{align}
From that we see that
\begin{align*}
    \frac{1}{2}\left\|\sigma_1 -\rho_1\right\|_1 &\le \left\|q_1\sigma_1 -\frac{1}{2}\rho_1\right\|_1 + \left|q_1-\frac{1}{2}\right| \le 2\left\|q_1\sigma_1 -\frac{1}{2}\rho_1\right\|_1 \\&\le 2\left\|\sigma_{AB}-\rho_{AB}\right\|_1 < \frac{1}{2}
\end{align*}
where for the second to last inequality we have again used block diagonality of $\rho_{AB}$ and $\sigma_{AB},$ and by the same argument
\begin{align*}
     \frac{1}{2}\left\|\sigma_2 -\rho_2\right\|_1 < \frac{1}{2}.
\end{align*}
Note furthermore, from a similar argument it is easy to see that $q_1,q_2>0$ as otherwise \eqref{eq:CQStateClosenessCounter} would be violated. The fixed point equation \eqref{eq:FIXEDPOINTCOUNTER} implies 
\begin{align*}
    \cM_{B}(\sigma_1) = \sigma_1\quad\quad\text{and}\quad\quad\cM_{B}(\sigma_2) = \sigma_2
\end{align*}
as $\cM_{B}(\sigma_1) = \frac{1}{q_1}\bra{1}\cM_B(\sigma_{AB})\ket{1}_A =\frac{1}{q_1}\bra{1}\sigma_{AB}\ket{1}_A =\sigma_1$ and similarly for $\sigma_2.$

Hence, by Theorem~\ref{thm:MultipleFixCounterEx} we see 
\begin{align*}
  \frac{1}{2}\left\|\cM_B-\cN_B\right\|_1 \ge \frac{1}{16d} =\frac{1}{16}\log_2(1/\eps_d)^{-1} = \frac{1}{8d}\eps_d^{1/d}.
\end{align*}
Hence, this shows that approximate fixed point equations for $\cS=\mathfrak{S}(\cH_{AB})$ and $\cC =\id_A\otimes{\rm{CPTP}}(\cH_B)$ cannot be rapidly fixable, i.e.~the approximation functions $f$ and $g$ from Definitions~\ref{def:FixingApproxFixPoints} and~\ref{def:RapidFixingApproxFixedPoint} cannot satisfy at the same time
\begin{align}
f(d,\eps) = c\,d^{b}\eps^{a}\quad\quad\text{and}\quad\quad g(d,\eps) = c'd^{b'}\eps^{a'},
\end{align}
for some $a,a'>0$ and $b',b',c,c'\ge 0$ all independent of $d$ and $\eps.$
\end{proof}

For the remainder of the section we construct the concrete example which is featured in the results of Theorems~\ref{thm:MultipleFixCounterEx} and~\ref{thm:MultipleFixCounterExQuant}. Consider for $d\in\N$ the non-negative vector
\begin{align*}
    \vec{v} \coloneqq \left(1,\frac{1}{2},\frac{1}{4},\cdots,\frac{1}{2^{d-2}},0\right),
\end{align*}
i.e.~$v_i = \frac{1}{2^{i-1}}$ for all $i\in[d-1]$ and $v_d=0.$
We then define two normalised probability vectors:
\begin{align}
\label{eq:DefP1Counter}
 \vec{P}_1  \coloneqq c\,\vec{v},
\end{align}
with normalisation constant $c\coloneqq\left(\sum_{i=1}^d v_i\right)^{-1} \le 1$ and
\begin{align}
\label{eq:DefP2Counter}
    \vec{P_2} \coloneqq \left(0,\cdots,0,1\right),
\end{align}
i.e.~$(P_2)_i = \delta_{id}$ for all $i\in[d].$ 
Furthermore, consider the stochastic matrix
\begin{align}
 T \coloneqq  \begin{pmatrix}
\frac{3}{4} & \frac{1}{2} & 0 & 0 &0&0&\cdots &0\\
\frac{1}{4} & \frac{1}{4} & \frac{1}{2} & 0&0&0& \cdots & 0 \\
0 & \frac{1}{4} & \frac{1}{4} & \frac{1}{2} &0&0&\cdots& 0  \\0 & 0 & \frac{1}{4} & \frac{1}{4} &\frac{1}{2}&0 &\cdots & 0\\
\vdots& & & &\ddots & & & \vdots\\
0& & \cdots & 0 & \frac{1}{4} & \frac{1}{4} & \frac{1}{2} &  0\\
0& &\cdots & 0& 0&  \frac{1}{4} &\frac{1}{4} &  0 \\
0& &\cdots & 0&0 &0& \frac{1}{4} &  1
\end{pmatrix}, \label{eq:TCounterEx}
\end{align}
i.e.~componentwise we have for the first column
$T_{11}= \frac{3}{4},$ $T_{21} =\frac{1}{4}$ and then for the $i^{th}$ column with $i=2,\cdots, d-1$
\begin{align*}
    T_{i-1\,i} = \frac{1}{2},\quad T_{i \,i} = \frac{1}{4}, \quad T_{i+1\, i} = \frac{1}{4}
\end{align*}
and finally for the $d^{th}$ column $T_{dd} =1$ and all other components of $T$ being equal to zero.

From that definition it is easy to see that $\vec P_2$ is the unique fixed point of $T$, \begin{align}
\label{eq:P2FixedCounter}
    T\vec P_2 = \vec P_2.
\end{align}
Moreover, we see that $P_1$ is an approximate fixed point of $T$:
\begin{align*}
    T\vec P_1 = c\,\left(1,\frac{1}{2},\frac{1}{4},\cdots,\frac{1}{2^{d-3}},\frac{3}{2^{d}},\frac{1}{2^{d}}\right) = \vec P_1 + \vec\Delta,
\end{align*}
where $\vec \Delta\in\R^d$ with \begin{align*}
\Delta_{d} = - \Delta_{d-1} \coloneqq \frac{c}{2^{d}}
\end{align*}
and all other components equal to zero and hence
\begin{align}
\label{eq:P1AprFixedCounter}
\frac{1}{2}\left\|T\vec P_1-\vec P_1\right\|_1 = \frac{1}{2}\left\|\vec \Delta\right\|_1 = \frac{c}{2^{d}} \le \frac{1}{2^{d}}.
\end{align}

In the following two lemmas we see that any stochastic matrix $S$ close to $T$ necessarily has an unique fixed point probability vector. From that Theorem~\ref{thm:MultipleFixCounterEx} is easy to conclude since, for $S$ being close to $T$, the fixed points $\vec Q_1$ and $\vec Q_2$ must be equal and hence far from the $\vec P_1$ and $\vec P_2$ respectively (cf. \eqref{eq:FixedPointBadApproxCounter}) as they are orthogonal. 

First we see in Lemma~\ref{lem:SPosiCounter} that any stochastic matrix $S$ sufficiently close to $T$ which has a fixed point $\vec Q\neq\vec P_2$ necessarily needs to have a positive flow from the entry $d$ to some entry $j\neq d$, with which we precisely mean
\begin{align*}
S_{j\,d} >0.
\end{align*}
From that we conclude in Lemma~\ref{lem:SUniqueFixCounter} that $S$ has in fact a one-dimensional fixed point space using the Perron-Frobenius theorem for irreducible matrices \cite[Theorem 8.4.4]{HornJohnson_MatrixAnalysis(Book)_1985}. This can be interpreted as a robustness result for the uniqueness of fixed points in a neighbourhood around $T.$ 
\begin{lemma}
\label{lem:SPosiCounter}
Let $S$ be a stochastic matrix in $d$ dimensions satisfying
\begin{align}
    \left\|S-T\right\| <\frac{1}{4},
\end{align}
with $T$ being the stochastic matrix defined in \eqref{eq:TCounterEx}, and furthermore $S\vec Q =\vec Q$ for some probability vector $\vec Q \neq (0,\cdots,0,1).$ Then 
\begin{align*}
S_{jd} >0
\end{align*}
for some $j\in[d-1]$.
\end{lemma}
\begin{proof}
Since 
\begin{align*}
\|S-T\| = \max_{j\in[d]}\sum_{i=1}^d|S_{ij} -  T_{ij}| <1/4  
\end{align*}
and $T_{i+1 \,i}=\frac{1}{4}$ for all $i\in[d-1],$ we see 
\begin{align}
\label{eq:S>0}
    S_{i+1 \,i}> 0
\end{align}
for all $i\in[d-1].$

Assume now for contradiction $S_{j\,d} = 0 $ for all $j\in[d-1]$ and hence $S_{dd}=1.$ Then, since $\vec Q$ is a fixed point of $S$ we know 
\begin{align*}
   Q_d=  \left(S\vec Q\right)_d = \sum_{j=1}^d S_{d\, j}\,Q_j \ge S_{d\,d-1}\,Q_{d-1}+Q_d \ge Q_d
\end{align*}
and therefore $Q_{d-1} =0$ since $S_{d\,d-1}>0$ by \eqref{eq:S>0}. From that we see
\begin{align*}
   0 = Q_{d-1} = \left(S\vec Q\right)_{d-1} = \sum_{j=1}^d S_{d-1\, j}\,Q_j \ge S_{d-1\,d-2}\,Q_{d-2}
 \end{align*}
 and therefore $Q_{d-2}=0$ since $S_{d-1\,d-2}>0$. By iterating this we find $Q_j=0$ for all $j\in[d-1]$ and since $\vec Q$ is a probability vector this already implies
 \begin{align*}
     \vec Q =\left(0,\cdots,0,1\right),
 \end{align*}
 which is a contraction to the assumption of the lemma. Therefore, we conclude $S_{dd}<1$ and hence that there exists a $j\in[d-1]$ such that $S_{j\,d}>0.$
\end{proof}
\begin{lemma}[Robustness of uniqueness of fixed point in classical case]
\label{lem:SUniqueFixCounter}
Let $S$ be a stochastic matrix in $d$ dimensions satisfying
\begin{align}
    \left\|S-T\right\| <\frac{1}{4},
\end{align}
with $T$ being the stochastic matrix defined in \eqref{eq:TCounterEx}.
Then $\cF(S)$ is one-dimensional, i.e.~$S$ has an unique fixed point probability vector.
\end{lemma}
\begin{proof}
Similarly to the argument in the proof of Lemma~\ref{lem:SPosiCounter} we get that since 
\begin{align*}
\|S-T\| = \max_{j\in[d]}\sum_{i=1}^d|S_{ij} -  T_{ij}| <1/4  
\end{align*}
we have \\
\begin{align}
\label{eq:Sposi1InProof}
    S_{i+1\,i} &>0,\quad\quad\text{for all $i=1,\dots,d-1,$}\\
    S_{i-1 \, i}&>0,\quad\quad\text{for all $i=2,\dots,d-1$,} \label{eq:Sposi1InProof2}
\end{align}
because $T_{i+1 \,i} =\frac{1}{4}$ for all $i=1,\dots,d-1$ and $T_{i-1 \,i} =\frac{1}{2}$ for all $i=2,\dots,d-1.$

Furthermore, assume that there exists a probability vector $\vec Q\neq (0,\cdots,0,1),$ which is a fixed point of $S$ (if such a $\vec Q$ does not exist we know that $(0,\cdots,0,1)$ must be, up to multiplication with a scalar factor, the unique fixed point of $S$ and we therefore finished the proof.) By Lemma~\ref{lem:SPosiCounter}, we therefore know that there exists $j\in[d-1]$ such that
\begin{align}
\label{eq:Sposi2InProof}
    S_{j\,d}>0.
\end{align}
 From \eqref{eq:Sposi1InProof} and \eqref{eq:Sposi2InProof} we see that for every $i,j\in[d]$ with $i\neq j$ there exists a \emph{path} connecting $i$ and $j$ and such that $S$ is componentwise positive along this path, with which we mean precisely the following:\footnote{To convince ourselves that for $i,j\in[d]$ with $i\neq j$ such a path exists, consider the following construction in different cases: \emph{Case 1) $i<j$:} Simply take the path $i_1 =i,$ $i_2 =i+1,\cdots ,i_{n} =j$ which is possible by \eqref{eq:Sposi1InProof2}. \emph{Case 2) $i>j,\ i\neq d$:} Take as path $i_1 =i,$ $i_2=i-1,\cdots i_{n}=j$ which is possible by \eqref{eq:Sposi1InProof}. \emph{Case 3) $i=d>j:$} Let $k\in[d-1]$ be such that $S_{kd}>0$ which exists as argued in \eqref{eq:Sposi2InProof}. Then start the path with $i_1=d,$ $i_2=k$ and then continue with case 1) or case 2) depending on $k\ge j$ or $k<j.$} For all $i,j\in[d]$ there exists $n\in\N$ and $i_2,i_3,\dots,i_{n-1}\in[d]$ such that denoting $i_1\equiv i$ and $i_{n}\equiv j$ we have for all $l\in[n]$
\begin{align*}
S_{i_{l+1}\,i_{l}} >0.
\end{align*}
In other words, $S$ is \emph{irreducible} (cf. \cite[Theorem 6.2.24]{HornJohnson_MatrixAnalysis(Book)_1985} for other characterisations of irreducible matrices). Therefore, we can apply the Perron-Frobenius theorem for irreducible matrices \cite[Theorem 8.4.4]{HornJohnson_MatrixAnalysis(Book)_1985} giving that $1$ is a simple eigenvalue of $S$, i.e.~$\cF(S)$ is one-dimensional.
\end{proof}

We are now ready to apply Lemma~\ref{lem:SUniqueFixCounter} to give the proof of Theoerm~\ref{thm:MultipleFixCounterEx}.

\begin{proof}[Proof of Theorem~\ref{thm:MultipleFixCounterEx}]
For every dimension $d\in\N$ we pick the probability vectors $\vec P_1$, $\vec P_2$ and the stochastic matrix $T$ as defined in \eqref{eq:DefP1Counter}, \eqref{eq:DefP2Counter} and \eqref{eq:TCounterEx}. Since $\vec P_1$ and $\vec P_2$ have orthogonal support we see
\begin{align}
\label{eq:P1P2Dist}
    \frac{1}{2}\left\|\vec P_1 -\vec P_2\right\|_1 = 1.
\end{align}
Moreover, from \eqref{eq:P2FixedCounter} and \eqref{eq:P1AprFixedCounter} we know 
\begin{align*}
  \frac{1}{2}\left\|T\vec P_1 - \vec P_1 \right\|_1 \le \frac{1}{2^d}, \quad\quad \frac{1}{2}\left\|T\vec P_2 - \vec P_2 \right\|_1 = 0. 
\end{align*}
Hence, $\vec P_1$, $\vec P_2$ and $T$ satisfy the assumptions of the theorem. To prove the final statement of the theorem \eqref{eq:FixedPointBadApproxCounter}, assume that $S$ is a stochastic matrix satisfying $\|S-T\|<\frac{1}{4}$. Then by Lemma~\ref{lem:SUniqueFixCounter}, we see that if $\vec Q_1, \vec Q_2$ are probability vectors and fixed points of $S,$ we necessarily have $ \vec Q \coloneqq\vec Q_1= \vec Q_2.$ Therefore, using \eqref{eq:P1P2Dist} we have
\begin{align*}
   1= \frac{1}{2}\left\|\vec P_1 -\vec P_2\right\|_1 \le \frac{1}{2}\left\|\vec P_1 -\vec Q\right\|_1 + \frac{1}{2}\left\|\vec Q -\vec P_2\right\|_1 
\end{align*}
and hence either 
\begin{align*}
   \frac{1}{2}\left\|\vec P_1 -\vec Q\right\|_1 \ge \frac{1}{2}\quad\text{or}\quad\left\|\vec P_2 -\vec Q\right\|_1 \ge \frac{1}{2}.
\end{align*}
\end{proof}

To prove Theorem~\ref{thm:MultipleFixCounterExQuant} for the quantum setting, we need to prove an analouge of the robustness result in Lemma~\ref{lem:SUniqueFixCounter} above. That is, embedding the stochastic matrix $T$ into the channel $\cN_T$ using the natural construction \eqref{eq:StochMatrixClassicalChannel}, also all quantum channels in a neighbourhood around $\cN_T$ have a unique fixed point state. Since the set of quantum channels around $\cN_T$ is obviously larger than the set of classical channels around $\cN_T$ (which corresponds to stochastic matrices around $T$) the desired robustness result in the quantum setting does not immediately follow from Lemma~\ref{lem:SUniqueFixCounter}. Therefore, we prove it independently in the next lemma. Here, the idea is that given a quantum channel $\cM$ close to $\cN_T,$ which has two fixed points, it necessarily needs to have two orthogonal invariant subspaces in the sense of \eqref{eq:InvarianteSubspaceFix}. But therefore also $\cN_T$ needs to leave named subspaces approximately invariant, which quickly leads to a contradiction given the precise form of the stochastic matrix $T.$ 

\begin{lemma}[Robustness of uniqueness of fixed point in quantum case]
\label{lem:RobUniqQuantCounter}
Let $\cM$ be a quantum channel on a $d$-dimensional Hilbert space $\cH$ satisfying
\begin{align}
\label{eq:QuantumToClassChanClosCounter}
    \frac{1}{2}\left\|\cM-\cN_T\right\|_{\diamond}\le \frac{1}{16d},
\end{align}
with $\cN_T$ being the channel defined trough \eqref{eq:StochMatrixClassicalChannel} and $T$ being the stochastic matrix defined in \eqref{eq:TCounterEx}.
Then $\cF(\cM)$ is one-dimensional, i.e.~$\cM$ has an unique fixed point state. 
\end{lemma}
\begin{proof}
Assume for contradiction $\dim(\cF(M))\ge2.$ Then let $A,B\in\cF(\cM)$ be non-zero with $A\neq B.$ Without loss of generality we can assume that $A$ and $B$ are both self-adjoint, as otherwise we could simply take $\Re(A)$ and $\Im(A)$ which, by \eqref{eq:ReIMPosNegFix}, are also fixed points. Moreover, taking hence $[A-B]_+$ and $[A-B]_-$ gives, again by \eqref{eq:ReIMPosNegFix}, two non-zero, positive semi-definite operators which are fixed points of $\cM.$ Therefore, by \eqref{eq:InvarianteSubspaceFix} we find two orthogonal invariant subspaces, i.e.~the orthogonal projections $\pi_1$ and $\pi_2$ onto the supports of $[A-B]_+$ and $[A-B]_-$ respectively satisfy 
\begin{align*}
   \Tr\left((\1-\pi_1)\cM(\pi_1)\right) = \Tr\left((\1-\pi_2)\cM(\pi_2)\right)=0
\end{align*}
and furthermore by definition $\pi_1\pi_2=0.$
Hence, by assumption \eqref{eq:QuantumToClassChanClosCounter} and the variational expression of the trace distance \eqref{eq:VarExpTraceDistance} we also have
 \begin{align}
 \label{eq:QuantumCounterContr}
     \nn\Tr\left((\1-\pi_1)\cN_T(\pi_1)\right) &= \Tr\left((\1-\pi_1)\left(\cN_T(\pi_1)-\cM(\pi_1)\right)\right) \\&\le 
     \frac{\Tr(\pi_1) }{2}\left\|\cN_T-\cM\right\|_\diamond \le \frac{1}{16}
 \end{align}
 and similarly
\begin{align}
 \label{eq:QuantumCounterContr2}
    \Tr\left((\1-\pi_2)\cN_T(\pi_2)\right)\le\frac{1}{16}.
\end{align}

Now note that this gives that either $\bra{d} \pi_1\ket{d}<\frac{1}{2}$ or $\bra{d} \pi_2\ket{d}< \frac{1}{2}$ as otherwise by definition of the stochastic matrix $T$ in \eqref{eq:TCounterEx}
\begin{align*}
     \Tr\left((\1-\pi_1)\cN_T(\pi_1)\right)\ge  \Tr\left(\pi_2\cN_T(\pi_1)\right) \ge T_{dd}\bra{d} \pi_2\ket{d}\bra{d} \pi_1\ket{d}\ge \frac{1}{4}.
\end{align*}
Without loss of generality we hence assume \begin{align}
\label{eq:dpi1Counter}
 \bra{d}\pi_1\ket{d}<1/2
\end{align}
This gives that there exists $i_*\in[d-1]$ such that \begin{align}
\label{eq:pi1i*CounterQuantum}
    \bra{i_*} \pi_1\ket{i_*}\ge \frac{1}{2}
\end{align} as otherwise $\bra{j}(\1- \pi_1)\ket{j}>\frac{1}{2}$ for all $j\in[d]$ and hence
\begin{align*}
     \Tr\left((\1-\pi_1)\cN_T(\pi_1)\right) \ge \sum_{j=1}^d T_{jj}\bra{j}(\1- \pi_1)\ket{j}\bra{j} \pi_1\ket{j} >\frac{1}{8}\sum_{j=1}^d\bra{j} \pi_1\ket{j} =\frac{\Tr(\pi_1)}{8},
\end{align*}
which is a contradiction to \eqref{eq:QuantumCounterContr} since $\pi_1\neq 0$. Note that since $\bra{i_*} \pi_1\ket{i_*}\ge 1/2$ also $\bra{i_*+1} \pi_1\ket{i_*+1}\ge 1/2$ as otherwise we have by definition of the stochastic matrix $T$ in \eqref{eq:TCounterEx} 
\begin{align*}
    \Tr\left((\1-\pi_1)\cN_T(\pi_1)\right) \ge T_{i_*+1\, i_*}\bra{i_*+1} (\1-\pi_1)\ket{i_*+1}\bra{i_*} \pi_1\ket{i_*} > \frac{1}{16},
\end{align*}
which is a contradiction to \eqref{eq:QuantumCounterContr}. However, by iterating this argument we see $\bra{j} \pi_1\ket{j}\ge 1/2$ for all $j\in\{i_*,\cdots,d\},$ which is a contradiction to \eqref{eq:dpi1Counter}. To conclude, we have shown that $\cM$ cannot have two orthogonal invariant subspaces, and therefore $\dim(\cF(\cM))=1$, as this would otherwise contradict assumption \eqref{eq:QuantumCounterContr}.

 \comment{We want to show that \eqref{eq:pi1i*CounterQuantum} implies $\bra{j} \pi_1\ket{j}\ge 1/2$ for all $j\in[d].$ For that note that since $\bra{i_*} \pi_1\ket{i_*}\ge 1/2$ also $\bra{i_*+1} \pi_1\ket{i_*+1}\ge 1/2$ as otherwise we have by definition of the stochastic matrix $T$ in \eqref{eq:TCounterEx} 
\begin{align*}
    \Tr\left((\1-\pi_1)\cN_T(\pi_1)\right) \ge T_{i_*+1\, i_*}\bra{i_*+1} (\1-\pi_1)\ket{i_*+1}\bra{i_*} \pi_1\ket{i_*} > \frac{1}{16},
\end{align*}
which is a contradiction to \eqref{eq:QuantumCounterContr}. Iterating this argument gives $\bra{j} \pi_1\ket{j}\ge 1/2$ for all $j\in\{i_*,\cdots,d\}.$ Moreover, if $i_*\ge 2$ we see by the same argument $\bra{i_*-1} \pi_1\ket{i_*-1}\ge 1/2$ as otherwise we have by definition of the stochastic matrix $T$ in \eqref{eq:TCounterEx} 
\begin{align*}
    \Tr\left((\1-\pi_1)\cN_T(\pi_1)\right) \ge T_{i_*-1\, i_*}\bra{i_*-1} (\1-\pi_1)\ket{i_*-1}\bra{i_*} \pi_1\ket{i_*} > \frac{1}{8},
\end{align*}
which is a contraction to \eqref{eq:QuantumCounterContr}. By iterating this argument again, we have therefore shown \begin{align*}
    \bra{j} \pi_1\ket{j}\ge 1/2
\end{align*} for all $j\in[d].$ However, again from the definition of the stochastic matrix $T$ in \eqref{eq:TCounterEx} this implies
\begin{align*}
    \Tr\left((\1-\pi_2)\cN_T(\pi_2)\right)\ge \Tr\left(\pi_1\cN_T(\pi_2)\right) \ge \sum_{j=1}^d T_{jj}\bra{j} \pi_1\ket{j}\bra{j} \pi_2\ket{j} \ge \frac{\Tr(\pi_2)}{8}
\end{align*}
which is a contraction to and \eqref{eq:QuantumCounterContr2} since $\pi_2\neq 0$. }

\end{proof}

\begin{proof}[Proof of Theorem~\ref{thm:MultipleFixCounterExQuant}]
The proof of Theorem~\ref{thm:MultipleFixCounterExQuant} follows from exactly the same lines as the one of Theorem~\ref{thm:MultipleFixCounterEx}, with the usage of Lemma~\ref{lem:RobUniqQuantCounter} instead of Lemma~\ref{lem:SUniqueFixCounter}. We still give it here explicitly for completeness.

We use the natural embedding \eqref{eq:ProbClassicalState} and pick for every dimension $d\in\N$ the states $\rho_1 \coloneqq \rho_{\vec P_1}$ and $\rho_2 \coloneqq \rho_{\vec P_2}$ with probability vectors $\vec P_1$ and $\vec P_2$ defined in  \eqref{eq:DefP1Counter} and \eqref{eq:DefP2Counter}. 
We again have by orthogonality
\begin{align}
\label{eq:P1P2DistQuant}
    \frac{1}{2}\left\|\rho_1 -\rho_2\right\|_1 = 1.
\end{align}

Furthermore, using the embedding \eqref{eq:StochMatrixClassicalChannel}, we pick as channel $\cN\coloneqq\cN_T$ with $T$ being the stochastic matrix defined in \eqref{eq:TCounterEx} and see from \eqref{eq:P2FixedCounter} and \eqref{eq:P1AprFixedCounter} that
\begin{align*}
  \frac{1}{2}\left\|\cN(\rho_1) - \rho_1 \right\|_1 =\frac{1}{2}\left\|T\vec P_1 - \vec P_1 \right\|_1 \le \frac{1}{2^d}, \quad\quad  \frac{1}{2}\left\|\cN(\rho_2) - \rho_2 \right\|_1= 0. 
\end{align*}
Hence, $\rho_1$, $\rho_2$ and $\cN$ satisfy the assumptions of the theorem. To prove the final statement of the theorem, i.e.~\eqref{eq:FixedPointBadApproxCounterQuantum}, assume that $\cM$ is a channel satisfying $\frac{1}{2}\|\cM-\cN\|\le\frac{1}{16d}$. Then by Lemma~\ref{lem:RobUniqQuantCounter}, we see that if $\sigma_1, \sigma_2$ are states and fixed points of $\cM,$ we necessarily have $ \sigma \coloneqq\sigma_1= \sigma_2.$ Therefore, using \eqref{eq:P1P2DistQuant} we have
\begin{align*}
   1= \frac{1}{2}\left\|\rho_1 -\rho_2\right\|_1 \le \frac{1}{2}\left\|\rho_1 -\sigma\right\|_1 + \frac{1}{2}\left\|\sigma -\rho_2\right\|_1 
\end{align*}
and hence either 
\begin{align*}
   \frac{1}{2}\left\|\rho_1 -\sigma\right\|_1 \ge \frac{1}{2}\quad\text{or}\quad\frac{1}{2}\left\|\rho_2 -\sigma\right\|_1 \ge \frac{1}{2}.
\end{align*}
\end{proof}

\section{Summary and open problems}
\label{sec:SumandOpen}
 We have defined the notion of fixability and rapid fixability of approximate fixed point equations of states and channels, which are, respectively, elements of the fixed sets $\cS\subseteq\mathfrak{S}(\cH)$ and $\cC\subseteq\CPTP(\cH)$ (Definition~\ref{def:FixingApproxFixPoints} and~\ref{def:RapidFixingApproxFixedPoint}). For finite dimensional Hilbert spaces and under the assumption that $\cS$ and $\cC$ are closed and contain a fixed point pair, we have shown that approximate fixed point equations are always fixable in the sense of Definition~\ref{def:FixingApproxFixPoints} (Proposition~\ref{prop:AbstractFixingApproxFix}). We have applied this result to show that for any approximate quantum Markov chain $\rho_{ABC}$ there exists an exact quantum Markov chain close to $\rho_{ABC}.$ Here, the provided upper bound on the distance between $\rho_{ABC}$ and $\sigma_{ABC}$ necessarily depends on the dimensions of the systems and decays for vanishing conditional mutual information $I(A:C|B)_\rho$ (Proposition~\ref{prop:RobustnessQMC}).
 
 Furthermore, we have established rapid fixability of approximate fixed point equations for the choices 
\begin{enumerate}
    \item $\cS=\mathfrak{S}(\cH)$ and $\cC={\rm{CPTP}}(\cH)$ (Theorem~\ref{thm:FixedPointQuantum}),
    \item $\cS={\rm{Prob}}(\cX)$ and $\cC={\rm{Stoch}}(\cX)$ (Theorem~\ref{thm:FixClassical}),
    \item $\cS=\mathfrak{S}(\cH)$ and $\cC$ \ being the set of unitary channels (Theorem~\ref{thm:FixUnitary}),
    \item $\cS=\mathfrak{S}(\cH)$ and $\cC$ \ being the set of mixed-unitary channels (Theorem~\ref{thm:FixingMixtureUnitaries}),
    \item $\cS=\mathfrak{S}(\cH)$ and $\cC$ \ being the set of unital channels (Theorem~\ref{thm:FixingUnitalChannels}),
    \item $\cS$ being the set of pure states on a bipartite Hilbert space $\cH_{AB}=\cH_A\otimes\cH_B$ and $\cC=\id_A\otimes{\rm{CPTP}}(\cH_B)$ (Theorem~\ref{thm:ApproxLocalFixPure}). 
\end{enumerate}
Lastly, in Corollary~\ref{cor:ImpossiLocalFix} we have shown that approximate fixed point equations are not rapidly fixable for
\begin{enumerate}
    \item $\cS=\mathfrak{S}(\cH_{AB})$ and $\cC=\id_A\otimes{\rm{CPTP}}(\cH_B),$
    \item $\cS={\rm{Prob}}(\cX\times\cY)$ and $\cC=\id_X\otimes{\rm{Stoch}}(\cY).$
\end{enumerate}
For the last result, we generalised the notion of rapid fixability to the situation of multiple approximate fixed point states of one specific channel.  In this context, we show that these multiple approximate fixed point equations are not rapidly fixable at once, by providing an explicit counterexample (Theorems~\ref{thm:MultipleFixCounterEx} and Theorem~\ref{thm:MultipleFixCounterExQuant}).

\bigskip

\noindent In the following we list some interesting open problems which can be subject to future research:

\noindent 

 It would be interesting to obtain optimal bounds on the approximation functions $f$ and $g$ 
introduced in Definitions~\ref{def:FixingApproxFixPoints} and~\ref{def:RapidFixingApproxFixedPoint} for different choices of $\cS$ and $\cC.$ Here, to establish optimality one needs to prove the corresponding lower bounds on the possible approximation functions.   For that, we have already seen in Remark~\ref{rem:LowerBoundApprox} that at least the scaling of $\max\{f,g\}$ for the cases $(\cS,\cC)=(\mathfrak{S}(\cH),\CPTP(\cH))$ and $(\cS,\cC)=(\Prob(\cX),\Stoch(\cX))$ given in Theorems~\ref{thm:FixedPointQuantum} and~\ref{thm:FixClassical} is optimal. However, providing optimal bounds on both $f$ and $g$ individually, including the correct constants is still open.
Furthermore, the constructions carried out for the above mentioned results on rapid fixability for all other choices of $\cS$ and $\cC$ led to bounds on the approximation functions which explicitly depend on the dimension of the underlying Hilbert space. It would be of great interest to determine when this dimension dependence is necessary. 
   
   Probably the most significant open problem, which lies outside the immediate scope of fixability of approximate fixed point equations, is to establish an improved version of Proposition~\ref{prop:RobustnessQMC}, with an explicit control on the robustness of quantum Markov chains. As outlined at the end of Section~\ref{sec:ClosedFixAbstract} (in particular around~\eqref{eq:QMCROBPLEASE}) it would be highly desirable to prove a bound for tripartite states $\rho_{ABC}$ of the form
    \begin{align}
    \min_{\sigma_{ABC}\  \text{QMC}}\,\frac{1}{2}\left\|\rho_{ABC} - \sigma_{ABC}\right\|_1 \stackrel{?}{\le} c\,d^{\,b_A\,}_Ad^{\,b_B\,}_Bd^{\,b_C}_C \,I(A:C|B)^a_\rho,
\end{align}
with constants $a>0$ and $b_A,b_B,b_C,c\ge 0$ independent of the dimensions $d_A,d_B,d_C$ and $\rho_{ABC}.$

\bigskip

\noindent\textbf{Acknowledgments.} The authors would like to thank  Guillaume Aubrun, Hamza Fawzi, Omar Fawzi, Niklas Galke, Lauritz van Luijk, Mizanur Rahaman, Satvik Singh, Mark M. Wilde and Michael M. Wolf for helpful discussions. RS acknowledges funding from the Cambridge Commonwealth,
European and International Trust and from the European Research Council (ERC Grant AlgoQIP, Agreement No. 851716).
BB acknowledges support from the UK Engineering and Physical Sciences Research Council (EPSRC)
under grant number EP/V52024X/1.

\bibliography{ref}

\newcommand{\etalchar}[1]{$^{#1}$}
\begin{thebibliography}{BaHOS15}

\bibitem[AF04]{alicki_continuity_2004}
R.~Alicki and M.~Fannes.
\newblock Continuity of quantum conditional information.
\newblock {\em Journal of Physics A: Mathematical and General}, 37(5):L55, 2004.

\bibitem[BaHOS15]{Brandao_CMI_2015}
Fernando G. S.~L. Brand\~ao, Aram~W. Harrow, Jonathan Oppenheim, and Sergii Strelchuk.
\newblock Quantum conditional mutual information, reconstructed states, and state redistribution.
\newblock {\em Phys. Rev. Lett.}, 115:050501, Jul 2015.

\bibitem[BCG{\etalchar{+}}13]{Burgarth_2013}
D~Burgarth, G~Chiribella, V~Giovannetti, P~Perinotti, and K~Yuasa.
\newblock Ergodic and mixing quantum channels in finite dimensions.
\newblock {\em New Journal of Physics}, 15(7):073045, jul 2013.

\bibitem[Bha97]{Bhatia_MatrixAnalysis_1997}
Rajendra Bhatia.
\newblock {\em Matrix Analysis}, volume 169.
\newblock Springer, 1997.

\bibitem[Car11]{caratheodory_uber_1911}
C.~Carath{\'e}odory.
\newblock {\"U}ber den variabilit{\"a}tsbereich der fourier'schen konstanten von positiven harmonischen funktionen.
\newblock {\em Rendiconti del Circolo Matematico di Palermo (1884-1940)}, 32(1):193--217, 1911.

\bibitem[CK11]{csiszar_korner_2011}
Imre Csiszár and János Körner.
\newblock {\em Information Theory: Coding Theorems for Discrete Memoryless Systems}.
\newblock Cambridge University Press, 2 edition, 2011.

\bibitem[CSW12]{christandl_entanglement_2012}
Matthias Christandl, Norbert Schuch, and Andreas Winter.
\newblock Entanglement of the antisymmetric state.
\newblock {\em Communications in Mathematical Physics}, 311(2):397--422, 2012.

\bibitem[FR15]{FawziRenner_CMIRecovery_2015}
Omar Fawzi and Renato Renner.
\newblock Quantum conditional mutual information and approximate markov chains.
\newblock {\em Communications in Mathematical Physics}, 340(2):575--611, 2015.

\bibitem[HHHO]{horodecki_information_2005}
Karol Horodecki, Micha{\l} Horodecki, Pawel Horodecki, and Jonathan Oppenheim.
\newblock Information theories with adversaries, intrinsic information, and entanglement.
\newblock 35(12):2027--2040.

\bibitem[HJ85]{HornJohnson_MatrixAnalysis(Book)_1985}
Roger~A. Horn and Charles~R. Johnson.
\newblock {\em Matrix Analysis}.
\newblock Cambridge University Press, 1985.

\bibitem[HJPW04]{hayden_structure_2004}
Patrick Hayden, Richard Jozsa, D{\'e}nes Petz, and Andreas Winter.
\newblock Structure of states which satisfy strong subadditivity of quantum entropy with equality.
\newblock {\em Communications in Mathematical Physics}, 246(2):359--374, 2004.

\bibitem[Hol01]{holevo_statistical_2001}
Alexander~S. Holevo.
\newblock {\em Statistical Structure of Quantum Theory}, volume~67 of {\em Lecture Notes in Physics Monographs}.
\newblock Springer, 2001.

\bibitem[ILW08]{IbinsonWinter_RobustnessQMC_2008}
Ben Ibinson, Noah Linden, and Andreas Winter.
\newblock Robustness of quantum markov chains.
\newblock {\em Communications in Mathematical Physics}, 277(2):289--304, 2008.

\bibitem[JRS{\etalchar{+}}18]{JungeSutterwilde_UniversalRecovery_2018}
Marius Junge, Renato Renner, David Sutter, Mark~M. Wilde, and Andreas Winter.
\newblock Annales henri poincar{\'e}.
\newblock 19(10):2955--2978, 2018.

\bibitem[KB19]{KatoBrandao_AQMCThermal_19}
Kohtaro Kato and Fernando G. S.~L. Brand{\~a}o.
\newblock Quantum approximate markov chains are thermal.
\newblock {\em Communications in Mathematical Physics}, 370(1):117--149, 2019.

\bibitem[KI02]{KoashiImoto}
Masato Koashi and Nobuyuki Imoto.
\newblock Operations that do not disturb partially known quantum states.
\newblock {\em Phys. Rev. A}, 66:022318, Aug 2002.

\bibitem[KW20]{khatri2020principles}
Sumeet Khatri and Mark~M. Wilde.
\newblock Principles of quantum communication theory: A modern approach.
\newblock arXiv:2011.04672, 2020.

\bibitem[LR73]{LiebRuskai_StrongSubadittivity_1973}
Elliott~H. Lieb and Mary~Beth Ruskai.
\newblock Proof of the strong subadditivity of quantum‐mechanical entropy.
\newblock {\em Journal of Mathematical Physics}, 14(12):1938--1941, 1973.
\newblock \_eprint: https://pubs.aip.org/aip/jmp/article-pdf/14/12/1938/8146113/1938\_1\_online.pdf.

\bibitem[MC76]{Morozova_StatofStoch_1976}
E.~A. Morozova and N.~N. Chentsov.
\newblock Stationary matrices of probabilities for stochastic supermatrix.
\newblock {\em Lecture Notes in Math., Vol. 550, Springer}, pages 379--418, 1976.

\bibitem[Pet03]{Petz_Recovery_2003}
D\'{e}nes Petz.
\newblock Monotonicity of quantum relative entropy revisited.
\newblock {\em Reviews in Mathematical Physics}, 15(01):79--91, 2003.

\bibitem[RS80]{ReedSimon_FunctionalAnalysis_1976}
Michael Reed and Barry Simon.
\newblock {\em Methods of Modern Mathematical Physics I: Functional Analysis}.
\newblock Academic Press, 1980.

\bibitem[Sal23]{Salzmann_PhDthesis_2023}
Robert Salzmann.
\newblock {\em Robustness of Fixed Points of Quantum Processes}.
\newblock PhD thesis, Apollo - University of Cambridge Repository, 2023.

\bibitem[SBT17]{Sutter_Multivariate_2017}
David Sutter, Mario Berta, and Marco Tomamichel.
\newblock Multivariate trace inequalities.
\newblock {\em Communications in Mathematical Physics}, 352(1):37--58, 2017.

\bibitem[SFR16]{Sutter_UniversalRecoveryCMI_2016}
David Sutter, Omar Fawzi, and Renato Renner.
\newblock Universal recovery map for approximate markov chains.
\newblock {\em Proceedings of the Royal Society A: Mathematical, Physical and Engineering Sciences}, 472(2186):20150623, 2016.

\bibitem[Sut18]{Sutter_approximateThesis_2018}
David Sutter.
\newblock Approximate quantum markov chains.
\newblock In {\em Approximate Quantum Markov Chains}, pages 75--100. Springer International Publishing, 2018.

\bibitem[Wat18]{watrous_2018}
John Watrous.
\newblock {\em The Theory of Quantum Information}.
\newblock Cambridge University Press, 2018.

\bibitem[Wil13]{wilde_2013}
Mark~M. Wilde.
\newblock {\em Quantum Information Theory}.
\newblock Cambridge University Press, 2013.

\bibitem[Wil15]{Wilde_RecoveryInterpolation_2015}
Mark~M. Wilde.
\newblock Recoverability in quantum information theory.
\newblock {\em Proceedings of the Royal Society A: Mathematical, Physical and Engineering Sciences}, 471(2182):20150338, 2015.

\bibitem[Wol12]{Wolf12}
Michael Wolf.
\newblock Quantum channels \& operations guided tour.
\newblock {\em Available at \url{https://mediatum.ub.tum.de/doc/1701036/1701036.pdf}}, 2012.

\end{thebibliography}
\bibliographystyle{alpha}

\end{document}